\documentclass[aps,prd,superscriptaddress,longbibliography,nofootinbib,preprintnumbers,preprint]{revtex4-2}
\usepackage[utf8]{inputenc}
\usepackage[T1]{fontenc}
\usepackage{lmodern} % readable font in pdf
\usepackage{microtype} % better font spacing
\usepackage{amsmath,amsfonts,amssymb,physics,subcaption}
\usepackage{mathtools}
\usepackage{setspace}
\usepackage{verbatim}
\usepackage{graphicx}
\usepackage[usenames,dvipsnames]{xcolor}
\usepackage[colorlinks=true, linkcolor=RoyalBlue, citecolor=RoyalBlue, urlcolor=RoyalBlue]{hyperref}
\usepackage{pifont}

\def\f{\frac}
\def\mc{\mathcal}

\def\v[#1]{\textbf{#1}}
\def\w[#1]{\widehat{#1}}
\def\vs[#1,#2]{\boldsymbol{{#1}_{#2}}}
\def\mes[#1]{d^{3}{#1}}
\def\del{\partial}
\def\pa{\partial}
\def\<{\langle}
\def\>{\rangle}
\def\vecs[#1,#2]{\boldsymbol{{#1}_{#2}}}

\newcommand{\be}{\begin{equation}}
\newcommand{\ee}{\end{equation}}
\newcommand{\bes}{\begin{subequations}}
\newcommand{\ees}{\end{subequations}}

\usepackage{scalerel}
\newcommand{\ost}[2]{\overset{\makebox[0pt]{$\scriptscriptstyle{\scaleto{\{\hspace{-0.06em}#2\hspace{-0.06em}\}}{6pt}}$}}{#1}{}}

\def\a{\alpha}
\def\b{\beta}
\def\d{\delta}
\def\D{\Delta}
\def\e{\epsilon}

\def\g{\gamma}
\def\G{\Gamma}
\def\k{\kappa}
\def\l{\lambda}

\def\m{\mu}
\def\n{\nu}
\def\N{\nabla}
\def\o{\omega}
\def\O{\Omega}

\def\s{\sigma}
\def\t{\tau}
\def\th{\theta}

\def\z{\zeta}

\begin{document}
	
\title{Hydrodynamics in the Carrollian regime}

\author{Kedar S.~Kolekar}
\email{kedarsk@mail.tsinghua.edu.cn}
\affiliation{Yau Mathematical Sciences Center, Tsinghua University, Beijing 100084, China.}

\author{Taniya Mandal}
\email{taniya.mandal@niser.ac.in}
\affiliation{School of Physical Sciences, National Institute of Science Education and Research, An OCC of Homi Bhabha National Institute, Bhubaneswar 752050, India.}

\author{Ashish Shukla}
\email{ashish.shukla@saha.ac.in}
\affiliation{Theory Division, Saha Institute of Nuclear Physics, 1/AF Bidhan Nagar, Kolkata 700064, India.}
\affiliation{Homi Bhabha National Institute, Anushakti Nagar, Mumbai 400094,
India.}
\affiliation{CPHT, CNRS, \'{E}cole polytechnique, Institut Polytechnique de Paris, 91120 Palaiseau, France.}

\author{Pushkar Soni}
\email{pushkars21@iitk.ac.in}
\affiliation{Indian Institute of Technology Kanpur, Kanpur 208016, India.}

\preprint{CPHT-RR061.082024}

\begin{abstract}
Carroll hydrodynamics arises in the $c\to 0$ limit of relativistic hydrodynamics. Instances of its relevance include the Bjorken and Gubser flow models of heavy-ion collisions, where the ultrarelativistic nature of the flow makes the physics effectively Carrollian. In this paper, we explore the structure of hydrodynamics in what can be termed as the \emph{Carrollian regime}, where instead of keeping only the leading terms in the $c\to 0$ limit of relativistic hydrodynamics, we perform a small-$c$ expansion and retain the subleading terms as well. We do so both for perfect fluids as well as viscous fluids incorporating first order derivative corrections. As apposite applications of the formalism, we utilize the subleading terms to compute modifications to the Bjorken and Gubser flow equations which bring in, in particular, dependence on rapidity.
\end{abstract}

%%%%%%%%%%
\maketitle
%%%%%%%%%%

\newpage
{\hypersetup{linkcolor=black}\tableofcontents}

%%%%%%%%%%%%%%%%%%%%%%%%%%%%%%%%%%%
\section{Introduction}
\label{sec:intro}
The physics of systems near thermal equilibrium is remarkably simple, when observed on macroscopic time and length scales. On these scales, it is the global symmetries of the system, and the associated conservation equations, that govern the relaxation processes. This effective low-energy long-wavelength description, which goes by the name of \emph{hydrodynamics}, has been immensely successful in investigating out-of-equilibrium phenomena, which can be notoriously difficult to study in full generality.

Depending upon the underlying symmetries, hydrodynamics can have various avatars. The one most familiar to us is the extensively studied Galilean hydrodynamics, applicable to ordinary fluids such as air and water, with the underlying symmetry generators satisfying the Galilean algebra \cite{Landau-Lifshitz}. On the other hand, for systems which respect Lorentz invariance i.e., their symmetries satisfy the Poincaré algebra, relativistic hydrodynamics gives the correct description of macroscopic out-of-equilibrium phenomena \cite{Landau-Lifshitz, Kovtun:2012rj, Jeon:2015dfa, Romatschke:2017ejr}. As one would expect, it is possible to arrive at the equations of Galilean hydrodynamics by taking a careful speed of light $c\to\infty$ limit of relativistic hydrodynamics \cite{Romatschke:2017ejr, Ciambelli:2018xat, Petkou:2022bmz}. 

A more exotic limit that can be taken starting from relativistic hydrodynamics is the Carroll limit, where opposite to the Galilean limit, one takes the speed of light $c\to 0$, under which the physics of the system acquires an ultralocal nature due to the closing up of local lightcones. The underlying symmetries  of such a system satisfy the Carroll algebra, which arises by taking the $c\to 0$ limit of the Poincaré algebra \cite{LevyLeblond, NDS}.\footnote{Also see \cite{Majumdar:2024rxg, Bagchi:2024epw} for unconventional contractions of the Poincaré algebra along null directions, leading to two copies of codimension-1 Carroll subalgebras.} The resulting hydrodynamic equations are said to govern the dynamics of Carroll fluids \cite{Ciambelli:2018xat, Petkou:2022bmz, Armas:2023dcz}. The subject has gained a significant traction of late because of the discovery of a mapping between boost-invariant models of relativistic hydrodynamics which are useful for modeling the spacetime evolution of the quark-gluon plasma (QGP) produced in heavy-ion collisions, such as the Bjorken flow \cite{Bjorken:1982qr} and Gubser flow models \cite{Gubser:2010ze, Gubser:2010ui}, and equations of Carroll hydrodynamics on degenerate Carroll manifolds with specific geometric properties \cite{Bagchi:2023ysc, Bagchi:2023rwd}. 
It should be emphasized that the speed of light in the ambient flat spacetime in which the ultrarelativistic QGP flows is not being taken to zero. The mapping works rather due to the fact that the strict $c\to 0$ limit that yields Carroll hydrodynamics practically lands one in the near-horizon region of the Milne patch in flat spacetime, where the ultrarelativistic fluid exists - see fig.~\ref{lightcone}. In particular, the small parameter $c$ in Carroll hydrodynamics should be thought of as measuring departures from ultrarelativistic nature on the QGP side, thus being proportional to $\mc{O}(1-\f{\rm v}{\hat{\rm c}})$, with ${\rm v}$ denoting the characteristic speed of the QGP flow, which is close to the speed of light $\hat{\rm c}$ in flat spacetime. This makes clear that computing subleading terms in $c\to 0$ on the Carroll side is akin to computing departures from an ultrarelativistic nature on the QGP side i.e.~incorporating corrections proportional to higher powers of $1-\f{\rm v}{\hat{c}}$.

The symmetries constituting the Carroll algebra have also been found to dictate the properties of several other systems. Of particular relevance is the emergence of Carrollian symmetries at future/past null infinity in an asymptotically flat spacetime, which play a vital role in the proper functioning of flat-space holography. To be more precise, the symmetries of asymptotically flat spacetimes constitute the Bondi-van der Burg-Metzner-Sachs (BMS) algebra \cite{Bondi:1962px,Sachs:1962zza}, which turns out to be isomorphic to the conformal Carroll algebra living on the null boundary \cite{Bagchi:2010zz, Duval:2014uva}. This isomorphism between the asymptotic symmetry algebra in the bulk with the symmetry algebra on the boundary is one of the key criteria for a holographic correspondence to exist in the first place, as does also happen in the more celebrated example of the anti de Sitter/conformal field theory (AdS/CFT) correspondence \cite{Maldacena:1997re}, where the symmetry algebra involved is the $\mathfrak{so}(d+1,2)$ algebra for $d$ boundary spatial dimensions. References delving into flat-space holography with emphasis on the Carrollian nature of the symmetries include \cite{Bagchi:2012cy, Barnich:2012aw, Barnich:2012xq, Bagchi:2012xr, Bagchi:2016bcd, Donnay:2022aba, Bagchi:2022emh, Donnay:2022wvx, Bagchi:2023fbj, Saha:2023hsl, Adami:2023fbm, Nguyen:2023vfz, Bagchi:2023cen}. Carroll symmetry is also relevant in cosmological settings \cite{deBoer:2021jej}, aspects of condensed matter systems \cite{Bidussi:2021nmp, Bagchi:2022eui, Kasikci:2023zdn}, string theory \cite{Bagchi:2013bga, Bagchi:2015nca, Bagchi:2020fpr, Bagchi:2024qsb}, and for the physics of black holes \cite{Penna:2018gfx, Donnay:2019jiz, Freidel:2022vjq, Redondo-Yuste:2022czg, Bagchi:2022iqb, Ecker:2023uwm, Bagchi:2023cfp, Aggarwal:2024yxy, Bagchi:2024rje, Tadros:2024bev}. The appearance of Carroll symmetry in such diverse setups is a testimony to its fundamental importance, and convinces one that the subject merits further exploration.\footnote{See \cite{Bagchi:2025vri} for a recent review of Carrollian physics.} 

In this paper, we explore hydrodynamics in the \emph{Carrollian regime}. The strict $c\to 0$ limit of relativistic hydrodynamics was the subject of discussion in \cite{Ciambelli:2018xat, Petkou:2022bmz, Bagchi:2023ysc, Bagchi:2023rwd}, which lead to the equations of Carroll hydrodynamics. At an operational level, the approach was to consider the constitutive relation for the relativistic energy-momentum tensor up to the desired order in derivatives in the hydrodynamic derivative expansion, subsequently imposing the $c\to 0$ limit on each of the terms present, and extracting the leading behaviour while dropping all subleading corrections. Our agenda in the present work is more general. We envision a physical situation where there can be an interesting interplay between the subleading terms in the $c\to 0$ limit at any particular order in the hydrodynamic derivative expansion, versus the leading term in the derivative expansion itself at the subsequent order. For instance, there might be a physical situation in which the subleading term that arises in the $c\to 0$ limit from the non-dissipative (ideal) part of the relativistic energy-momentum tensor is comparable, or even larger, than the leading term in the $c\to 0$ limit that follows from the first-order dissipative term in the energy-momentum tensor. If this be the case, then it is sensible to discard the dissipative corrections, while keeping the subleading terms in the $c\to 0$ limit from the non-dissipative part and obtain the correct dynamical equations for the fluid. Notice that we are departing from a strict $c\to 0$ limit, which would demand from us to keep only the leading term in the $c\to 0$ limit at each order in the derivative expansion, and are including the subleading terms as well, comparing their magnitudes to subsequent terms in the derivative expansion. Thus, the approach should be thought of as studying hydrodynamics in the Carrollian regime, where terms beyond the strict $c\to 0$ limit can play an important role, similar in spirit to the discussion in \cite{deBoer:2023fnj}.  

With the advent of the understanding gained in \cite{Bagchi:2023ysc, Bagchi:2023rwd}, it is now known that Carroll symmetries hold relevance for the physics of QGP produced in ultrarelativistic heavy-ion collisions. More specifically, the phenomenological assumption of boost-invariance along the beam axis that underpins the Bjorken and Gubser flow models for the QGP arises naturally in Carroll hydrodynamics without any particular assumption about the fluid velocity profile. However, demanding exact boost-invariance is quite restrictive, and is certainly an over-simplification as far as the dynamics of the QGP is concerned. In the present work, by computing the subleading terms for hydrodynamics in the Carrollian regime, and utilizing the geometric structure on the Carroll manifold that captures the mapping with Bjorken/Gubser flow, we are able to compute corrections to the hydrodynamic equations governing the spacetime evolution of the QGP which now capture departures from exact boost-invariance, along with relaxing other phenomenological constraints that are imposed by hand in these models. The equations thus obtained are expected to model heavy-ion collisions better than the Bjorken/Gubser flow models.    

This paper is organized as follows. In section \ref{sec:interplay}, we provide more details of the physical setup and explain the intricate interplay between the hydrodynamic derivative expansion and the subleading terms in the $c\to 0$ Carroll expansion. Section \ref{sec:PR} provides an overview of the Papapetrou-Randers (PR) parametrization for a pseudo-Riemannian manifold, as well as for the four-velocity profile of a relativistic fluid, which turns out to be suitable for imposing the $c\to 0$ limit. Section \ref{sec:ideal_fluid} then delves into computing the subleading corrections in the Carrollian regime for an ideal fluid, while section \ref{sec:viscous_fluid} reports the dissipative terms, focusing on the first order in derivatives. Building upon the results of the previous two sections, subsections \ref{sec:Bjorken} and \ref{sec:Gubser} then specialize to computing the corrections that follow from the newly computed subleading terms to the Bjorken and Gubser flow models. We conclude the paper with a discussion and an outlook towards future directions in section \ref{sec:discussion}. \ref{CBI-FE} discusses the invariance of the hydrodynamic equations for a Carroll fluid, as well as the invariance of their mapping to Bjorken or Gubser flow, under tangent space Carroll boosts in detail. \ref{sec:second_der} contains a discussion about the second order derivative corrections that arise for a conformal Carroll fluid.

\noindent \emph{\underline{Notation}:} We work with spacetime dimensionality $d+1$, except in section \ref{sec:Applications}, where $d=3$. The Greek indices $\m, \n, \ldots$ run over $d+1$ spacetime dimensions, while the Latin indices $i,j,\ldots$ run over the $d$ spatial directions. Spatial components of a quantity are collectively denoted by a vector sign e.g.~$\vec{x}$ stands for $x^i$.

%%%%%%%%%%%%%%%%%%%%%%%%%%%%%%%%%%
\section{Basic setup}
\label{sec:basic}
\subsection{The hydrodynamic derivative expansion vs.~the $c\to 0$ limit}
\label{sec:interplay}
The basic premise underlying hydrodynamics is the existence of a local expansion for the conserved currents of the system in terms of derivatives of the macroscopic hydrodynamic degrees of freedom, with successive terms in the expansion carrying more and more derivatives. For a system with Poincaré symmetry, the conserved current is the energy-momentum tensor $T^{\m\n}$,\footnote{Spacetime translation invariance of the Poincaré invariant system leads to the conservation of energy and momentum, $\del_\m T^{\m\n} = 0$, whilst invariance under rotations implies $T^{ij} =T^{ji}$, and invariance under Lorentz boosts leads to the energy current being equal to the momentum density, $T^{0i} = T^{i0}$.} while the hydrodynamic degrees of freedom comprise the fluid velocity $u^\m$ (normalized such that $u^\m u_\m = -c^2$) and the temperature $T$.\footnote{In general, the system can have additional conserved charges with associated conservation equations. In this paper, we restrict attention to the case where the only conserved quantities are energy and momentum.} The derivative expansion for the energy-momentum tensor has the schematic form
\be
\label{eq:stress_schematic_1}
T^{\m\n} =  T^{\m\n}_{(0)} + T^{\m\n}_{(1)} + T^{\m\n}_{(2)} + \ldots\, ,
\ee
with the subscript denoting the number of derivatives acting on hydrodynamic degrees of freedom in the corresponding term. Thus $T^{\m\n}_{(0)}$ denotes the ideal/perfect part with no derivatives and thereby corresponding to the system in equilibrium, $T^{\m\n}_{(1)}$ comprises the viscous terms involving for instance the shear and bulk viscosity of the fluid, and so on. Since the system is near equilibrium, implying that the hydrodynamic degrees of freedom vary slowly, one can expect the successive terms in the expansion eq.~\eqref{eq:stress_schematic_1} to become smaller, and hence less relevant. For bookkeeping purposes, we introduce a dimensionless parameter $\varepsilon$, with $0<\varepsilon \ll 1$, and re-express the derivative expansion for the energy-momentum tensor as
\be
\label{eq:stress_schematic_2}
T^{\m\n} =  T^{\m\n}_{(0)} + \varepsilon^2 T^{\m\n}_{(1)} + \varepsilon^4 T^{\m\n}_{(2)} + \ldots\, ,
\ee
making explicit the decreasing magnitude of successive terms in the derivative expansion.\footnote{The choice to use only even powers of $\varepsilon$ is purely aesthetic and facilitates comparison with the Carroll expansion later, which is performed in even powers of the speed of light $c$ in the limit $c\to 0$.} In microscopic terms, the cutoff above which the continuum approximation of hydrodynamics is reasonable is set by the mean free path $\l_{\rm mfp}$ of the system. As already mentioned, for the derivative expansion to make sense, the hydrodynamic degrees of freedom must vary slowly and on much longer length scales compared to the mean free path i.e., $\del_i u \ll u/\l_{\rm mfp}$. Denoting the length scale of variation by $L$, which can be taken to be comparable to the system size, this requirement can be rephrased in terms of the Knudsen number ${\rm Kn} \equiv \l_{\rm mfp}/L$, with the hydrodynamic derivative expansion sensible when ${\rm Kn} \ll 1$ \cite{Jeon:2015dfa, Romatschke:2017ejr}. One can think of the small parameter $\varepsilon$ in eq.~\eqref{eq:stress_schematic_2} as being of the order of the Knudsen number, $\varepsilon \simeq {\rm Kn}$. 

In the Carroll limit, the Lorentz boost invariance of relativistic hydrodynamics gets replaced by invariance under Carroll boosts. There are two equivalent ways of viewing the Carroll limit. The intrinsic perspective is to consider it as an ultralocal limit where lightcones close, and can be reached by taking the limit of vanishing speed of light i.e., $c\to 0$. On the other hand, the extrinsic perspective implies an ultrarelativistic explanation, where the velocity scales associated to the system become comparable to the speed of light. The equivalence between the two perspectives was established via the realization that ultrarelativistic boost-invariant models of fluid dynamics satisfy the same equations as does a Carroll fluid constructed using the $c\to 0$ limit of relativistic hydrodynamics on a suitable degenerate manifold \cite{Bagchi:2023ysc, Bagchi:2023rwd}.

To impose the Carroll limit on eq.~\eqref{eq:stress_schematic_2}, one should expand each of the terms present in powers of $c^2$ as $c\to 0$. This leads to
\be
\label{eq:stress_schematic_3}
T^{\m\n} =  \left(\ost{T}{0}^{\m\n}_{(0)} + c^2  \ost{T}{2}_{(0)}^{\m\n} + c^4  \ost{T}{4}^{\m\n}_{(0)}+ \ldots\right) + \varepsilon^2 \left(\ost{T}{0}^{\m\n}_{(1)} + c^2 \ost{T}{2}^{\m\n}_{(1)} + c^4 \ost{T}{4}^{\m\n}_{(1)} + \ldots \right) + \ldots\, ,
\ee
where we have expanded terms at the $i^{th}$ order in the derivative expansion for the energy-momentum tensor as $T^{\m\n}_{(i)} = \sum_{a=0}^{\infty} c^{2a} \, \ost{T}{2a}^{\m\n}_{(i)}$, with the overscript denoting the number of factors of $c$ in the coefficient of the term. 

Now, as discussed in the Introduction section \ref{sec:intro}, the strict Carroll limit corresponds to keeping only the leading term in $c\to 0$ limit at each order in the derivative expansion i.e.,  
\be 
\label{eq:stress_schematic_4}
T^{\m\n} \approx \ost{T}{0}^{\m\n} = \ost{T}{0}^{\m\n}_{(0)} + \varepsilon^2 \ost{T}{0}^{\m\n}_{(1)} + \varepsilon^4 \ost{T}{0}^{\m\n}_{(2)} + \ldots,
\ee 
which has been the approach for deriving equations of Carroll hydrodynamics for a dissipative Carroll fluid in previous works \cite{Ciambelli:2018xat, Petkou:2022bmz, Bagchi:2023ysc, Bagchi:2023rwd}. 
Our focus presently is to illustrate the plethora of possibilities that can arise from the double expansion in eq.~\eqref{eq:stress_schematic_3}. For instance, one can keep $\mathcal{O}(c^2)$ corrections at each order in the derivative expansion to get hydrodynamic equations which capture the leading departures from the strict Carroll limit. These subleading corrections when $c\to 0$ are captured by
\be 
\label{eq:stress_schematic_5}
\ost{T}{2}^{\m\n} = \ost{T}{2}^{\m\n}_{(0)} + \varepsilon^2 \ost{T}{2}^{\m\n}_{(1)} + \varepsilon^4 \ost{T}{2}^{\m\n}_{(2)} + \ldots.
\ee
The computation of the subleading terms appearing at $\mathcal{O}(c^2)$ is one of the major objectives of this paper. An immediate utility of computing these terms lies in the fact that they give rise to corrections in equations for the Bjorken and Gubser flow models, in particular bringing-in dependence on rapidity, as discussed in detail in sections \ref{sec:Bjorken} and \ref{sec:Gubser}. 

Another possible scenario is when the derivative corrections are small enough to be completely neglected. Here, one can approximate the double expansion of eq.~\eqref{eq:stress_schematic_3} with 
\be
\label{eq:stress_schematic_6}
T^{\m\n} \approx  \ost{T}{0}^{\m\n}_{(0)} + c^2  \ost{T}{2}_{(0)}^{\m\n} + c^4  \ost{T}{4}^{\m\n}_{(0)} + \ldots.
\ee
The hydrodynamic equations that will follow from this approximation will capture the dynamics of an ideal fluid in the Carrollian regime, where subleading corrections in the $c\to 0$ limit are non-negligible.  

One can also imagine a situation where the two expansions carry equal relevance, in which case the terms in eq.~\eqref{eq:stress_schematic_3} should be rearranged as
\be
\label{eq:stress_schematic_7}
T^{\m\n} = \ost{T}{0}^{\m\n}_{(0)} + \left(\varepsilon^2 \ost{T}{0}^{\m\n}_{(1)} + c^2 \ost{T}{2}^{\m\n}_{(0)}\right) +  \left(\varepsilon^4 \ost{T}{0}^{\m\n}_{(2)} + \varepsilon^2 c^2 \ost{T}{2}^{\m\n}_{(1)} + c^4 \ost{T}{4}^{\m\n}_{(0)}\right) + \ldots 
\ee
The hydrodynamic equations that follow from the expansion above will involve a non-trivial interplay between dissipative effects and physics in the Carrollian regime. We explore the structure of these terms in greater detail in the following sections.

\subsection{The Papapetrou-Randers parametrization}
\label{sec:PR}
Before we delve into the details of hydrodynamics in the Carrollian regime, it is imperative to begin with an appropriate parametrization for the pseudo-Riemannian manifold on which the original relativistic fluid exists, whose $c\to 0$ limit we are interested in computing. The parametrization should have various factors of $c$ manifest and should make the Carroll limit transparent. The Papapetrou-Randers (PR) parametrization (also referred alternatively to as the PR gauge) for a pseudo-Riemannian manifold serves this purpose well, in terms of which the background metric takes the form \cite{Ciambelli:2018xat, Petkou:2022bmz, Bagchi:2023ysc, Bagchi:2023rwd}
\begin{equation}
\label{PRmetric}
 {\rm d}s^2 = g_{\mu\nu}{\rm d}x^{\mu}{\rm d}x^{\nu} = -c^2 (\Omega \,{\rm dt} - b_i {\rm d}x^i)^2 + a_{ij} {\rm d} x^i {\rm d} x^j,
\end{equation}
implying that
\be
\begin{split}
 \label{PRmetric-components}
g_{\rm tt} &= -c^2 \Omega^2, \quad g_{{\rm t} i} = c^2 \Omega b_i, \quad g_{ij} = a_{ij}-c^2 b_i b_j,  \\
g^{\rm tt} &= -\frac{1}{c^2 \Omega^2} + \frac{b^2}{\Omega^2}, \quad g^{{\rm t}i} = \frac{b^i}{\Omega}, \quad g^{ij} = a^{ij}, \quad \sqrt{-g} = c\Omega\sqrt{a}\, .
\end{split}
\ee
In the PR parametrization, $\Omega, b_i$ and $a_{ij}$ are all functions of the coordinates $({\rm t}, \vec{x})$, with ${\vec{x}} \equiv x^i$. The indices on $b_i$ are raised using the inverse spatial metric $a^{ij}$ i.e., $b^i = a^{ij} b_j$, and $b^2 \equiv b^i b_i$. Also, $g \equiv {\rm det}(g_{\m\n})$ and $a \equiv {\rm det}(a_{ij})$. 

On taking the $c\to 0$ limit of the metric in eq.~\eqref{PRmetric}, one ends up with a degenerate manifold $\mathcal{C}$ called a Carrollian manifold, which has the structure of a fibre bundle, with time fibred over a base spatial manifold. The fibre bundle structure is mathematically described in terms of the degenerate metric $h_{\m\n}$ on $\mc{C}$, along with the kernel $k^\m$ of $h_{\m\n}$, i.e.~$h_{\m\n} k^\n = 0$. In terms of the coordinates $({\rm t}, \vec{x})$, one has the line element and the kernel
\be
\label{Carroll-structure}
{\rm d}\ell^2 = h_{\m\n} {\rm d}x^\m {\rm d}x^\n = a_{ij} {\rm d}x^i {\rm d}x^j \, , \quad k = \frac{1}{\Omega} \del_{\,\rm t}\, .
\ee
The dual form of the kernel is given by $\vartheta = \O \, {\rm dt} - b_i {\rm d}x^i$, with $k^\m \vartheta_\m = 1$. The quantity $b_i$ is referred to as the Ehresmann connection on the Carroll manifold $\mc{C}$. 
In this parametrization, a flat Carroll manifold corresponds to $\Omega = 1, a_{ij} = \d_{ij}, b_i = {\rm constant}$. Apart from spatial translations and rotations, the flat Carroll manifold admits supertranslations ${\rm t} \to {\rm t} + f({\rm t}, x^i)$ as isometries. Under the requirement that the isometries also preserve the connection and commute with $\del_{\, \rm t}$ the supertranslations reduce to Carroll boosts \cite{Penna:2018gfx}. These when combined with spatial translations and rotations generate the finite dimensional Carroll algebra, which can also be obtained from the $c\to 0$ contraction of the Poincaré algebra. 

Let us also introduce several quantities which will be useful for the subsequent discussion \cite{Ciambelli:2018xat, Petkou:2022bmz}.  
On the Carroll manifold eq.~\eqref{Carroll-structure}, one can construct objects that transform covariantly under the residual diffeomorphisms ${\rm t} \to {\rm t}'({\rm t},\vec{x}), {\vec{x}} \to {\vec{x}}{\,'}({\vec{x}})$ that preserve the PR gauge. For instance, the Carroll covariant temporal and spatial derivatives are given by
\begin{equation}
    \hat{\partial}_{\,\rm t} \equiv \frac{1}{\Omega}\partial_{\,\rm t}\, , \quad \hat{\partial}_i \equiv \partial_i + \frac{b_i}{\Omega}\partial_{\,\rm t}\, .
\end{equation}
One can also define temporal and spatial Levi-Civita-Carroll connections via
\begin{equation}
    \hat{\gamma}_{ij} \equiv \frac{1}{2\Omega}\partial_{\,\rm t} a_{ij}, \quad \hat{\gamma}^i_{jk} \equiv \frac{a^{il}}{2}(\hat{\partial}_j a_{kl} + \hat{\partial}_k a_{jl} - \hat{\partial}_l a_{jk}).
    \label{Carroll-Christoffel}
\end{equation}
The indices on $\hat{\gamma}_{ij}$ are also raised using $a^{ij}$, since it too transforms like a Carrollian tensor under the residual diffeomorphisms. One can utilize the above definitions for the Levi-Civita-Carroll connections to define temporal and spatial Levi-Civita-Carroll covariant derivatives, $\hat{\nabla}_{\rm t}$ and $\hat{\nabla}_i$, respectively, whose action on Carrollian vectors takes the form 
\be
\begin{split}
&\hat{\nabla}_{\rm t} V^j = \hat{\partial}_{\,\rm t} V^j + \hat{\gamma}^j_k V^k, \quad \hat{\nabla}_{\rm t} V_j = \hat{\partial}_{\,\rm t} V_j - \hat{\gamma}_j^k V_k, \\
& \hat{\nabla}_i V^j = \hat{\partial}_i V^j + \hat{\gamma}^j_{ik}V^k, \quad \hat{\nabla}_i V_j = \hat{\partial}_i V_j - \hat{\gamma}^k_{ij}V_k .
\end{split}
\ee
The Levi-Civita-Carroll connections defined above are metric compatible i.e., $\hat{\nabla}_{\rm t} a_{jk} = 0$, $\hat{\nabla}_i a_{jk} = 0$. Further, one can also define a Carrollian expansion $\theta$, a Carrollian acceleration $\varphi_i$ and an anti-symmetric Carrollian tensor $f_{ij}$  via\footnote{We use the convention $A_{(ab)} = \frac{1}{2}(A_{ab} + A_{ba})$ and $A_{[ab]} = \frac{1}{2}(A_{ab} - A_{ba})$.}
\begin{equation}
\theta \equiv \hat{\gamma}^i_i = \frac{1}{\Omega}\partial_{\,\rm t} \operatorname{log} \sqrt{a}, \quad \varphi_i \equiv \frac{1}{\Omega}(\partial_i \Omega+\partial_{\,\rm t} b_i), \quad f_{ij}\equiv 2(\partial_{[i}b_{j]}+b_{[i}\varphi_{j]}).
\label{SomeDefs}
\end{equation}

The temporal and spatial Levi-Civita-Carroll connections $\hat{\gamma}_{ij}$ and $\hat{\gamma}^i_{jk}$ appear naturally in the leading order terms in the  small-$c$ expansion of the Christoffel connection for the PR metric eq.~\eqref{PRmetric}. Explicitly, the components of the Christoffel connection 
\begin{equation}
    \Gamma^{\mu}_{\nu\rho} = \frac{g^{\mu\sigma}}{2}(\partial_{\nu}g_{\rho\sigma} + \partial_{\rho}g_{\nu\sigma} - \partial_{\sigma}g_{\nu\rho})
\end{equation}
for the PR metric eq.~\eqref{PRmetric} are
{\allowdisplaybreaks
\begin{align}
& \Gamma^{\rm t}_{\rm tt}=\hat{\partial}_{\,\rm t}\Omega+ c^2 \Omega \vec{b}\cdot\vec{\varphi}\, ,
\quad \Gamma^{\rm t}_{{\rm t} j}= \varphi_j-\hat{\nabla}_{\rm t} b_j+c^2 \left( -\frac{1}{2} f_{ij}b^i -\Vec{b}\cdot\Vec{\varphi} \, b_j\right), \nonumber \\
& \Gamma^{\rm t}_{ij} = \frac{1}{c^2}\frac{\hat{\gamma}_{ij}}{\Omega} - \frac{1}{\Omega}\Big(\hat{\nabla}_{(i}b_{j)}-b_{(i}\hat{\nabla}_{\rm t} b_{j)}+b_{(i}\varphi_{j)}+b_{(i}\hat{\gamma}^l_{j)}b_l\Big)+c^2\frac{b^k}{\Omega}(\varphi_k b_ib_j-b_{(i}f_{j)k}), \nonumber \\
& \Gamma^k_{\rm tt}=c^2 \Omega^2 \varphi^k, \quad \Gamma^k_{{\rm t} j}=\Omega \hat{\gamma}^k_j+c^2\left(\frac{\Omega}{2}f^{~k}_j-\Omega b_j\varphi^k\right), \nonumber \\
&\Gamma^k_{ij}=\hat{\gamma}^k_{ij}-2\hat{\gamma}^k_{(i}b_{j)}+c^2 \Big(\varphi^k b_ib_j-b_{(i}f_{j)}^{~k}\Big), 
\label{PR-Christoffel}
\end{align}}
where $\vec{b}\cdot\vec{\varphi} = a^{ij}b_i\varphi_j$. The expressions above are exact and not truncated in the small-$c$ expansion.

It is also useful to introduce a parametrization for the fluid four-velocity $u^\m$, which satisfies the normalization $u^\m u_\m = -c^2$, in terms of the geometric data entering the PR parametrization. Defining $u\equiv \g\del_{\,\rm t}+\g v^i\del_i$, one can parametrize $\g, v^i$ in terms of a vector field $\b^i({\rm t},\vec{x})$ via
\be
\label{eq:velo_para}
 \gamma= \frac{1+ c^2 \vec{\beta}\cdot\vec{b}}{\Omega\sqrt{1-c^2 \beta^2}}\, , \quad v^i=\frac{c^2 \Omega \beta^i}{1+c^2\vec{\beta}\cdot\vec{b}}\, .
\ee
It turns out that under the residual diffeomorphisms $\b^i$ transforms like a Carrollian vector field, whose indices can thus be lowered using the spatial metric $a_{ij}$ i.e., $\b_i = a_{ij} \b^j$, as well as $\b^2 \equiv \b^i \b_i$, $\vec{\b}\cdot\vec{b} = \b^i b_i = \b_i b^i$. More explicitly, the components of $u^{\mu}$ are
\begin{equation}
u^{\rm t} = \frac{1+ c^2 \vec{\beta}\cdot\vec{b}}{\Omega\sqrt{1-c^2 \beta^2}}\, , \quad u_{\rm t} = -\frac{c^2\Omega}{\sqrt{1-c^2 \beta^2}}\, , \quad u^i = \frac{c^2 \beta^i}{\sqrt{1-c^2 \beta^2}}\, , \quad u_i = \frac{c^2 (b_i+\beta_i)}{\sqrt{1-c^2 \beta^2}}\, . 
\label{PR-fluid-velocity-components}
\end{equation}
With this choice of parametrization it is straight forward to take the Carroll limit $c\to 0$, on which we focus our attention now.

%%%%%%%%%%%%%%%%%%%%%%%%%%%%%%%%%%
\section{Perfect fluids in the Carrollian regime}
\label{sec:ideal_fluid}
We begin our discussion by first looking at an ideal/perfect fluid in the Carrollian regime. The energy-momentum tensor for a relativistic perfect fluid has the form
\be
T^{\m\n}_{(0)} = \left(\e+P\right) \frac{u^\m u^\n}{c^2} + Pg^{\m\n},
\label{TPerfect}
\ee
where $\e, P$ respectively denote the energy density and pressure of the fluid, while the subscript $``0"$ denotes the absence of any derivative corrections, following the notation introduced in section \ref{sec:interplay}. We take the following ansatz for the $c\to 0$ limit of the thermodynamic quantities \cite{Ciambelli:2018xat},\footnote{For the thermodynamic quantities $\e, P$, the subscript denotes the corresponding power of $c$ associated with the term in the $c\to 0$ expansion. This is unlike the case for $T^{\m\n}$ as a whole, where the subscript denotes the number of derivatives while the power of $c$ in the coefficient is denoted by an overscript.}
\be
\e = \e_{(0)} + c^2 \e_{(2)} + c^4 \e_{(4)} + \mc{O}(c^6)\, ,\quad P = p_{(0)} + c^2 p_{(2)} + c^4 p_{(4)} + \mc{O}(c^6)\, .
\label{ePexps}
\ee
On the other hand, the $c\to 0$ behaviour of $g_{\m\n}, u^\m$ entering eq.~\eqref{TPerfect} can be extracted to the desired order using the PR parametrization introduced in section \ref{sec:PR}. For instance, the components of the fluid velocity have the expansion
\begin{equation}
\begin{split}
&u^{\rm t} = \frac{1}{\Omega}(1+c^2\vec{\b}\cdot\vec{b}) \mc{F}, \hspace{14mm} u^i = c^2 \beta^i \mc{F}, \\
&u_{\rm t} = -c^2\Omega \mc{F}, \hspace{29mm} u_i = c^2 (b_i+\b_i)\mc{F}, 
\label{4-velocity-series}
\end{split}
\end{equation}
where $\mc{F} = \sum_{n=0}^{\infty}C_n\beta^{2n} c^{2n}$ and $C_n \equiv \frac{(2n-1)!!}{2^nn!}$.\footnote{The double factorial is defined as follows. For a positive even integer $n$, $n!! = n(n-2)(n-4)\ldots 4\cdot 2$, while for a positive odd integer $n$, $n!! = n(n-2)(n-4)\ldots 3\cdot 1$. The double factorial can be extended to negative integers by using the relation $n!!=\frac{(n+2)!!}{n+2}$, leading to $(-1)!! = 1, (-3)!! = -1$, and so on.}  With these ingredients put in, the perfect fluid energy-momentum tensor in the $c\to 0$ limit has the expansion
\begin{equation}
    T^{\mu\nu}_{(0)} = \ost{T}{0}^{\m\n}_{(0)} + c^2  \ost{T}{2}_{(0)}^{\m\n} + c^4  \ost{T}{4}^{\m\n}_{(0)} + {\cal O}(c^6), 
    \label{perf-EMT-expansion}
\end{equation}
where we have once again borrowed notation from section \ref{sec:interplay}, with the overscript denoting the power of $c$ associated with the term in the $c\to 0$ expansion.    
More explicitly, at the leading order we have\footnote{Note that the vanishing of the energy current $\ost{T}{0}_{(0)}\mbox{}^{\!i}_{\;\rm t}$ in eq.~\eqref{LOT1} i.e., in the strict Carroll limit, is a consequence of the Carroll boost Ward identity.}
\be
\ost{T}{0}_{(0)}\mbox{}^{\!\rm t}_{\;\rm t} = -\e_{(0)}\, ,\quad \ost{T}{0}_{(0)}\mbox{}^{\!i}_{\,j} = p_{(0)} \d^i_{\, j}\, , \quad\ost{T}{0}_{(0)}\mbox{}^{\!\rm t}_{\;i} = \f{b_i+\b_i}{\O} \left(\e_{(0)} + p_{(0)}\right)\, , \quad \ost{T}{0}_{(0)}\mbox{}^{\!i}_{\;\rm t} = 0 \, ,
\label{LOT1}
\ee
while at the next to leading order we have
\begin{subequations}
\label{NLOT1}
\begin{align}
\ost{T}{2}_{(0)}\mbox{}^{\!\rm t}_{\;\rm t} &= - \epsilon_{(2)} - \left(\epsilon_{(0)} + p_{(0)}\right)\big(\beta^2 + \vec{\beta}\cdot\vec{b}\big), \\
\ost{T}{2}_{(0)}\mbox{}^{\!i}_{\,j} &= p_{(2)}\delta^i_j + \beta^i(b_j + \beta_j)\left(\epsilon_{(0)} + p_{(0)}\right), \\
\ost{T}{2}_{(0)}\mbox{}^{\!\rm t}_{\;i} &= \frac{(b_i + \beta_i)}{\Omega}\left[\left(\epsilon_{(2)} + p_{(2)}\right) + \left(\epsilon_{(0)} + p_{(0)}\right)\big(\beta^2 + \vec{\beta}\cdot\vec{b}\big)\right], \\
\ost{T}{2}_{(0)}\mbox{}^{\!i}_{\;\rm t} &= -\Omega\beta^i\left(\epsilon_{(0)} + p_{(0)}\right).
\end{align}
\end{subequations}
The leading order terms for a perfect Carroll fluid, eq.~\eqref{LOT1}, have been the subject of study in earlier works \cite{Ciambelli:2018xat, Petkou:2022bmz, Bagchi:2023rwd, Bagchi:2023ysc}, whereas the subleading terms given in eq.~\eqref{NLOT1} for a perfect fluid in the Carrollian regime are entirely new. One can in principle compute even higher order terms in the expansion eq.~\eqref{perf-EMT-expansion} depending upon the sensitivity required. 

Let us now look at the hydrodynamic equations, which for an uncharged relativistic fluid are simply given by the conservation equation for the energy-momentum tensor,
\be
\N_\m T^{\m}_{\;\;\n} = 0.
\label{Hydro1}
\ee
Using the $c\to 0$ expansion for the perfect fluid energy-momentum tensor, eqs.~\eqref{LOT1} and \eqref{NLOT1}, along with the $c\to 0$ expansion of the Christoffel symbols given in PR parametrization in eq.~\eqref{PR-Christoffel}, we can now expand the LHS of eq.~\eqref{Hydro1} order-by-order in small $c$. To do so, let us first write the expansion of the Christoffel symbols in eq.~\eqref{PR-Christoffel} as
\be
\G^\m_{\n\rho} = \f{1}{c^2}\, \ost{\G}{-2}^\m_{\n\rho} + \ost{\G}{0}^\m_{\n\rho} + c^2 \ost{\G}{2}^\m_{\n\rho},
\label{PR-Christ-exp}
\ee
with $\ost{\G}{n}^\m_{\n\rho} = 0$ when $n > 2$. Making use of eqs.~\eqref{perf-EMT-expansion} and \eqref{PR-Christ-exp}, we can express the $c\to 0$ expansion of the LHS of the hydrodynamic equations \eqref{Hydro1} for a perfect fluid i.e.~$\nabla_{\mu}T_{(0)}\mbox{}^{\!\mu}_{\,\nu}$ as
\begin{equation}
    \nabla_{\mu}T_{(0)}\mbox{}^{\!\mu}_{\,\nu} = \frac{1}{c^2}\big(\nabla_{\mu}T_{(0)}\mbox{}^{\!\mu}_{\,\nu}\big)^{\{-2\}} + \big(\nabla_{\mu}T_{(0)}\mbox{}^{\!\mu}_{\,\nu}\big)^{\{0\}} + c^2 \big(\nabla_{\mu}T_{(0)}\mbox{}^{\!\mu}_{\,\nu}\big)^{\{2\}} + {\cal O}(c^4),
    \label{Hydro2}
\end{equation}
where
\begin{align}
\big(\nabla_{\mu}T_{(0)}\mbox{}^{\!\mu}_{\,\nu}\big)^{\{-2\}} &= \ost{\G}{-2}^\m_{\m\rho} \ost{T}{0}_{(0)}\mbox{}^{\!\rho}_{\,\n} - \ost{\G}{-2}^\rho_{\m\n} \ost{T}{0}_{(0)}\mbox{}^{\!\m}_{\,\rho} \, ,\nonumber\\
\big(\nabla_{\mu}T_{(0)}\mbox{}^{\!\mu}_{\,\nu}\big)^{\{0\}} &= \del_\m\ost{T}{0}_{(0)}\mbox{}^{\!\m}_{\,\n} + \ost{\G}{-2}^\m_{\m\rho} \ost{T}{2}_{(0)}\mbox{}^{\!\rho}_{\,\n} + \ost{\G}{0}^\m_{\m\rho} \ost{T}{0}_{(0)}\mbox{}^{\!\rho}_{\,\n} - \ost{\G}{-2}^\rho_{\m\n} \ost{T}{2}_{(0)}\mbox{}^{\!\m}_{\,\rho} - \ost{\G}{0}^\rho_{\m\n} \ost{T}{0}_{(0)}\mbox{}^{\!\m}_{\,\rho}\, ,\\
\big(\nabla_{\mu}T_{(0)}\mbox{}^{\!\mu}_{\,\nu}\big)^{\{2\}} &= \del_\m\ost{T}{2}_{(0)}\mbox{}^{\!\m}_{\,\n} + \ost{\G}{-2}^\m_{\m\rho} \ost{T}{4}_{(0)}\mbox{}^{\!\rho}_{\,\n} + \ost{\G}{0}^\m_{\m\rho} \ost{T}{2}_{(0)}\mbox{}^{\!\rho}_{\,\n} + \ost{\G}{2}^\m_{\m\rho} \ost{T}{0}_{(0)}\mbox{}^{\!\rho}_{\,\n} - \ost{\G}{-2}^\rho_{\m\n} \ost{T}{4}_{(0)}\mbox{}^{\!\m}_{\,\rho} \nonumber\\
&\,\quad- \ost{\G}{0}^\rho_{\m\n} \ost{T}{2}_{(0)}\mbox{}^{\!\m}_{\,\rho} - \ost{\G}{2}^\rho_{\m\n} \ost{T}{0}_{(0)}\mbox{}^{\!\m}_{\,\rho}\, . \nonumber
\end{align}
In the Carrollian regime, meeting the hydrodynamic equations \eqref{Hydro1} amounts to each of the terms in the expansion eq.~\eqref{Hydro2} vanishing individually. From eq.~\eqref{PR-Christoffel}, we see that only $\ost{\G}{-2}^{\rm t}_{ij} \neq 0$ while the other components of $\ost{\G}{-2}^\m_{\n\rho}$ vanish, leading to $\ost{\G}{-2}^\m_{\m\rho} = 0$. This, along with $\ost{T}{0}_{(0)}\mbox{}^{\!i}_{\;\rm t} = 0$,\footnote{In later sections, we will see that $\ost{T}{0}_{(n)}\mbox{}^{\!i}_{\;\rm t} = 0$ holds true for $n>0$ as well i.e.~the energy current in the strict Carroll limit will continue to vanish even after including dissipative corrections as a consequence of the Carroll boost Ward identity.} leads to $\big(\nabla_{\mu}T_{(0)}\mbox{}^{\!\mu}_{\,\nu}\big)^{\{-2\}} = 0$ being trivially satisfied, while the ${\cal O}(c^0)$ and ${\cal O}(c^2)$ energy-momentum conservation equations reduce respectively to
\begin{equation}
\big(\nabla_{\mu}T_{(0)}\mbox{}^{\!\mu}_{\,\nu}\big)^{\{0\}} = 0 \Rightarrow \del_\m\ost{T}{0}_{(0)}\mbox{}^{\!\m}_{\,\n} + \ost{\G}{0}^\m_{\m\rho} \ost{T}{0}_{(0)}\mbox{}^{\!\rho}_{\,\n} - \ost{\G}{-2}^\rho_{\m\n} \ost{T}{2}_{(0)}\mbox{}^{\!\m}_{\,\rho} - \ost{\G}{0}^\rho_{\m\n} \ost{T}{0}_{(0)}\mbox{}^{\!\m}_{\,\rho} = 0 \, ,
\label{Hydro3}
\end{equation}
and
\begin{align}
&\big(\nabla_{\mu}T_{(0)}\mbox{}^{\!\mu}_{\,\nu}\big)^{\{2\}} = 0 \nonumber\\
&\Rightarrow \del_\m\ost{T}{2}_{(0)}\mbox{}^{\!\m}_{\,\n} + \ost{\G}{0}^\m_{\m\rho} \ost{T}{2}_{(0)}\mbox{}^{\!\rho}_{\,\n} + \ost{\G}{2}^\m_{\m\rho} \ost{T}{0}_{(0)}\mbox{}^{\!\rho}_{\,\n} - \ost{\G}{-2}^\rho_{\m\n} \ost{T}{4}_{(0)}\mbox{}^{\!\m}_{\,\rho} - \ost{\G}{0}^\rho_{\m\n} \ost{T}{2}_{(0)}\mbox{}^{\!\m}_{\,\rho} - \ost{\G}{2}^\rho_{\m\n} \ost{T}{0}_{(0)}\mbox{}^{\!\m}_{\,\rho} = 0 \, .
\label{Hydro4}
\end{align}
For $\nu = \rm{t}$, the conservation equations $\big(\nabla_{\mu}T_{(0)}\mbox{}^{\!\mu}_{\;\rm t}\big)^{\{0\}} = 0$ and $\big(\nabla_{\mu}T_{(0)}\mbox{}^{\!\mu}_{\;\rm t}\big)^{\{2\}} = 0$ give the \emph{energy equations} or the \emph{continuity equations} at the leading order (LO) and the next-to-leading order (NLO) in the $c\to 0$ expansion, respectively. The $\nu = i$ components of the conservation equations i.e.~$\big(\nabla_{\mu}T_{(0)}\mbox{}^{\!\mu}_{\,i}\big)^{\{0\}} = 0$ and $\big(\nabla_{\mu}T_{(0)}\mbox{}^{\!\mu}_{\,i}\big)^{\{2\}} = 0$ contain, respectively, the LO and NLO continuity equations in addition to the LO and NLO \emph{momentum equations}. One can remove the contribution from the continuity equations to get the independent LO and NLO momentum equations. To be precise, for the relativistic metric in the PR parametrization eq.~\eqref{PRmetric-components}, we can write\footnote{The following argument is generically true at each order in the derivative expansion for the energy-momentum tensor as well as for each order in the $c\to 0$ expansion. We have therefore suppressed the over-/sub-scripts on $T^{\m\n}$ here.}
\begin{equation}
    \nabla_{\mu}T^{\mu i} = \nabla_{\mu}(T^{\mu}_{\;\;\nu}g^{\nu i}) = g^{\nu i}\nabla_{\mu}T^{\mu}_{~\nu} = g^{{\rm t}i}\nabla_{\mu}T^{\mu}_{~~\rm t} + g^{ji}\nabla_{\mu}T^{\mu}_{~j} = \frac{b^i}{\Omega}\nabla_{\mu}T^{\mu}_{~~\rm t} + a^{ji}\nabla_{\mu}T^{\mu}_{~j},
\end{equation}
which gives
\be
a_{ik}\nabla_{\mu}T^{\mu k} = \frac{b_i}{\Omega}\nabla_{\mu}T^{\mu}_{~~\rm t} + \nabla_{\mu}T^{\mu}_{\;\;i}\, ,
\ee
implying that
\be
\nabla_{\mu}T^{\mu}_{\;\;i} = a_{ik}\nabla_{\mu}T^{\mu k} - \frac{b_i}{\Omega}\nabla_{\mu}T^{\mu}_{~~\rm t}.
\ee 
Thus, as mentioned, $\N_\m T^\m_{\;\;i} = 0$ can be simplified by using the continuity equation $\N_\m T^\m_{~~\rm t} = 0$, leaving behind the momentum equations $a_{ik}\nabla_{\mu}T^{\mu k} = 0$.

Let us now state the perfect fluid LO and NLO hydrodynamic equations in the Carrollian regime explicitly. The LO equations \eqref{Hydro3} correspond to the strict Carroll limit, and take the form
\begin{subequations}
\label{LOPF}
\begin{align}
\hat{\partial}_{\,\rm t} \epsilon_{(0)} &= - \theta(\epsilon_{(0)} + p_{(0)}), \label{LOPF1} \\
\hat{\partial}_i p_{(0)} &= - \varphi_i(\epsilon_{(0)} + p_{(0)}) - (\hat{\partial}_{\,\rm t} + \theta)\big((\epsilon_{(0)} + p_{(0)})\beta_i\big),\label{LOPF2}
\end{align}
\end{subequations}
with the Carrollian expansion $\th$ and acceleration $\varphi_i$ defined in eq.~\eqref{SomeDefs}. These equations have been the subject of close examination in \cite{Bagchi:2023ysc, Bagchi:2023rwd}, wherein it was identified that the hydrodynamic equations governing boost-invariant flows in models of heavy-ion collisions and quark-gluon plasma viz.~Bjorken flow \cite{Bjorken:1982qr} and Gubser flow \cite{Gubser:2010ui, Gubser:2010ze}, are in fact equations for Carroll hydrodynamics, with appropriately chosen geometric data $(\O, b_i, a_{ij})$ on the Carroll manifold (more on this in sec.~\ref{sec:Applications}). Further, as discussed in \ref{CBI-FE}, these equations respect Carroll boost invariance in the tangent space.

The NLO hydrodynamic equations \eqref{Hydro4} for a perfect fluid in the Carrollian regime are
\begin{subequations}
\label{NLOPF}
\begin{align}
\hat{\partial}_{\,\rm t} \epsilon_{(2)} &= - \theta(\epsilon_{(2)} + p_{(2)}) - (\hat{\partial}_{\,\rm t} + \theta)\big[(\epsilon_{(0)} + p_{(0)})\beta^2\big] - \hat{\nabla}_i\big[(\epsilon_{(0)} + p_{(0)})\beta^i\big]\nonumber\\
&\quad- (\e_{(0)} + p_{(0)}) (2\vec{\b}\cdot\vec{\varphi} + \hat{\g}_{ij} \b^i \b^j), 
\label{NLOPF1} \\
\hat{\partial}_i p_{(2)} &= -\varphi_i(\epsilon_{(2)} + p_{(2)}) - (\hat{\partial}_{\,\rm t} + \theta)\big((\epsilon_{(2)} + p_{(2)})\beta_i\big) - (\hat{\nabla}_j + \varphi_j)\big[(\epsilon_{(0)} + p_{(0)})\beta^j\beta_i\big]\nonumber \\
&\quad- (\hat{\partial}_{\,\rm t} + \theta)\big[(\epsilon_{(0)} + p_{(0)})\beta^2\beta_i\big] - (\epsilon_{(0)} + p_{(0)})(\varphi_i\beta^2 + \beta^j f_{ji}).
\label{NLOPF2}
\end{align}
\end{subequations}
As is evident from the equations above, to obtain the behaviour of $\e_{(2)}, p_{(2)}$ in the Carrollian regime, we will need to first solve for the strictly Carrollian quantities $\e_{(0)}, p_{(0)}$ using eq.~\eqref{LOPF} and use their solutions in eq.~\eqref{NLOPF}. In fact, making use of eq.~\eqref{LOPF2} in eqs.~\eqref{NLOPF}, we can write down the NLO equations in a slightly simpler form as
\begin{subequations}
\label{NLOPFF}
\begin{align}
\hat{\partial}_{\,\rm t} \epsilon_{(2)} &= - \theta(\epsilon_{(2)} + p_{(2)}) + (\epsilon_{(0)} + p_{(0)})\Big(\frac{\theta\beta^2}{2} - \tilde{\th}\Big) - \beta^i\hat{\partial}_i\epsilon_{(0)}, \label{NLOPF3} \\
\hat{\partial}_i p_{(2)} &= -\varphi_i(\epsilon_{(2)} + p_{(2)}) - (\hat{\partial}_{\,\rm t} + \theta)\big((\epsilon_{(2)} + p_{(2)})\beta_i\big) - \beta_i\beta^j\hat{\partial}_j\epsilon_{(0)} \nonumber \\
&\quad- (\epsilon_{(0)} + p_{(0)})\Big[\hat{\nabla}_j(\beta^j\beta_i) + \beta^2\hat{\partial}_{\,\rm t}\beta_i + \beta_i \beta_j \hat{\partial}_{\,\rm t}\beta^j + \varphi_i\beta^2 + \beta^j f_{ji}\Big]. \label{NLOPF4}
\end{align}
\end{subequations}
Here $\tilde{\th}$ is the $c^2$ term in the $c\to 0$ expansion of the relativistic expansion parameter $\Theta \equiv \nabla_{\mu}u^{\mu}$, i.e.~$\Theta = \th + c^2 \tilde{\th} + \mc{O}(c^4)$, and is given by
\begin{equation}
\tilde{\th} = \hat{\nabla}_i\beta^i + \vec{\beta}\cdot\vec{\varphi} + \big(\hat{\partial}_{\,\rm t} + \theta\big)\frac{\beta^2}{2}.
\label{tiltheta}
\end{equation}
The NLO eqs.~\eqref{NLOPFF} are some of the key results of this paper. Following the procedure outlined above, one can in principle compute the hydrodynamic equations satisfied by higher order thermodynamic quantities $\e_{(2n)},p_{(2n)}$, i.e.~$(\nabla_\m T_{(0)}\mbox{}^{\!\m}_{\,\n})^{\{2n\}} = 0$ as well. It turns out that there is a compact way to represent these higher order equations, which we now illustrate.

\subsection{Perfect fluid equations to all orders in the $c\to 0$ expansion}
\label{sec:all_orders}
Consider first the continuity equation $\nabla_\m T_{(0)}\mbox{}^{\!\m}_{\;\rm t} = 0$ for the perfect fluid energy-momentum tensor eq.~\eqref{TPerfect}. Then, by only expanding the fluid velocity eq.~\eqref{4-velocity-series} and the Christoffel symbols eq.~\eqref{PR-Christoffel} in the $c\to 0$ limit, the continuity equation admits the following ``partial'' expansion,
\be
\hat{\del}_{\,\rm t} \e + \th(\e+P) + \sum^\infty_{n=1} \mathbb{F}(2n)\, c^{2n} = 0\, ,
\ee
where the function $\mathbb{F}$ is defined via
\be
\mathbb{F}(2n) \equiv \hat{\nabla}_i[\beta^{2n-2}(\epsilon + P)\beta^i] + (\hat{\partial}_{\,\rm t} + \theta)[\beta^{2n}(\epsilon + P)] + (\epsilon + P)(2\vec{\beta}\cdot\vec{\varphi} + \hat{\gamma}_{ij}\beta^i\beta^j)\beta^{2n-2}\,,
\label{partial_cont}
\ee
with $n=0,1,2,\ldots$ being non-negative integers. Next, inserting the $c\to 0$ expansion eq.~\eqref{ePexps} for the energy density and pressure in eq.~\eqref{partial_cont} above, we get the all orders expansion for the continuity equation, given by 
\begin{equation}
    \sum^\infty_{n=0}c^{2n}\Big(\mathbb{G}^{(2n)} + \sum_{m=0}^{n-1} \mathbb{F}^{(2m)}(2n-2m)\Big) = 0\,,
    \label{full_cont}
\end{equation}
where we have defined
\begin{subequations}
\begin{align}
    \mathbb{G}^{(2m)} &\equiv \partial_{\,\rm t} \epsilon_{(2m)}+\Omega\theta(\epsilon_{(2m)} + p_{(2m)}), \\
    \mathbb{F}^{(2m)}(2n) &\equiv \hat{\nabla}_i[\beta^{2n-2}(\epsilon_{(2m)} + p_{(2m)})\beta^i] + (\hat{\partial}_{\,\rm t} + \theta)[\beta^{2n}(\epsilon_{(2m)} + p_{(2m)})] \nonumber \\
    &\quad+ (\epsilon_{(2m)} + p_{(2m)})(2\vec{\beta}\cdot\vec{\varphi} + \hat{\gamma}_{ij}\beta^i\beta^j)\beta^{2n-2}.
\end{align}
\end{subequations}
The leading term in eq.~\eqref{full_cont}, $\mathbb{G}^{(0)} = 0$, corresponds to the continuity equation \eqref{LOPF1} in the strict Carroll limit, while the next-to-leading term, $\mathbb{G}^{(2)} + \mathbb{F}^{(0)}(2) = 0$, corresponds to the NLO continuity equation \eqref{NLOPF1}. Of course, when taken in its entirety, eq.~\eqref{full_cont} is simply a rewriting of the (Lorentzian) relativistic continuity equation $\nabla_\m T_{(0)}\mbox{}^{\!\m}_{~\rm t} = 0$. One enters the Carrollian regime when the series in eq.~\eqref{full_cont} is truncated to any finite order $n$ in the $c\to 0$ expansion.

A similar procedure can be carried out with the momentum equation $a_{ij} \nabla_\m T_{(0)}\mbox{}^{\m j} = 0$ as well. Once again, by only  expanding the fluid velocity and the Christoffel symbols in powers of $c\to 0$, the momentum equation can be expressed as
\begin{equation}
   \mathbb{X}_i+\sum_{n=1}^{\infty} c^{2n} \mathbb{Y}_i(2n) = 0,
\end{equation}
where we have defined
\begin{subequations}
\begin{align}
\mathbb{X}_i &\equiv \hat{\partial}_i P + (\epsilon + P) \varphi_i + (\hat{\partial}_{\,\rm t} +\theta)[(\epsilon + P)\beta_i],\\
\mathbb{Y}_i(2n) &\equiv  (\hat{\nabla}_j + \varphi_j)[\beta^j(\epsilon + P)\beta_i\beta^{2n-2}] + (\hat{\partial}_{\,\rm t}+\theta)[\beta^{2n}(\epsilon + P)\beta_i], \nonumber \\
&\quad+ \beta^{2n-2}(\epsilon + P)(\beta^2\varphi_i + \beta^j f_{ji}).
\end{align}
\end{subequations}
Inserting now the ansatz eq.~\eqref{ePexps} for the $c\to 0$ behaviour of the energy density and pressure, we get the following representation for the momentum equation,
\begin{equation}
    \sum_{n=0}^\infty c^{2n} \Big(\mathbb{X}_i^{(2n)} + \sum_{m=0}^{n-1} \mathbb{Y}_i^{(2m)}(2n-2m)\Big)  = 0,
\label{full_moment}
\end{equation}
where
\begin{subequations}
\begin{align}
\mathbb{X}_i^{(2m)} &\equiv \hat{\partial}_i p_{(2m)} + (\epsilon_{(2m)} + p_{(2m)}) \varphi_i + (\hat{\partial}_{\,\rm t} +\theta)(\epsilon_{(2m)} + p_{(2m)})\beta_i, \\
\mathbb{Y}_i^{(2m)}(2n) &= (\hat{\nabla}_j + \varphi_j)[\beta^j(\epsilon_{(2m)} + p_{(2m)})\beta_i\beta^{2n-2}] + (\hat{\partial}_{\,\rm t}+\theta)[\beta^{2n}(\epsilon_{(2m)} + p_{(2m)})\beta_i] \nonumber \\
&\quad+ \beta^{2n-2}(\epsilon_{(2m)} + p_{(2m)})(\beta^2\varphi_i + \beta^j f_{ji}).
\end{align}
\end{subequations}
Once again, the leading term in eq.~\eqref{full_moment}, $\mathbb{X}_i^{(0)} = 0$, corresponds to the momentum equation \eqref{LOPF2}  in the strict Carroll limit, while the next-to-leading term, $\mathbb{X}_i^{(2)} + \mathbb{Y}_i^{(0)}(2) = 0$, gives the NLO momentum equation \eqref{NLOPF2} in the $c\to 0$ expansion.

It is quite interesting to witness the relativistic energy-momentum conservation equations admit such compact representations in a $c\to 0$ expansion as is illustrated by eqs.~\eqref{full_cont} and \eqref{full_moment}. Depending upon the sensitivity required, one can truncate this expansion at the desired order. Whilst the very leading term corresponds to the strict Carroll limit, the subleading terms provide a measure of departure from this extreme case and might be better suited for practical applications such as the physics of QGP (more on this in sec.~\ref{sec:Applications}).

%%%%%%%%%%%%%%%%%%%%%%%%%%%%%%%%%%%

\section{Viscous effects in the Carrollian regime}
\label{sec:viscous_fluid}
So far our discussion has focused on perfect fluids, describing equilibrium (more precisely, stationary) states without dissipation. However, in the real world, dissipative effects are omnipresent and are in fact responsible for driving the system from an out-of-equilibrium state towards thermal equilibrium.\footnote{Quite interestingly, the QGP produced in heavy-ion collisions is an almost perfect fluid, with the ratio of shear viscosity to entropy density $\eta/s$ being $\le 0.2$ at the QCD deconfinement temperature \cite{Parkkila:2021tqq}. The bulk viscosity to entropy density ratio $\zeta/s$ is further smaller by an order of magnitude.} As alluded to earlier, the key assumption of hydrodynamics is to consider the gradients quantifying departures from the equilibrium state, or equivalently the strength of dissipative effects, to be small, justifying the derivative expansion for the microscopically well-defined conserved currents of the system on macroscopically large length and time scales. With this in mind, we now turn our attention to investigate hydrodynamic dissipative effects in the Carrollian regime. In section \ref{sec:first_der}, we consider first order derivative corrections that arise for a relativistic fluid, and carefully implement the $c\to 0$ limit on these terms to extract their behaviour in the Carrollian regime. As always, the leading terms correspond to the strict Carroll limit, while subleading terms encapsulate corrections beyond this extreme case. The first order viscous effects will play an important role when we discuss the applications of the formalism to the Bjorken and Gubser flow models of heavy-ion collisions in section \ref{sec:Applications}. 

We consider second order derivative corrections that arise for a relativistic conformal fluid in appendix \ref{sec:second_der},\footnote{For a neutral relativistic fluid, the number of independent transport parameters at the second order in derivative expansion is ten \cite{Bhattacharyya:2012nq}, whilst in the conformal limit only five of them survive \cite{Baier:2007ix}, thereby making it simpler to illustrate the $c\to 0$ limiting procedure. Besides, with an eye for application towards the physics of QGP, it is not entirely unjustified to consider the fluid as conformal, given that $\zeta\ll\eta$ in such settings, while the quark masses can be neglected too at the energy scales involved.} and implement the $c\to 0$ limit on these terms, retaining only the ones that survive in the strict Carroll limit. Rather than being exhaustive, the purpose of this appendix is to illustrate the structure of the second order derivative corrections for a conformal Carroll fluid, which can be used for further generalization of the Bjorken/Gubser flow models beyond what we discuss in section \ref{sec:Applications}.

\subsection{First order derivative corrections in the Carrollian regime}
\label{sec:first_der}
At the first order in the hydrodynamic derivative expansion eq.~\eqref{eq:stress_schematic_1}, one has two independent derivative corrections to the energy-momentum tensor, given by\footnote{We will work in the Landau frame to avoid the hydrodynamic frame choice ambiguity, thus demanding the derivative corrections to satisfy $T^{\m\n}_{(i)} u_\n = 0$.}
\begin{equation}
T^{\mu\nu}_{(1)} = -\eta \sigma^{\mu\nu} - \zeta \Theta \Delta^{\mu\nu}.
\label{visc-EMT}
\end{equation}
Here, $\D_{\m\n} \equiv g_{\m\n} + \f{u_\m u_\n}{c^2}$ is the projector orthogonal to the fluid velocity $u^\m$. As before, $\Theta \equiv \nabla_\m u^\m$ is the fluid expansion parameter, while $\s^{\m\n}$ is the shear tensor (which is symmetric, transverse to the fluid velocity, and traceless), given by
\begin{equation}
\sigma^{\mu\nu} = \Delta^{\mu\alpha}\Delta^{\nu\beta} \nabla_{(\a} u_{\b)} - \frac{1}{d}\Delta^{\mu\nu}\Theta.
\label{shear}
\end{equation}
Also, $\eta, \zeta$ are the shear and bulk viscosities of the fluid, respectively, which satisfy $\eta, \zeta \ge 0$.\footnote{This condition follows by demanding the (onshell) divergence of the entropy current to be non-negative.}

Let us now embark upon computing the $c\to 0$ behaviour of the above quantities, utilizing the PR parametrization of the metric and the fluid velocity, section \ref{sec:PR}. As mentioned earlier, the fluid expansion parameter in the Carrollian regime behaves as $\Theta = \th + c^2 \tilde{\th} + \mc{O}(c^4)$, where the Carrollian expansion $\th$ is defined in eq.~\eqref{SomeDefs}, while the subleading term $\tilde{\th}$ is defined in eq.~\eqref{tiltheta}. The components of the shear tensor eq.~\eqref{shear} in the $c\to 0$ limit then become
{\allowdisplaybreaks
\begin{subequations}
\begin{align}
\sigma^{\rm t}_{~\rm t} &= - c^2\beta^j(b^i + \beta^i)\xi_{ij} + {\cal O}(c^4), \\
\sigma^{\rm t}_{~i} &= \frac{(b^j + \beta^j)}{\Omega}\xi_{ji} + c^2\Bigg(\frac{(b^j + \beta^j)}{\Omega}(\tilde{\xi}_{ji} + \a_{ji}) - \frac{(b^2 + \vec{\beta}\cdot\vec{b})}{\Omega}\xi_{ij}\beta^j\Bigg) + {\cal O}(c^4),  \\
\sigma^i_{~\rm t} &= - c^2\Omega\xi^i_{\ j}\beta^j - c^4\Omega\Big[\big(\tilde{\xi}^{i}_{\ j} + \a^{i}_{\ j}\big)\beta^j - b^i\xi_{jk}\beta^j\beta^k - \vec{\beta}\cdot\vec{b} \, \xi^i_{\ j}\beta^j \Big] + {\cal O}({c^6}), \\
\sigma^i_{\ j} &= \xi^i_{\ j} + c^2\Big[ \tilde{\xi}^{i}_{\ j} + \a^{i}_{\ j} - b^i\xi_{jk}\beta^k \Big] + {\cal O}(c^4). 
\end{align}
\label{shear-tensor-expansion}
\end{subequations}}
Note that the quantities $\xi_{ij}, \tilde{\xi}_{ij}$ arise in the $c\to 0$ expansion of $\s_{ij}$ i.e.
\be
\s_{ij} = \xi_{ij} + c^2 \tilde{\xi}_{ij} + \mc{O}(c^4)\, ,
\ee
where we have
\begin{subequations}
\begin{align}
\xi_{ij} &= \hat{\gamma}_{ij} - \frac{\theta}{d}a_{ij}, \quad a^{ij}\xi_{ij} = 0, \label{Carrollian-shear-tensor}\\
\tilde{\xi}_{ij} &= \hat{\nabla}_{(i}\beta_{j)} + \beta_{(i}\hat{\nabla}_{\rm t}\beta_{j)} + \beta_{(i}\varphi_{j)} + \frac{\beta^2}{2}\hat{\gamma}_{ij} - \frac{\tilde{\theta}}{d}a_{ij}, \quad a^{ij}\tilde{\xi}_{ij} = 0. \label{NLO-shear-tensor}
\end{align}
\end{subequations}
In particular, the quantity $\xi_{ij}$ which survives in the strict Carroll limit is termed the Carrollian shear tensor.
Also, in eq.~\eqref{shear-tensor-expansion}, the quantity $\a_{ij}$ is given by
\begin{equation}
    \a_{ij} \equiv [\beta_{(i} + 2b_{(i}]\xi_{j)k}\beta^k, \quad \a \equiv a^{ij}\a_{ij} = \xi_{ij}\beta^i(\beta^j + 2b^j).
\end{equation}

To compute the behaviour of $T^{\m\n}_{(1)}$ when $c\to 0$, we also need to make an ansatz for the behaviour of the transport parameters $\eta, \z$. Following \cite{Bagchi:2023ysc, Bagchi:2023rwd}, we assume
\begin{equation}
\eta = \eta_{(0)} + c^2 \eta_{(2)} + {\cal O}(c^4), \quad \zeta = \zeta_{(0)} + c^2 \zeta_{(2)} + {\cal O}(c^4)
\label{eta-zeta}
\end{equation}
for the $c\to 0$ behaviour of the shear and bulk viscosities. Using this ansatz along with eq.~\eqref{shear-tensor-expansion}, the first order viscous stress tensor eq.~\eqref{visc-EMT} has the $c\to 0$ expansion
{\allowdisplaybreaks
\begin{subequations}
\begin{align}
T_{(1)}\mbox{}^{\!\rm t}_{\;\rm t} &= c^2\beta^i(b^j+\beta^j)\Xi_{ij} + {\cal O}(c^4),\\
T_{(1)}\mbox{}^{\!\rm t}_{~i} &= -\frac{(b^j + \beta^j)}{\Omega}\Xi_{ij} - \frac{c^2}{\Omega}\Big[(b^j + \beta^j)\tilde{\Xi}_{ij} + \frac{(\vec{\beta}\cdot\vec{b} + \beta^2)}{2}\Xi_{ij}\beta^j \nonumber \\
&\quad + \Big(b_i + \frac{\beta_i}{2}\Big)\Xi_{jk}\beta^j(b^k+\beta^k)  \Big] + {\cal O}(c^4), \\
T_{(1)}\mbox{}^{\!i}_{\;\rm t} &= c^2\Omega\,\Xi^i_{\ j}\beta^j + c^4\Omega\beta^j\Big[ \tilde{\Xi}^{i}_{\ j} + \frac{1}{2}\big(\beta^i\beta^k\Xi_{kj} + \beta^2\Xi^i_{\ j}\big) \Big] + {\cal O}(c^6), \\
T_{(1)}\mbox{}^{\!i}_{~j} &= -\Xi^i_{\ j} - c^2\Big[\tilde{\Xi}^{i}_{\ j} + \Big(b_j+\frac{\beta_j}{2}\Big)\Xi^i_{\ k}\beta^k + \frac{\beta^i}{2}\Xi_{jk}\beta^k \Big] + {\cal O}(c^4).
\end{align}
\label{visc-EMT-expansion}
\end{subequations}}
Here $\Xi_{ij}$ is the Carrollian viscous stress tensor, defined via
\begin{equation}
    \Xi_{ij} \equiv \eta_{(0)}\xi_{ij} + \zeta_{(0)}\theta a_{ij}, \quad \Xi \equiv a^{ij}\Xi_{ij} = \zeta_{(0)}\theta d,
\end{equation}
and $\tilde{\Xi}_{ij}$ is the ${\cal O}(c^2)$ viscous stress tensor, defined via
\begin{equation}
    \tilde{\Xi}_{ij} = \eta_{(0)}\tilde{\xi}_{ij} + \eta_{(2)}\xi_{ij} + (\zeta_{(0)}\tilde{\theta} + \zeta_{(2)}\theta)a_{ij}, \quad \tilde{\Xi} \equiv a^{ij}\tilde{\Xi}_{ij} = d(\zeta_{(0)}\tilde{\theta} + \zeta_{(2)} \theta).
\end{equation}

With the above results in hand, we can now elucidate the structure of the viscous hydrodynamic equations with first order derivative corrections in the Carrollian regime. The equations are of course given by eq.~\eqref{Hydro1}, but now with the energy-momentum tensor being $T_{(0)}\mbox{}^{\!\m}_{~\n} + T_{(1)}\mbox{}^{\!\m}_{~\n}$. The equations admit a $c\to 0$ expansion akin to the one for the perfect fluid case, eq.~\eqref{Hydro2}. Once again, the $\mc{O}(c^{-2})$ equations, i.e.~$[\nabla_\m (T_{(0)}\mbox{}^{\!\m}_{~\n} + T_{(1)}\mbox{}^{\!\m}_{~\n})]^{\{-2\}} = 0$ are trivially satisfied. The LO equations thus follow from $[\nabla_\m (T_{(0)}\mbox{}^{\!\m}_{~\n} + T_{(1)}\mbox{}^{\!\m}_{~\n})]^{\{0\}} = 0$, and for $\n = {\rm t}, i$ are respectively given by
\begin{subequations}
\label{LOVF}
\begin{align}
    \hat{\partial}_{\,\rm t}\epsilon_{(0)} &= - \theta\Big(\epsilon_{(0)} + p_{(0)} - \f{\Xi}{d}\Big) + \xi^{ij}\Xi_{ij}, \label{LOVF1}\\
    \hat{\partial}_i p_{(0)} &=  - \varphi_i(\epsilon_{(0)} + p_{(0)}) - (\hat{\partial}_{\,\rm t} + \theta)[(\epsilon_{(0)} + p_{(0)})\beta_i - \Xi_{ij}\beta^j] + (\hat{\nabla}_j + \varphi_j)\Xi^{j}_{\ i}. \label{LOVF2}
\end{align}
\end{subequations}
The LO equations above govern the dynamics of a viscous Carroll fluid, and were the focus of investigation in \cite{Bagchi:2023ysc, Bagchi:2023rwd}, where an equivalence was established between them and the equations for viscous Bjorken and Gubser flow models of heavy-ion collisions for appropriate choices of the geometric data $(\O, b_i, a_{ij})$ on the Carroll manifold. Just like their ideal fluid analogs eq.~\eqref{LOPF}, the hydrodynamic equations \eqref{LOVF} respect Carroll boost invariance in the tangent space, as the case should be - see \ref{CBI-FE} for a discussion.

The viscous NLO equations which take us away from the strict Carroll limit captured by eqs.~\eqref{LOVF} into the Carrollian regime are given by $[\nabla_\m (T_{(0)}\mbox{}^{\!\m}_{~\n} + T_{(1)}\mbox{}^{\!\m}_{~\n})]^{\{2\}} = 0$. On substituting the appropriate terms, they take the form
\begin{subequations}
\label{NLOVF}
\begin{align}
\hat{\partial}_{\,\rm t}\epsilon_{(2)} &= - \theta\Big(\epsilon_{(2)} + p_{(2)}-\f{\tilde{\Xi}}{d}\Big) + \xi^{ij}\tilde{\Xi}_{ij} - (\hat{\partial}_{\,\rm t} + \theta)[\beta^2(\epsilon_{(0)} + p_{(0)})-\beta^i\beta^j\Xi_{ij}]\nonumber \\
&\quad- \hat{\nabla}_i[(\epsilon_{(0)} + p_{(0)})\beta^i-\Xi^i_{\ j}\beta^j] - (\epsilon_{(0)} + p_{(0)})\Big(2\vec{\beta}\cdot\vec{\varphi} + \hat{\gamma}_{ij}\beta^i\beta^j\Big)  \nonumber \\
&\quad+ \Xi_{ij}\Big(2\beta^i\varphi^j + \beta^i\hat{\gamma}^j_k\beta^k\Big), \label{NLOVF1}\\
\hat{\partial}_i p_{(2)} &= - \varphi_i(\epsilon_{(2)} + p_{(2)}) - (\hat{\nabla}_j + \varphi_j)\Big[\beta^j\beta_i(\epsilon_{(0)} + p_{(0)})-\tilde{\Xi}^{j}_{\ i} -\f{1}{2}\beta_i\Xi^j_k\beta^k -\f{1}{2} \beta^j\Xi_{ik}\beta^k\Big] \nonumber \\
& - (\hat{\partial}_{\,\rm t} + \theta)\Big[\beta_i\beta^2(\epsilon_{(0)} + p_{(0)})+(\epsilon_{(2)} + p_{(2)})\beta_i-\tilde{\Xi}_{ij}\beta^j - \f{1}{2}\beta_i\Xi_{jk}\beta^j\beta^k - \f{\beta^2}{2}\Xi_{ik}\beta^k\Big] \nonumber \\
&- (\epsilon_{(0)} + p_{(0)})(\varphi_i\beta^2 + \beta^j f_{ji}) + \Xi_{jk}(\varphi_i\beta^j\beta^k + \beta^j f^k_{\ \ i}). \label{NLOVF2}
\end{align} 
\end{subequations}
The viscous hydrodynamic equations above are some of the main results of this paper. Needless to say, one can continue this process and compute the equations satisfied by even higher order terms in the $c\to 0$ expansion. In what follows, we will now turn towards some applications of these abstract equations to hydrodynamic flows relevant for heavy-ion collisions and the spacetime evolution of the QGP.

%%%%%%%%%%%%%%%%%%%%%%%%%%%%%%%%%%%
\section{Applications}
\label{sec:Applications}
Ultrarelativistic heavy-ion collisions are able to recreate the circumstances present at the big bang birth of our universe for a tiny fraction of a second. The study of these collisions over the last couple of decades has shed enormous light on a new state of matter, the quark-gluon plasma, where strongly coupled quarks and gluons exist momentarily in a deconfined state \cite{Busza:2018rrf}. Surprisingly enough, the QGP behaves like a fluid, with its observed behaviour well described by the equations of relativistic hydrodynamics after choosing appropriate initial conditions. In line with this, to gain a better analytic understanding which often gets obfuscated in a numerical fitting of the data, people have also developed simple models of relativistic hydrodynamics for the spacetime evolution of the QGP. The most celebrated amongst these models was developed by Bjorken \cite{Bjorken:1982qr}, making simplifying assumptions such as invariance of the flow under boosts along the beam axis, as well as translation invariance in the transverse plane. Subsequently, Gubser proposed a more general model \cite{Gubser:2010ze, Gubser:2010ui}, albeit applicable only to conformal fluids, which retains boost invariance along the beam axis while allowing for a nontrivial radial profile for the flow in the transverse directions.

\begin{figure}[t]
\centering
\includegraphics[width=10.2cm]{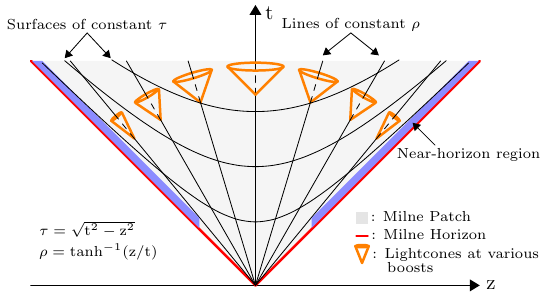}
\caption{A pictorial representation for the emergence of Carrollian physics in the Bjorken and Gubser flow models of QGP. The collision event takes place at the origin, with the Milne patch representing its forward light cone. The ultrarelativistic nature of the flow lands one in the near-horizon region, where local lightcones collapse, or effectively $c\to 0$, thereby making the dynamics Carrollian.}
\label{lightcone}
\end{figure}

Interestingly, the ultrarelativistic nature of fluid flow in both Bjorken and Gubser models for the QGP essentially restricts it to the near-horizon region in terms of Milne coordinates i.e.~proper time $\tau$ and rapidity $\rho$, as shown in figure \ref{lightcone}. Given the proximity of this region with the Milne horizon, where local lightcones close up, one may expect the effective dynamics to become Carrollian in nature. This physical intuition was first made precise in \cite{Bagchi:2023ysc}, where starting from a generic Carroll fluid on a Carroll manifold described in terms of the PR parametrization of sec.~\ref{sec:PR}, an appropriate choice for the geometric data $(\Omega, b_i, a_{ij})$ reduced the Carroll fluid equations to the equations for Bjorken flow. In \cite{Bagchi:2023rwd}, this mapping between boost-invariant models of heavy-ion collisions and Carroll fluids was further generalized to Gubser flow, with a somewhat more complicated choice for the Carroll geometric data. Put differently, the construction in \cite{Bagchi:2023ysc, Bagchi:2023rwd} provided a geometrization of the phenomenological assumptions that underlie the Bjorken and Gubser flow models in terms of Carroll fluids on specific Carroll manifolds.

In the present section, our aim is to utilize the mapping between boost-invariant models of heavy-ion collisions and Carroll hydrodynamics to compute corrections to these models that follow from the newly found subleading terms that arise in the Carrollian regime. More precisely, in terms of the terminology of the present work, the mapping of interest was established in \cite{Bagchi:2023ysc, Bagchi:2023rwd} in the strict Carroll limit i.e.~by keeping only the leading terms in the $c\to 0$ expansion, which indeed provide Carroll covariant hydrodynamic equations (see comments on symmetries below). In sections \ref{sec:ideal_fluid} and \ref{sec:viscous_fluid}, we have now found the equations for hydrodynamics in the Carrollian regime, going beyond the strict Carroll limit by including subleading terms in a $c\to 0$ expansion. We would now like to specialize these subleading terms to the specific geometric data $(\O, b_i, a_{ij})$ that maps to the corresponding QGP model. In subsection \ref{sec:Bjorken}, we compute the modifications in the hydrodynamic equations for Bjorken flow, while in subsection \ref{sec:Gubser} we do so for Gubser flow. The inclusion of subleading terms in the hydrodynamic equations, that now bring in rapidity dependence as well, is expected to lead to a better analytic understanding of the spacetime evolution of the QGP.

A few comments about various symmetries involved are in order. It is important to clearly distinguish between the symmetries of the background geometry (i.e.~the isometries) versus the local tangent space symmetries versus any symmetries imposed by hand on the flow. On a flat Carroll manifold, Carroll symmetries such as Carroll boosts are realized as isometries. However, on a curved Carroll manifold, including the ones to which Bjorken/Gubser flow map to, the isometries could be more complicated, while Carroll boosts are realized as a symmetry in the local tangent space. This is analogous to the fact that Lorentz boosts, which are isometries of Minkowski spacetime, are realized as local tangent space symmetries for a general curved pseudo-Riemannian manifold. Equations of Carroll hydrodynamics, which arise as leading-order terms in the small-$c$ limit of relativistic hydrodynamics, respect Carroll boosts as a symmetry in the tangent space (we discuss this extensively in \ref{CBI-FE}). By including more and more terms in the small-$c$ expansion, i.e.~going from the strict Carroll limit into the Carrollian regime, one is then essentially restoring Lorentz boosts as a symmetry for the resulting equations, also in the tangent space. It is important to distinguish this from the spacetime Lorentz boost invariance imposed along the beam axis in Bjorken/Gubser flow by hand. The first is purely a statement about the underlying geometry, while the second is a phenomenological choice. What was shown in \cite{Bagchi:2023ysc, Bagchi:2023rwd} was this phenomenological assumption for Bjorken/Gubser flow maps to specific choices of geometric data (more specifically, the Ehresmann connection) on the Carroll side. With subleading in small-$c$ terms now included in the Carrollian regime, we are departing from the phenomenological assumption of Lorentz boost invariance on the Bjorken/Gubser side. In other words, the subleading terms now bring in dependence on the rapidity $\rho$ for the fluid velocity, temperature etc., thus providing corrections to Bjorken/Gubser flow models. 

\subsection{Computing corrections to Bjorken flow}
\label{sec:Bjorken}
Let us begin our discussion with Bjorken flow. In \cite{Bagchi:2023ysc}, it was found that the choice of the geometric data\footnote{The mapping in eq.~\eqref{Bjorken_Data} respects local Carroll boost invariance - see \ref{CBI-FE} for a discussion.}
\be
\O = 1, \quad b_i = -\b_i, \quad a_{ij} dx^i dx^j = \tau^2 d\rho^2 + {\rm dx}^2 + {\rm dy}^2,
\label{Bjorken_Data}
\ee
with the identification of the coordinate time ${\rm t}$ on the Carroll manifold with the proper time $\tau$, along with identifying the coordinates $x^i$ with $(\rho, {\rm x}, {\rm y})$,\footnote{Here $(\rm x,y,z)$ denote Cartesian coordinates in flat space.} the equations for a Carroll fluid reduced to the equations that govern Bjorken flow. To wit, the choice eq.~\eqref{Bjorken_Data} for the perfect Carroll fluid eq.~\eqref{LOPF} implies
\be
\del_\tau \e_{(0)} = - \f{\e_{(0)}+p_{(0)}}{\tau}, \quad \del_i p_{(0)} = 0.
\label{Bjorken_Ideal}
\ee
We can identify the Carrollian energy density and pressure $(\e_{(0)}, p_{(0)})$ with the energy density and pressure $(\e, P)$ of the quark-gluon plasma simply via $\e=\e_{(0)}, P=p_{(0)}$, as is also evident from eq.~\eqref{ePexps}. The second equation in eq.~\eqref{Bjorken_Ideal} implies that the pressure is independent of the transverse directions ${\rm x},{\rm y}$, as well as the rapidity $\rho$, thereby bringing out the key phenomenological assumptions of Bjorken flow simply as a consequence of the underlying Carrollian nature of the dynamics. Invoking the equation of state $P = P(\e)$, the rapidity independence of pressure automatically implies the rapidity independence of energy density as well.\footnote{As well as independence of $\epsilon$ from the transverse directions ${\rm x}, {\rm y}$.} Given the initial conditions, the first equation in eq.~\eqref{Bjorken_Ideal} then computes the evolution of the energy density of the QGP as a function of proper time.

With the advent of eqs.~\eqref{NLOPFF}, let us now compute the modifications to the Bjorken flow equations for a perfect fluid once we include the effects of the subleading terms in the $c\to 0$ expansion. Specializing to the geometric data eq.~\eqref{Bjorken_Data} in eqs.~\eqref{NLOPFF}, we get
\begin{subequations}
\begin{align}
\partial_{\tau}\epsilon_{(2)} &= - \frac{(\epsilon_{(2)} + p_{(2)})}{\tau} - (\epsilon_{(0)} + p_{(0)})\Big( \partial_i\beta^i - \f{1}{2}\partial_{\tau}{\beta^2}\Big), \label{Bjorken_e2}\\
\partial_i p_{(2)} &= - \f{1}{2} (\epsilon_{(0)} + p_{(0)})\, \partial_i{\beta^2}. \label{Bjorken_p2}
\end{align}
\label{Bjorken_sub}
\end{subequations}
Now, with the subleading corrections to the strict Carroll limit included, the energy density and pressure of the QGP are $\e \simeq \e_{(0)} + c^2 \e_{(2)} , P \simeq p_{(0)} + c^2 p_{(2)}$, where we have dropped $\mc{O}(c^4)$ corrections. By combining eqs.~\eqref{Bjorken_Ideal} and \eqref{Bjorken_sub}, we can arrive at equations governing $(\e, P)$ as follows. Firstly, multiply eqs.~\eqref{Bjorken_e2} and \eqref{Bjorken_p2} with an overall $c^2$. Secondly, replace $\e_{(0)} + p_{(0)}$ with $\e+P$ on the RHS of the resulting equations, which incurs an error of $\mc{O}(c^4)$, negligible at the present order. Finally, combine with eq.~\eqref{Bjorken_Ideal} to get
\begin{subequations}
\label{Mod_Bjorken_Ideal}
\begin{align}
\del_\t \e &= - \f{\e+P}{\t} -c^2 (\e+P) \Big(\del_i\b^i - \f{1}{2}\del_\t\b^2\Big),\label{Mod_Bjorken_1}\\
\del_i P &= - \f{c^2}{2} (\epsilon + P)\, \partial_i{\beta^2}\, .\label{Mod_Bjorken_2}
\end{align}
\end{subequations}
We see that the equations governing $(\e, P)$ depend nontrivially on the functions $\b^i(\t,{\rm x},{\rm y},\rho)$, with $i = ({\rm x},{\rm y},\rho)$, bringing in explicit dependence on the rapidity parameter. Eq.~\eqref{Mod_Bjorken_Ideal} thus incorporates leading departures from the exact Bjorken flow model by including rapidity dependence, arrived at by exploiting its mapping to Carroll fluids.

Let us pause for a moment and contemplate how one would go about utilizing eq.~\eqref{Mod_Bjorken_Ideal} for heavy-ion collisions. The energy density $\e$ and pressure $P$ are functions of the temperature $T$ of the QGP, implying that eq.~\eqref{Mod_Bjorken_Ideal} provides us with a set of four coupled differential equations to determine the four functions $(T, \b^i)$, given appropriate initial conditions. Determining $T$ as a function of spacetime gives one information about how $\e, P$ evolve in spacetime, now including dependence on rapidity, as mentioned above. Further, the functions $\b^i$ encapsulate information about the fluid velocity profile. \\

\noindent \ding{112} \emph{\underline{Including viscous effects}:}\\
One can also include first order viscous corrections into the hydrodynamic equations. To begin, for the strict Carroll limit, employing the geometric data eq.~\eqref{Bjorken_Data} in the LO eq.~\eqref{LOVF} leads to
\be
\label{Bjorken_Visc}
\del_\t\epsilon_{(0)} = - \f{\epsilon_{(0)} + p_{(0)}}{\t} + \f{1}{\t^2} \left(\f{2\eta_{(0)}}{3} + \z_{(0)}\right), \qquad \del_i p_{(0)} = 0.
\ee
As observed in \cite{Bagchi:2023ysc}, the above equations for a viscous Carroll fluid are identical to the equations that govern viscous Bjorken flow, once the identification $\e = \e_{(0)}, P = p_{(0)}, \eta = \eta_{(0)}, \z = \z_{(0)}$ is made, identifying the energy density, pressure, shear and bulk viscosity of the QGP with that of the Carroll fluid. 

Let us now consider the NLO equations \eqref{NLOVF} specialized to the geometric data eq.~\eqref{Bjorken_Data}, and compute the rapidity dependent corrections to the viscous Bjorken flow equations \eqref{Bjorken_Visc}, or in other words, generalize the hydrodynamic equations \eqref{Mod_Bjorken_Ideal} to include viscous effects. Eq.~\eqref{NLOVF1} with the choice eq.~\eqref{Bjorken_Data} yields 
\be
\begin{split}
    \partial_{\tau}&\epsilon_{(2)} = -\frac{(\epsilon_{(2)} + p_{(2)})}{\tau} -(\epsilon_{(0)} + p_{(0)})\Big(\partial_i\beta^i -\f{1}{2}\partial_{\tau}\beta^2\Big) +\frac{2\z_{(0)}}{\tau}\Big(\partial_i\beta^i -\f{1}{2}\partial_{\tau}\beta^2 - \f{\b^2}{4\t}\Big) \label{Mod_Bjor_Visc_e2}\\
    &+ \eta_{(0)}\Big[ \frac{1}{\tau}\Big(\frac{4}{3}\partial_{\rho}\beta^{\rho} - \frac{2}{3}\partial_{i_{\!\perp}}\beta^{i_{\!\perp}} - \partial_{\tau}\Big(\frac{2}{3}\beta_{\rho}\beta^{\rho} -\frac{1}{3}\beta_{i_{\!\perp}}\beta^{i_{\!\perp}}\!\Big) \Big) -\frac{\beta^2}{3\tau^2}\Big] + \frac{1}{\tau^2}\Big( \frac{2\eta_{(2)}}{3} + \zeta_{(2)}\Big),
\end{split}
\ee
where we have introduced the notation $i = (i_{\!\perp}, \rho)$ with $i_{\!\perp} = (\rm x, \rm y)$ collectively denoting the transverse directions. 
Combining eqs.~\eqref{Bjorken_Visc} and \eqref{Mod_Bjor_Visc_e2}, we arrive at an equation that governs the evolution of the QGP energy density beyond the Bjorken approximation including viscous effects, given by
\be
\begin{split}
    \partial_{\tau}\epsilon &= -\frac{(\epsilon + P)}{\tau} + \frac{1}{\tau^2}\Big(\frac{2\eta}{3} + \zeta\Big) \\
    &\quad- c^2(\epsilon + P)\Big(\partial_i\beta^i -\f{1}{2}\partial_{\tau}\beta^2\Big) +\frac{2c^2\z}{\tau}\Big(\partial_i\beta^i -\f{1}{2}\partial_{\tau}\beta^2 - \f{\b^2}{4\t}\Big) \label{MBV_e}\\
    &\quad+ c^2\eta\Big[ \frac{1}{\tau}\Big(\frac{4}{3}\partial_{\rho}\beta^{\rho} - \frac{2}{3}\partial_{i_{\!\perp}}\beta^{i_{\!\perp}} - \partial_{\tau}\Big(\frac{2}{3}\beta_{\rho}\beta^{\rho} -\frac{1}{3}\beta_{i_{\!\perp}}\beta^{i_{\!\perp}}\Big) \Big) -\frac{\beta^2}{3\tau^2}\Big].
\end{split}
\ee
Similarly, specializing the NLO eq.~\eqref{NLOVF2} to the data eq.~\eqref{Bjorken_Data} and combining with eq.~\eqref{Bjorken_Visc}, we get 
{\allowdisplaybreaks
\begin{subequations}
\label{MBV_P}
\begin{align}
\partial_{\rho}P &= - \f{c^2}{2}(\epsilon + P)\partial_{\rho}\beta^2 + c^2 \eta\Big[ \frac{2}{3}\partial_{\rho}^2\beta^{\rho} + \frac{1}{6}\partial_{\rho}\partial_{i_{\!\perp}}\beta^{i_{\!\perp}}\!+ \frac{1}{2}\partial_{i_{\!\perp}}\!\partial^{i_{\!\perp}}\!\beta_{\rho} - \partial_{\tau}\partial_{\rho}\Big(\frac{\beta_{\rho}\beta^{\rho}}{3} - \frac{\beta_{i_{\!\perp}}\!\beta^{i_{\!\perp}}}{6}\Big) \nonumber \\
&\quad- \frac{1}{2}\partial_{\tau}\partial_{i_{\!\perp}}\!(\beta^{i_{\!\perp}}\!\beta_{\rho}) - \frac{1}{2\tau}\partial_{\rho}(\beta_{i_{\!\perp}}\!\beta^{i_{\!\perp}}\!) + \tau\partial_{i_{\!\perp}}\!(\beta^{i_{\!\perp}}\!\beta^{\rho})\Big] + c^2 \zeta\partial_{\rho}\Big(\partial_i\beta^i - \f{1}{2}\partial_{\tau}\beta^2\Big),\label{MBV_P1}\\
\partial_{i_{\!\perp}}\!P &= - \f{c^2}{2}(\epsilon + P)\partial_{i_{\!\perp}}\!\beta^2 + c^2 \eta\Big[\frac{1}{2}\partial_{j}\partial^j\beta_{i_{\!\perp}}\! + \frac{1}{6}\partial_{i_{\!\perp}}\!\partial_{j}\beta^j - \frac{1}{2}\partial_{\tau}\partial_{j}(\beta^j\beta_{i_{\!\perp}}\!) + \frac{1}{6}\partial_{\tau}\partial_{i_{\!\perp}}\!\beta^2 \nonumber \\
&\quad- \frac{1}{\tau}\partial_{\rho}(\beta^{\rho}\beta_{i_{\!\perp}}\!) + \frac{1}{2\tau}\partial_{i_{\!\perp}}\!(\beta^{\rho}\beta_{\rho}) \Big] + c^2\zeta\partial_{i_{\!\perp}}\!\Big(\partial_j\beta^j - \f{1}{2}\partial_{\tau}\beta^2 \Big).\label{MBV_P2}
\end{align}
\end{subequations}}
Eqs.~\eqref{MBV_e} and \eqref{MBV_P} are some of the key results of this paper. We expect them to provide better fits to the QGP data compared to the Bjorken flow model. Interestingly, from eq.~\eqref{MBV_P2} (and also the ideal fluid version eq.~\eqref{Mod_Bjorken_2}), we see that the pressure gradient along the two transverse directions can be different, leading to a pressure anisotropy in the transverse plane itself, which might lead to nontrivial observable consequences. 

%%%%%%%%%%%%%%%%%%%%%%%%%%%%%%%%%%

\subsection{Computing corrections to Gubser flow}
\label{sec:Gubser}
Let us now consider another application of the hydrodynamic equations in the Carrollian regime. As was observed in \cite{Bagchi:2023rwd}, the Gubser flow model of heavy-ion collisions \cite{Gubser:2010ze, Gubser:2010ui}, a generalization of Bjorken flow that allows for nontrivial radial evolution of the QGP albeit assuming conformal invariance,\footnote{In $d=3$ spatial dimensions conformal invariance implies the equation of state $\e = 3P$.} also turns out to be an example of a Carroll fluid due to its ultrarelativistic nature. As discussed in \cite{Gubser:2010ui}, it turns out that the equations for Gubser flow take a particularly simple form when written on the global ${\rm dS}_3\times \mathbb{R}$ background, which is related to the flat background in Milne coordinates via a Weyl rescaling. The ${\rm dS}_3$ part of the product manifold is covered by the coordinates $(\varsigma, \psi, \phi)$, with $\varsigma$ being timelike and $(\psi, \phi)$ parametrizing the two-sphere, while the rapidity $\rho$ runs along the real line $\mathbb{R}$. In \cite{Bagchi:2023rwd}, it was found that by choosing the geometric data\footnote{Like the case for Bjorken flow, the mapping in eq.~\eqref{Gubser_Data} respects local Carroll boost invariance, as discussed in \ref{CBI-FE}.}
\be
\label{Gubser_Data}
\O = 1, \quad b_i = -\b_i, \quad a_{ij}dx^i dx^j = \cosh^2\varsigma\,(d\psi^2+\sin^2\psi \, d\phi^2) + d\rho^2,
\ee 
where the time ${\rm t}$ on the Carroll manifold is identified with $\varsigma$, while the spacelike coordinates $x^i$ are identified with $(\rho, \psi, \phi)$, the equations for a Carroll fluid reduced to the Gubser flow equations on the ${\rm dS}_3\times\mathbb{R}$ background. To wit, eq.~\eqref{LOPF} with the conformal equation of state and the geometric data eq.~\eqref{Gubser_Data} implies
\be
\del_\varsigma \epsilon_{(0)} = - \f{8\epsilon_{(0)}}{3} \tanh\varsigma\, , \quad \partial_i \epsilon_{(0)} = 0.
\label{Gubser_Ideal}
\ee
These are identical to the equations for Gubser flow on the ${\rm dS}_3 \times \mathbb{R}$ background, with the identification $\e = \e_{(0)}$ between the energy density of the QGP and that of the Carroll fluid. In particular, the second equation above implies the rapidity independence of the flow. 

As we did in section \ref{sec:Bjorken} for Bjorken flow, we can now compute corrections to the Gubser flow eq.~\eqref{Gubser_Ideal} by departing from the strict Carroll limit and keeping the subleading terms in the Carrollian regime intact. To do so, we specialize eq.~\eqref{NLOPFF} with the conformal equation of state to the geometric data eq.~\eqref{Gubser_Data} and arrive at the equations
\begin{subequations}
\label{Gubser_sub}
\begin{align}
\del_\varsigma \epsilon_{(2)} &= - \f{8\epsilon_{(2)}}{3} \tanh\varsigma - \f{4\e_{(0)}}{3} \Big(\del_i \b^i - \f{1}{2}\del_\varsigma \b^2 + \b^\psi \cot\psi \Big), \label{Gubser_e2_1}\\
\partial_i \epsilon_{(2)} &= - 2 \e_{(0)} \del_i \b^2. \label{Gubser_e2_2}
\end{align}
\end{subequations}
We can now combine eqs.~\eqref{Gubser_Ideal} and \eqref{Gubser_sub} to obtain equations governing the evolution of the QGP energy density $\e \simeq \e_{(0)} + c^2 \e_{(2)}$, where we have neglected $\mc{O}(c^4)$ terms. To do so, we follow the procedure laid out in section \ref{sec:Bjorken}. Namely, we first multiply equations \eqref{Gubser_e2_1} and \eqref{Gubser_e2_2} with a factor of $c^2$, followed by replacing $\e_{(0)}$ on the RHS of the resulting equations with $\e$, incurring an error of $\mc{O}(c^4)$ which can be neglected to the order we are working, and finally combine the resulting equations with eq.~\eqref{Gubser_Ideal} to arrive at 
\begin{subequations}
\label{Mod_Gubser_Ideal}
\begin{align}
\del_\varsigma \epsilon &= - \f{8\epsilon}{3} \tanh\varsigma - \f{4c^2\e}{3} \Big(\del_i \b^i - \f{1}{2}\del_\varsigma \b^2 + \b^\psi \cot\psi \Big) , \label{Mod_Gubser_1}\\
\partial_i \epsilon &= - 2 c^2 \e \, \del_i \b^2. \label{Mod_Gubser_2}
\end{align}
\end{subequations}
Once again, we see that the equations governing the energy density depend nontrivially on the functions $\b^i(\varsigma, \rho, \psi, \phi)$, with $i = (\rho, \psi,\phi)$ as mentioned above, bringing in explicit dependence on the rapidity parameter. Eq.~\eqref{Mod_Gubser_Ideal} thus captures leading departures from exact Gubser flow by including rapidity dependence, arrived at by using its mapping to Carroll fluids.\\

\noindent \ding{112} \emph{\underline{Including viscous effects}:} \\
Needless to say, a similar story holds if one wants to include viscous effects. With the geometric data eq.~\eqref{Gubser_Data}, eq.~\eqref{LOVF} becomes
\be
\del_\varsigma \epsilon_{(0)} = - \f{8\epsilon_{(0)}}{3} \tanh\varsigma + \f{2 \eta_{(0)}}{3} \tanh^2\varsigma\, , \quad \partial_i \epsilon_{(0)} = 0.
\label{Gubser_Visc}
\ee
As observed in \cite{Bagchi:2023rwd}, the equations above are identical to the viscous Gubser flow equations with the identification of the QGP energy density and shear viscosity with those of the Carroll fluid.\footnote{We presently follow a slightly different convention compared to \cite{Bagchi:2023rwd}. Whereas in \cite{Bagchi:2023rwd} the derivatives acted on the shear viscosity as well while computing the hydrodynamic equations, we now do not allow for such terms, in line with the usual approach taken in linearized hydrodynamics. The mapping between Gubser flow and Carroll hydrodynamics still works fine.} Next, to compute corrections to eq.~\eqref{Gubser_Visc}, we utilize the geometric data eq.~\eqref{Gubser_Data} in the NLO viscous equation \eqref{NLOVF}. Eq.~\eqref{NLOVF1} then yields
\begin{align}
\del_\varsigma \epsilon_{(2)} &= - \f{8\epsilon_{(2)}}{3} \tanh\varsigma - \f{4\e_{(0)}}{3} \Big(\del_i \b^i - \f{1}{2}\del_\varsigma \b^2 + \b^\psi \cot\psi\Big) + \f{2\eta_{(2)}}{3}\tanh^2\varsigma \label{Gubser_Visc_e2}\\
&+ \f{2\eta_{(0)}}{3}\tanh\varsigma \, \Big(\del_i \b^i - 3 \del_\rho \b^\rho - \f{1}{2}\del_\varsigma \b^2 +\f{3}{2}\del_\varsigma(\b^\rho\b_\rho) - \f{\b^2}{2} \tanh^2\varsigma + \b^\psi \cot\psi\Big). \nonumber
\end{align}
This can be combined with eq.~\eqref{Gubser_Visc} to get an equation for the $\varsigma$-derivative of the QGP energy density, i.e.
\begin{align}
\del_\varsigma \epsilon &= - \f{8\epsilon}{3} \tanh\varsigma - \f{4c^2\e}{3} \Big(\del_i \b^i - \f{1}{2}\del_\varsigma \b^2 + \b^\psi \cot\psi\Big) + \f{2\eta}{3}\tanh^2\varsigma\\
&\quad+ \f{2c^2\eta}{3}\tanh\varsigma \, \Big(\del_i \b^i - 3 \del_\rho \b^\rho - \f{1}{2}\del_\varsigma \b^2 +\f{3}{2}\del_\varsigma(\b^\rho\b_\rho) - \f{\b^2}{2} \tanh^2\varsigma + \b^\psi \cot\psi\Big). \nonumber
\end{align}
Similarly, making use of the data eq.~\eqref{Gubser_Data} in \eqref{NLOVF2} and combining with \eqref{Gubser_Visc} yields
{\allowdisplaybreaks
\begin{subequations}
\begin{align}
\partial_\psi \epsilon &= -2c^2 \epsilon \partial_\psi\beta^2 + \f{3 c^2 \eta}{2}\Bigg[\frac{2\beta^\psi}{3}+\mathfrak{d}_\psi\left(\frac{4}{3}\del_\psi\beta^\psi-\frac{\theta}{2}\beta_\psi\beta^\psi\right) + \pa_\rho\bigg(\frac{1}{3}\pa_{\psi}\beta^\rho+\pa_{\rho}\beta_\psi\bigg) \nonumber\\
&\quad + \pa_\phi\Big( \pa^\phi\beta_\psi+\frac{1}{3}\del_\psi\beta^\phi-\pa_\varsigma(\beta^\phi\beta_\psi)\Big) -\mathfrak{d}_\varsigma\bigg(\del_{\rho}(\beta_\psi\beta_\rho)-\frac{1}{3}\pa_{\psi}\beta^2 \nonumber\\
&\quad +\mathfrak{d}_\psi(\beta^\psi\beta_\psi)-\cot\psi\beta^\phi\beta_\phi\bigg) - 2\cot\psi\left(\pa_\phi\beta^\phi+\frac{2}{3}\beta^\psi\cot\psi-\frac{\theta}{4}\beta^\phi\beta_\phi\right) \nonumber\\
&\quad + \tanh\varsigma\left(\frac{1}{3}\partial_\psi(\beta^{i_\odot}\beta_{i_\odot}-2\beta_\rho\beta^\rho)+\partial_\rho(\beta_\psi\beta^\rho)\right)\!\Bigg]\, ,\\
\partial_\phi \epsilon &= -2c^2\epsilon \partial_\phi\beta^2 + \f{3c^2\eta}{2}\Bigg[\mathfrak{d}_\psi\left(\partial^{\psi}\beta_{\phi}+\partial_{\phi}\beta^{\psi}-\partial_\varsigma(\beta_{\phi}\beta^{\psi})- 2\beta_\phi \sech^2\varsigma \cot\psi+\frac{4}{3}\partial_\phi\beta^\psi\right)\nonumber\\
&\quad -2\partial_\phi\left(\partial_\psi\beta^\psi-\frac{1}{6}\pa_\rho\beta^\rho-\frac{2}{3}\pa_\phi\beta^\phi\right) - \mathfrak{d}_{\varsigma}\left(\pa_\phi(\beta^\phi\beta_\phi)-\frac{1}{3}\partial_{\phi}\beta^2+\partial_\rho(\beta_\rho\beta_\phi)\right)\nonumber \\
&\quad +\partial^2_{\rho}\beta_{\phi}+\tanh\varsigma \bigg(\f{1}{3}\partial_\phi(\beta^{\psi}\beta_{\psi}-2\beta^\phi\beta_\phi - 2\beta^\rho\beta_\rho)+\partial_\rho(\beta_\phi\beta^\rho) \bigg)\Bigg]\, ,\\
\partial_\rho \epsilon &= - 2c^2 \epsilon \partial_\rho\beta^2+\f{3c^2\eta}{2}\Bigg[\mathfrak{d}_{i_\odot}\!\left(\partial^{i_\odot}\!\beta_{\rho}-(\mathfrak{d}_\varsigma+\theta)(\beta^{i_\odot}\!\beta_{\rho})+\frac{1}{3}\partial_\rho\beta^{i_\odot}\right) \nonumber\\
&\quad +\frac{2}{3}\partial_\rho\Big(2\partial_{\rho}\beta^\rho+(\mathfrak{d}_\varsigma+\theta) \beta^2 - \frac{1}{2}\partial_{\varsigma}(\beta^{i_\odot}\!\beta_{i_\odot}+4\beta^\rho\beta_{\rho})\Big)\nonumber \\
&\quad +\tanh\varsigma\bigg(\frac{1}{3}\partial_\rho(\beta^{i_\odot}\!\beta_{i_\odot}-2\beta^\rho\beta_\rho)-\mathfrak{d}_{i_\odot}(\beta_\rho\beta^{i_\odot})\bigg)\Bigg]\, .
\end{align}
\end{subequations}}
Here we have used the notation $i_{\odot} \equiv \{\psi, \phi\}$ to denote the angular coordinates on $S^2$. Also, $\mathfrak{d}_\varsigma \equiv \del_\varsigma - \tanh\varsigma, \mathfrak{d}_\psi \equiv \del_\psi + \cot\psi$ and $\mathfrak{d}_\phi \equiv \del_\phi$.

To summarize, in computing the hydrodynamic equations above, we have illustrated the potential utility of the subleading terms that arise for a fluid in the Carrollian regime, and exploited the map between Bjorken/Gubser flow models and Carroll fluids to compute nontrivial correction terms that are expected to better model the spacetime dynamics of the QGP in heavy-ion collisions. Needless to say, the procedure detailed above can be carried on to include even more subleading terms in the Carrollian regime, which would then be expected to provide an even better description of the QGP dynamics. 

%%%%%%%%%%%%%%%%%%%%%%%%%%%%%%%%%%%
\section{Discussion and outlook}
\label{sec:discussion}
In this paper, we have explored hydrodynamics in the Carrollian regime. This amounts to keeping subleading terms in a systematic $c\to 0$ expansion of relativistic hydrodynamics, capturing departures from the strict Carroll limit, where one keeps only the leading order pieces. Several scenarios where these subleading terms might become relevant were envisioned in section \ref{sec:interplay}, including situations that may involve an interesting interplay between the $c\to 0$ expansion versus the hydrodynamic derivative expansion. As concrete applications exemplifying the utility of these subleading terms, we explored the mapping between Carroll hydrodynamics and certain boost-invariant models for the spacetime evolution of the QGP produced in heavy-ion collisions, namely Bjorken and Gubser flow \cite{Bagchi:2023ysc, Bagchi:2023rwd}. We have computed corrections to the hydrodynamic equations for these models that follow from the subleading terms in the Carrollian regime, which bring in dependence on rapidity (along with other coordinates in general), thereby taking us beyond the strict and somewhat oversimplifying assumption of exact boost invariance that underlies these models.  

There are several directions worth pursuing in the future. First and foremost, it would be exciting to look at how well the corrections we have computed describe the QGP data. There are numerous studies that fit the QGP data to Bjorken/Gubser flow in an attempt to extract its properties - see for e.g.~\cite{Dusling:2007gi, Kurkela:2019set, Nijs:2020roc}, including holographic approaches seeking to address the issue of rapidity dependence of the flow \cite{vanderSchee:2015rta}. We believe that the correction terms we have found will lead to improved fits for the data, and hence a more accurate determination of the QGP properties. 

In the present work, we have made use of the PR parametrization for the background geometry and performed the $c\to 0$ expansion to arrive at the hydrodynamic equations in the Carrollian regime. This builds upon the previous works \cite{Bagchi:2023ysc, Bagchi:2023rwd}, where the mapping between Carroll hydrodynamics and Bjorken/Gubser flow was established using the same machinery. A more general approach, however, would be to work with pre-ultra-local (PUL) variables \cite{Hansen:2021fxi, Armas:2023dcz} and perform a $c\to 0$ expansion, without choosing any specific parametrization for the metric from the outset, and arrive at the hydrodynamic equations in the Carrollian regime. The equations we have obtained in the present work via the PR parametrization scheme will follow from these more general equations simply by choosing the PR gauge \cite{Armas:2023dcz}. The distinct advantage offered by a formulation in terms of PUL variables is that it can help in demystifying and lead to a better understanding of several obscure aspects associated with the symmetries and coordinate transformation properties of the various quantities involved, in turn making the formulation of hydrodynamics in the Carrollian regime more transparent. 

Another notable question to ponder over is what happens to the entropy current in the Carrollian regime. In the construction of relativistic hydrodynamic theories, the entropy current plays an important role in imposing constraints on the various transport parameters present in the constitutive relations of the conserved currents \cite{Loganayagam:2008is, Romatschke:2009kr, Bhattacharyya:2012nq, Bhattacharyya:2013lha, Bhattacharyya:2014bha}. These constraints follow by demanding the positive semidefiniteness of the onshell divergence of the entropy current, which is equivalent to demanding the second law of thermodynamics being locally satisfied by the system. In the construction of Carroll hydrodynamics using the $c\to 0$ limit of relativistic hydrodynamics the transport parameters are automatically constrained, at least at the leading order. However, in a first principles construction of Carroll hydrodynamics, using for instance the PUL variables, one has to exploit the entropy current to constrain the transport parameters. Such an approach can in particular allow one to explore the possibility of purely Carrollian transport parameters, ones which may not be arrived at by the $c\to 0$ limit from a relativistic counterpart. The first steps along this direction have been taken in \cite{Armas:2023dcz}, where the authors construct a more general class of Carroll fluids, with 12 independent transport parameters at the first order in the hydrodynamic derivative expansion. It would further be interesting to think about the entropy current in light of the emergence of Carroll hydrodynamics on the horizons of black holes \cite{Donnay:2019jiz}, and possible connections of such scenarios with holography \cite{Bhattacharyya:2008xc}.

Finally, it would be interesting to work out the symmetry properties of various terms that arise in the Carrollian regime. In the strict Carroll limit, the symmetries are of course Carrollian. However, the subleading terms will exhibit departures from this. A systematic way to approach this problem is to look at the small-$c$ expansion of the Poincar\'{e} algebra itself, which in the strict Carroll limit gives rise to the Carroll algebra, but the subleading terms in the $c\to 0$ expansion of the Poincar\'{e} generators will have a different symmetry structure. This will have relevance beyond hydrodynamics as well. The procedure of Lie algebra expansion with relevance to the Galilean limit has already been discussed in \cite{Hansen:2019vqf, Bergshoeff:2019ctr, Gomis:2019fdh, Gomis:2019sqv, Hansen:2020pqs}.

The above discussion conveys the fact that there are many uncharted directions in the field of Carroll hydrodynamics, which need further thought and exploration. We hope to report upon some of these directions in the near future.

%%%%%%%%%%%%%%%%%%%%%%%%%%%%%%%%%%%
\acknowledgments
We would like to thank Arjun Bagchi for initial collaboration and several helpful discussions. PS is also grateful to Sudipta Dutta for useful discussions.

\medskip

The work of KK is partially supported by the China Postdoctoral Science Foundation from the 76th batch of general funding under grant 2024M761593, ``Quantum theories near null surfaces.'' The work of TM is supported by the grant RJF/2022/000130 from the Science and Engineering Research Board (SERB), India. The work of AS is partly supported by the European Research Council (ERC) under the European Union’s Horizon 2020 research and innovation programme (grant agreement no. 758759). PS is supported by an IIT Kanpur Institute Assistantship. AS would like to thank IIT Kanpur for hospitality during the last stages of this project.

%%%%%%%%%%%%%%%%%%%%%%%%%%%%%%%%%%%
\appendix
%%%%%%%%%%%%%%%%%%%%%%%%%%%%%%%%%%%
\section{Local Carroll boost invariance of the hydrodynamic equations}
\label{CBI-FE}
The hydrodynamic equations \eqref{LOVF} for a viscous Carroll fluid (or eqs.~\eqref{LOPF} for an ideal Carroll fluid), obtained at the leading order in a small-$c$ expansion from relativistic hydrodynamic equations, are expected to be invariant under the local tangent space symmetries of the underlying Carroll manifold. Of particular interest is invariance under local Carroll boosts. In this appendix, we illustrate the local Carroll boost invariance of the Carroll fluid equations. We further argue that the choice of geometric data on the Carroll manifold that maps these hydrodynamic equations to either Bjorken or Gubser flow also respects local Carroll boost invariance. To be more general, we demonstrate the invariance of the viscous equations \eqref{LOVF}; the invariance of the ideal fluid equations \eqref{LOPF} follows simply by turning off the viscous transport parameters. 

As mentioned above, Carroll boosts are realized as local symmetries acting in the tangent space on an arbitrary Carrollian manifold. In \cite{Hartong:2015xda}, using the formal procedure of gauging the Carroll algebra, it was computed how the degenerate metric $h_{\mu\nu}$, its kernel $k^\mu$, the dual form $\vartheta_\mu$ and the inverse metric $h^{\mu\nu}$ transform under local Carroll boosts. One has
\begin{equation}\label{transformations-local-c-boost}
    \delta_c h_{\mu\nu}=0, \quad \delta_c k^\mu=0,\quad \delta_c \vartheta_\mu=\lambda_\mu,\quad \delta_ch^{\mu\nu}=-2\lambda _\rho h^{\rho(\mu}k^{\nu)},
\end{equation}
where $\delta_c$ denotes variation under a local Carroll boost transformation with parameters $\lambda^i$ (the index $i$ runs over spatial directions), while $\lambda^\mu$ is defined in terms of $\lambda^i$ and the inverse spatial vielbein $e^\mu_{~i}$ via $\lambda^\mu \equiv e^\mu_{~i} \lambda^i$. The transformations above preserve the following orthogonality and completeness relations,
\begin{equation}
    k^\mu h_{\mu\nu}=0, \quad  \vartheta_\mu h^{\mu\nu}=0,\quad k^\mu \vartheta_\mu=1, \quad k^\mu \vartheta_\nu+h^{\mu\sigma}h_{\sigma\nu}=\delta^\mu_{\ \nu}.
\end{equation}
Now, for the PR parametrization discussed in section \ref{sec:PR}, one has
\begin{equation}
    \label{eq:LO-PRgauge}
	k^{\mu} = (\Omega^{-1},0), \quad \vartheta_{\mu} = (\Omega, - b_i), \quad h_{\mu\nu} = \begin{pmatrix}
		0 & \vec{0} \\
		\vec{0} & a_{ij}
	\end{pmatrix}, \quad h^{\mu\nu} = \begin{pmatrix}
		\Omega^{-2}b^2 & \Omega^{-1}b^j \\
		\Omega^{-1}b^i & a^{ij}
	\end{pmatrix}.
\end{equation}
Then, making use of eq.~\eqref{transformations-local-c-boost}, the various quantities entering the PR parametrization transform under a local Carroll boost via
\begin{equation}
\label{loc-transformation-PR}
    \delta_c \Omega=0,\quad \delta_c a_{ij}=0,\quad \delta_c a^{ij}=0,\quad \delta_cb_i=-\lambda_i.
\end{equation}
The transformations above further imply that under local Carroll boosts one has $\delta_c\hat{\gamma}_{ij}=0$, with $\hat{\gamma}_{ij}$ being the temporal Levi-Civita-Carroll connection, eq.~\eqref{Carroll-Christoffel}. Now, to determine how the fields $\epsilon_{(0)}, p_{(0)}$ and $\beta_i$ transform,  we use the fact that under local Carroll boosts the energy-momentum tensor eq.~\eqref{LOT1} for a Carroll fluid remains invariant \cite{Armas:2023dcz}, thereby implying
\begin{equation}
\label{goldstone}
    \delta_c \epsilon_{(0)}=0,\quad \delta_cp_{(0)}=0,\quad \delta_c \beta_i=-\delta_c b_i=\lambda_i.
\end{equation}
Equipped with these results, we are now prepared to investigate how the hydrodynamic equations transform under local Carroll boosts.\\

\noindent \ding{112} \underline{\textit{Energy equation}}: \\
We first consider the hydrodynamic equation \eqref{LOVF1}, which can be rewritten as
\begin{equation}
    \hat{\partial}_{\,\rm t}\epsilon_{(0)} + \theta\Big(\epsilon_{(0)} + p_{(0)} - \f{\Xi}{d}\Big) - \xi^{ij}\Xi_{ij} = 0.
    \label{CEE-II}
\end{equation}
Consider now the LHS of the equation above. From  eq.~\eqref{goldstone}, under a local Carroll boost we have $\delta_c \epsilon_{(0)}=\delta_c p_{(0)}=0$. Also, since $\theta=\Omega^{-1} a^{ij}\partial_{\rm t}a_{ij}$, using eq.~\eqref{loc-transformation-PR} it is also invariant under local Carroll boosts. Next, we consider the quantities $\xi^{ij},~\Xi_{ij}$ and $\Xi$, which are given by
\begin{equation}
    \xi^{ij}=a^{ik}a^{jl}\hat{\gamma}_{kl}-\frac{\theta}{d}a^{ij},\quad \Xi_{ij}=\eta_{(0)}\xi_{ij}+\zeta_{(0)}\theta a_{ij},\quad \Xi=\zeta_{(0)}\theta d.
\end{equation}
It is straightforward to see that\footnote{The viscosities $\eta_{(0)}, \zeta_{(0)}$ are functions of $\epsilon_{(0)}$, and are thus local Carroll boost invariant too.}
\begin{equation}
    \delta_c\xi^{ij}=0,\quad\delta_c \Xi_{ij}=0,\quad \delta_c\Xi=0.
\end{equation}
Since the variation of all the terms in eq.~\eqref{CEE-II} vanishes individually, this proves that the energy equation eq.~\eqref{LOVF1} is invariant under local Carroll boosts.\\

\noindent \ding{112} \underline{\textit{Momentum equation}}: \\
We next consider the hydrodynamic equation \eqref{LOVF2}, which can be rewritten as 
\begin{align}
     \hat{\partial}_i p_{(0)}   + \varphi_i(\epsilon_{(0)} + p_{(0)}) + (\hat{\partial}_{\,\rm t} + \theta)[(\epsilon_{(0)} + p_{(0)})\beta_i - \Xi_{ij}\beta^j] - (\hat{\nabla}_j + \varphi_j)\Xi^{j}_{\ i}=0.
     \label{CME-II}
\end{align}
Using the results obtained above, we can compute the variation of various terms in eq.~\eqref{CME-II} under a local Carroll boost. We get
\begin{align}
         &\delta_c[\hat{\partial}_ip_{(0)}]=-\lambda_i \hat{\partial}_{\rm t}p_{(0)}, \quad \delta_c[\varphi_i(\epsilon_{(0)}+p_{(0)})]=- (\epsilon_{(0)}+p_{(0)})\hat{\partial}_{\rm t}\lambda_i, \quad \delta_c[\varphi_j\Xi^{j}_{\ i}]=-\hat{\partial}_{\rm t}\lambda_j \Xi^j_{\ i}, \nonumber\\
         &\delta_c[(\hat{\partial}_{\rm t}+\theta)(\epsilon_{(0)}+p_{(0)})\beta_i]=(\hat{\partial}_{\rm t}+\theta)(\epsilon_{(0)}+p_{(0)})\lambda_i,\nonumber\\
         &\delta_c[(\hat{\partial}_{\rm t}+\theta)\Xi_{ij}\beta^j]=(\hat{\partial}_{\rm t}+\theta)\Xi_{ij}\lambda^j,\quad \delta_c[\hat{\nabla}_j\Xi^j_{\ i}]=-\lambda_j(\hat{\partial}_{\rm t}\Xi^j_{\ i}+\theta \Xi^j_{\ i})+\lambda_i\Xi^{jl}\hat{\gamma}_{jl}.
\end{align}
Thus the total variation of the LHS of eq.~\eqref{CME-II} turns out to be
\begin{equation}
    \lambda_i\left(\hat\partial_{\rm t}\epsilon_{(0)}+\theta (\epsilon_{(0)}+p_{(0)})-\Xi^{jl}\hat{\gamma}_{jl}\right).
\end{equation}
But this too vanishes, since the quantity in the parentheses above is nothing but the LHS of the energy equation \eqref{CEE-II} (use $\hat{\gamma}_{ij} = \xi_{ij} + \theta a_{ij}/d$). This proves that the momentum equation \eqref{LOVF2} is also local Carroll boost invariant.\\

\noindent \ding{112} \underline{\textit{Carroll boost invariance of Bjorken/Gubser mapping}}: \\
We now show that the geometric data one chooses to map the Carroll fluid equations to the equations for Bjorken or Gubser flow respects local Carroll boost invariance. 

Consider first the case for Bjorken flow. Apart from fixing $\Omega$ and $a_{ij}$ to the form given by eq.~\eqref{Bjorken_Data}, which are local Carroll boost invariant, eq.~\eqref{loc-transformation-PR}, we impose the relation $b_i = - \beta_i$. However, from eq.~\eqref{goldstone}, it is clear that this condition is also local Carroll boost invariant. In fact, one uses the same condition for the case of Gubser flow in ${\rm dS}_3\times \mathbb{R}$ coordinates, eq.~\eqref{Gubser_Data}, implying that the geometric data for Gubser flow too respects local Carroll boost invariance. Combined with the invariance of the Carroll fluid equations \eqref{LOVF}, this conclusively proves that eqs.~\eqref{Bjorken_Visc} and \eqref{Gubser_Visc} (as well as their ideal fluid analogs, eqs.~\eqref{Bjorken_Ideal} and \eqref{Gubser_Ideal}) are local Carroll boost invariant.

Though not discussed in the present work, one can show that the geometric data that maps the Carroll fluid equations \eqref{LOVF} to Gubser flow in Milne coordinates also respects local Carroll boost invariance. The conditions one imposes for this case are \cite{Bagchi:2023rwd},
\begin{eqnarray}
    b_r=-\beta_r +\operatorname{sinh}\kappa,\quad b_\rho=-\beta_\rho,\quad b_\phi=-\beta_\phi,
    \label{Gubser_Data_2}
\end{eqnarray}
with $(r, \phi)$ being the radial and angular coordinates in the plane transverse to the beam axis. Since $\kappa= \cosh^{-1}\Omega$, one has $\delta_c\kappa=0$ following from $\delta_c\Omega = 0$, eq.~\eqref{loc-transformation-PR}. Therefore, the mapping in eq.~\eqref{Gubser_Data_2} is local Carroll boost invariant too.

The discussion above clarifies the Carrollian nature of the fluid equations as well as the maps to Bjorken and Gubser flow. In particular, it abolishes any confusion that may be caused by thinking in terms of Carroll diffeomorphisms, under which $b_i$ and $\beta_i$ transform differently, and make it appear as if the maps only respect Aristotelian symmetries. Under a general diffeomorphism, both $b_i$ and $\beta_i$ transform like the spatial components of a one-form field. Since Carroll diffeomorphisms are only a subset of the most general diffeomorphisms, they may obscure certain features of the setup. On the other hand, the fact that local tangent space symmetries are Carrollian throughout ensures that the dynamics is indeed Carrollian in nature.

\section{Second order derivative corrections for a conformal Carroll fluid}
\label{sec:second_der}
In this appendix, we aim to establish an understanding of the second order derivative corrections in the hydrodynamic derivative expansion for a conformal Carroll fluid. As has been the strategy so far, we first list out the independent second order terms that arise for a conformal relativistic fluid, and then impose the $c\to 0$ limit on them to extract their behaviour in the Carrollian regime.

Let us begin by listing the independent second order derivative corrections that can contribute to the energy-momentum tensor of a conformal relativistic fluid in the Landau frame. Following the notation set out in \cite{Baier:2007ix}, these are
\be
T^{\m\n}_{(2)} = \tau_{\pi} \mathbb{S}_1^{\m\n} + \k \, \mathbb{S}^{\m\n}_2 + \l_1 \mathbb{S}_3^{\m\n} + \l_2 \mathbb{S}_4^{\m\n} + \l_3 \mathbb{S}_5^{\m\n} ,
\label{Tsecond}
\ee
with\footnote{Indices enclosed within angle brackets denote a symmetric-transverse-traceless projection i.e.
\begin{equation*}
A^{\langle\m\n\rangle} = \D^{\m\a} \D^{\n\b} A_{(\a\b)} -\f{1}{d} \D^{\m\n} \D^{\a\b} A_{\a\b}. 
\end{equation*}}
{\allowdisplaybreaks
\begin{align}
\mathbb{S}_1^{\m\n} &=  u\cdot\nabla \s^{\langle\m\n\rangle} + \f{\Theta}{d} \s^{\m\n}\, , \quad \mathbb{S}_2^{\m\n} = \left(c^2 R^{\langle\m\n\rangle} - (d-1) u_\a  R^{\a\langle\m\n\rangle\b}u_\b\right),\nonumber\\
\mathbb{S}_3^{\m\n} &= \s^{\langle\m}_{~~\l} \s^{\n\rangle\l} \, ,\quad \mathbb{S}_4^{\m\n} = \s^{\langle\m}_{~~\l} \o^{\n\rangle\l}\, , \quad \mathbb{S}_5^{\m\n} =  \o^{\langle\m}_{~~\l} \o^{\n\rangle\l}.
\end{align}}
Here $\s^{\m\n}$ is the shear tensor, eq.~\eqref{shear}, $R_{\m\n\a\b}$ and $R_{\m\n}$ are the Riemann and Ricci tensors of the background, respectively, while $\o^{\m\n}$ is the vorticity tensor, defined via
\be
\o^{\m\n} \equiv \D^{\m\a} \D^{\n\b} \nabla_{[\a} u_{\b]}.
\ee
Of the five terms that appear in eq.~\eqref{Tsecond}, $\mathbb{S}_2^{\m\n}$ and $\mathbb{S}_5^{\m\n}$ are thermodynamic in nature \cite{Kovtun:2018dvd, Kovtun:2019wjz}, implying that they can be non-vanishing even in the limit of thermal equilibrium (see for e.g.~\cite{Shukla:2019shf, Grieninger:2021rxd}), while the remaining are genuine out of equilibrium effects. 

On imposing the $c\to 0$ limit on the derivative corrections appearing in the constitutive relation eq.~\eqref{Tsecond}, we find the following behaviour for the various terms.
{\allowdisplaybreaks
\begin{align}
\mathbb{S}_1\mbox{}^{\!\rm{t}}_{~\rm{t}} &= \mc{O}(c^2)\, , \qquad \mathbb{S}_1\mbox{}^{\!i}_{~\rm{t}} = - c^2\O \b^j \left(\f{\th}{d}\, \xi^i_{~j} + \psi^i_{~j}\right) + \mc{O}(c^4), \nonumber \\
\mathbb{S}_1\mbox{}^{\!\rm{t}}_{~i} &=  \f{b_j+\b_j}{\O} \left(\f{\th}{d}\,\xi_{ji}+\psi_{ji}\right) + \mc{O}(c^2)\, , \quad \mathbb{S}_1\mbox{}^{\!i}_{~j} = \f{\th}{d}\,\xi^i_{~j} + \psi^i_{~j} + \mc{O}(c^2)\, ,\nonumber \\
\mathbb{S}_2\mbox{}^{\!\rm{t}}_{~\rm{t}} &= \mc{O}(c^2)\, , \qquad \mathbb{S}_2\mbox{}^{\!i}_{~\rm{t}} = -c^2 \O \b^j \mc{R}^i_{~j} + \mc{O}(c^4)\, ,\nonumber \\
\mathbb{S}_2\mbox{}^{\!\rm{t}}_{~i} &= \f{b_j+\b_j}{\O} \mc{R}^{j}_{~i} + \mc{O}(c^2) \, , \quad \mathbb{S}_2\mbox{}^{\!i}_{~j} =  \mc{R}^i_{~j} + \mc{O}(c^2)\, , \nonumber \\
\mathbb{S}_3\mbox{}^{\!\rm{t}}_{~\rm{t}} &= \mc{O}(c^2) \, , \quad \mathbb{S}_3\mbox{}^{\!i}_{~\rm{t}} = - c^2 \O \b^j \left(\xi^i_{~k} \xi^k_{~j} - \f{1}{d}\,\d^i_{~j}\xi^2\right) + \mc{O}(c^4)\, , \nonumber \\
\mathbb{S}_3\mbox{}^{\!\rm{t}}_{~i} &= \f{b_j+\b_j}{\O}\left(\xi_{jk}\xi^k_{~i} - \f{1}{d} a_{ij} \xi^2\right) + \mc{O}(c^2)\, , \quad \mathbb{S}_3\mbox{}^{\!i}_{~j} = \xi^i_{~k} \xi^k_{~j} - \f{1}{d}\,\d^i_{~j} \xi^2 + \mc{O}(c^2) \, ,\nonumber \\
\mathbb{S}_4\mbox{}^{\!\rm{t}}_{~\rm{t}} &= \mc{O}(c^4)\, , \quad \mathbb{S}_4\mbox{}^{\!i}_{~\rm{t}} = -c^4\O\b^j\mc{A}^i_{~j}+\mc{O}(c^6)\, , \nonumber \\
\mathbb{S}_4\mbox{}^{\!\rm{t}}_{~i} &= \f{c^2(b_j+\b_j)}{\O} \,\mc{A}^j_{~i} + \mc{O}(c^4)\, , \quad \mathbb{S}_4\mbox{}^{\!i}_{~j} = c^2\mc{A}_{~j}^i + \mc{O}(c^4)\, , \nonumber \\
\mathbb{S}_5\mbox{}^{\!\rm{t}}_{~\rm{t}} &= \mc{O}(c^6)\, , \quad \mathbb{S}_5\mbox{}^{\!i}_{~\rm{t}} = -c^6\O\b^j\mc{B}^i_{~j} + \mc{O}(c^8)\, , \nonumber \\
\mathbb{S}_5\mbox{}^{\!\rm{t}}_{~i} &= \f{c^4(b_j+\b_j)}{\O} \, \mc{B}_{~i}^j + \mc{O}(c^6)\, , \quad \mathbb{S}_5\mbox{}^{\!i}_{~j} = c^4 \mc{B}^i_{~j} + \mc{O}(c^6)\, . \label{2ordbeh}
\end{align}}
In writing eq.~\eqref{2ordbeh}, we have introduced the notation
{\allowdisplaybreaks
\begin{align}
\psi_{ij} &\equiv \hat{\nabla}_{\rm t} \xi_{ij}\, , \quad
\mc{R}^i_{~j} \equiv \left(a^{ik}\, \ost{R}{-2}_{kj} - \f{1}{d}\, \d^i_{\,j} a^{kl} \; \ost{R}{-2}_{kl}\right) + (d-1)\left(a^{ik} \, \ost{R}{-2}^{\,\rm t}_{~kj\rm{t}} - \f{1}{d} \d^i_{\,j} a^{kl} \; \ost{R}{-2}^{\,\rm t}_{~kl\rm{t}}\right), \nonumber \\
\mc{A}^i_{~j} &\equiv a^{ik} \tilde{\o}_{kl}\xi^l_{~j} - \f{1}{d} \, \d^i_{~j} \tilde{\o}_{kl} \xi^{kl}\, , \quad \mc{B}^i_{~j} \equiv a^{ik}\tilde{\o}_{kl}\tilde{\o}^l_{~j} - \f{1}{d} \, \d^i_{~j} \tilde{\o}_{kl} \tilde{\o}^{kl}\, ,
\end{align}}
where $\ost{R}{-2}_{ij} = \hat{\nabla}_{\rm t} \hat{\g}_{ij} + \th \hat{\g}_{ij}$, $\ost{R}{-2}^{\,\rm t}_{~ij\rm{t}} = - \hat{\nabla}_{\rm t} \hat{\g}_{ij} - \hat{\g}_{ik} \hat{\g}^k_{~j}$ and $\tilde{\o}_{ij} = \hat{\N}_{[i} \b_{j]} + \f{1}{2} f_{ij} + \b_{[i} \phi_{j]} + \b_{[i} \hat{\del}_{\,\rm t} \b_{j]}$ are the leading terms in the $c\to 0$ expansion of the background Ricci, Riemann and vorticity tensors, respectively. 

As applications of the $c\to 0$ behaviour of the second order derivative corrections computed above, we can once again look at the (conformal) Bjorken and Gubser flow models. It turns out that the correct ansatz for the second order transport parameters which ensures the mapping between these models and Carroll hydrodynamics continues to work is
\be
\begin{split}
\tau_\pi &\xrightarrow{c\to 0} \tau_{\pi(0)} + c^2 \tau_{\pi(2)} + \mc{O}(c^4),\\
\k &\xrightarrow{c\to 0} c^2 \k_{(2)} + c^4 \k_{(4)} + \mc{O}(c^6), \\
\l_1 &\xrightarrow{c\to 0} \l_{1(0)} + c^2 \l_{1(2)} + \mc{O}(c^4).
\end{split}
\label{2ordansatz}
\ee
Further, both Bjorken and Gubser flow models are non-vortical, thereby the terms $\mathbb{S}_4$ and $\mathbb{S}_5$ do not play any role. With this, one finds that the equations of Carroll hydrodynamics with the choice of data eq.~\eqref{Bjorken_Data} yield the conformal Bjorken flow equations with the inclusion of second order derivative corrections,
\be
\del_\tau \epsilon_{(0)} = - \f{4\epsilon_{(0)}}{3\t} + \f{2\eta_{(0)}}{3\tau^2} + \f{4\tau_{\pi(0)}}{9\t^3} - \f{2\l_{1(0)}}{9\t^3}\, , \quad \del_{i}\epsilon_{(0)} = 0\, ,
\label{2ord_Bjorken}
\ee 
and with the choice of data eq.~\eqref{Gubser_Data} yield the Gubser flow equations with second order derivative corrections,
\begin{align}
\del_\varsigma\epsilon_{(0)} &= - \f{8\epsilon_{(0)}}{3}\tanh\varsigma+\f{2\eta_{(0)}}{3} \tanh^2\varsigma - \f{2\tau_{\pi(0)}}{3}\tanh\varsigma\Big(\sech^2\varsigma+\f{2}{3} \tanh^2\varsigma\Big)  \nonumber \\
&\,\,\quad+ \f{2\l_{1(0)}}{9} \tanh^3\varsigma\, , \label{2ord_Gubser}\\
\del_i \epsilon_{(0)} &= 0. \nonumber 
\end{align}
As expected, in the strict Carroll limit, the independence from rapidity is maintained. In the spirit of this paper, one can compute the subleading terms in $c\to 0$ that follow from eq.~\eqref{2ordbeh} and find new terms beyond the leading ones in eqs.~\eqref{2ord_Bjorken} and \eqref{2ord_Gubser}, which will bring in dependence on rapidity, now including second order derivative corrections as well.

%%%%%%%%%%%%%%%%%%%%%%%%%%%%%%%%%%%
\bibliography{refs}

\begin{thebibliography}{78}%
\makeatletter
\providecommand \@ifxundefined [1]{%
 \@ifx{#1\undefined}
}%
\providecommand \@ifnum [1]{%
 \ifnum #1\expandafter \@firstoftwo
 \else \expandafter \@secondoftwo
 \fi
}%
\providecommand \@ifx [1]{%
 \ifx #1\expandafter \@firstoftwo
 \else \expandafter \@secondoftwo
 \fi
}%
\providecommand \natexlab [1]{#1}%
\providecommand \enquote  [1]{``#1''}%
\providecommand \bibnamefont  [1]{#1}%
\providecommand \bibfnamefont [1]{#1}%
\providecommand \citenamefont [1]{#1}%
\providecommand \href@noop [0]{\@secondoftwo}%
\providecommand \href [0]{\begingroup \@sanitize@url \@href}%
\providecommand \@href[1]{\@@startlink{#1}\@@href}%
\providecommand \@@href[1]{\endgroup#1\@@endlink}%
\providecommand \@sanitize@url [0]{\catcode `\\12\catcode `\$12\catcode
  `\&12\catcode `\#12\catcode `\^12\catcode `\_12\catcode `\%12\relax}%
\providecommand \@@startlink[1]{}%
\providecommand \@@endlink[0]{}%
\providecommand \url  [0]{\begingroup\@sanitize@url \@url }%
\providecommand \@url [1]{\endgroup\@href {#1}{\urlprefix }}%
\providecommand \urlprefix  [0]{URL }%
\providecommand \Eprint [0]{\href }%
\providecommand \doibase [0]{https://doi.org/}%
\providecommand \selectlanguage [0]{\@gobble}%
\providecommand \bibinfo  [0]{\@secondoftwo}%
\providecommand \bibfield  [0]{\@secondoftwo}%
\providecommand \translation [1]{[#1]}%
\providecommand \BibitemOpen [0]{}%
\providecommand \bibitemStop [0]{}%
\providecommand \bibitemNoStop [0]{.\EOS\space}%
\providecommand \EOS [0]{\spacefactor3000\relax}%
\providecommand \BibitemShut  [1]{\csname bibitem#1\endcsname}%
\let\auto@bib@innerbib\@empty
%</preamble>
\bibitem [{\citenamefont {Landau}\ and\ \citenamefont
  {Lifshitz}(1987)}]{Landau-Lifshitz}%
  \BibitemOpen
  \bibfield  {author} {\bibinfo {author} {\bibfnamefont {L.~D.}\ \bibnamefont
  {Landau}}\ and\ \bibinfo {author} {\bibfnamefont {E.~M.}\ \bibnamefont
  {Lifshitz}},\ }\href@noop {} {\emph {\bibinfo {title} {Fluid mechanics}}},\
  \bibinfo {edition} {2nd}\ ed.,\ \bibinfo {series} {Course of Theoretical
  Physics}, Vol.~\bibinfo {volume} {6}\ (\bibinfo  {publisher} {Elsevier},\
  \bibinfo {year} {1987})\BibitemShut {NoStop}%
\bibitem [{\citenamefont {Kovtun}(2012)}]{Kovtun:2012rj}%
  \BibitemOpen
  \bibfield  {author} {\bibinfo {author} {\bibfnamefont {P.}~\bibnamefont
  {Kovtun}},\ }\bibfield  {title} {\bibinfo {title} {{Lectures on hydrodynamic
  fluctuations in relativistic theories}},\ }\href
  {https://doi.org/10.1088/1751-8113/45/47/473001} {\bibfield  {journal}
  {\bibinfo  {journal} {J. Phys. A}\ }\textbf {\bibinfo {volume} {45}},\
  \bibinfo {pages} {473001} (\bibinfo {year} {2012})},\ \Eprint
  {https://arxiv.org/abs/1205.5040} {arXiv:1205.5040 [hep-th]} \BibitemShut
  {NoStop}%
\bibitem [{\citenamefont {Jeon}\ and\ \citenamefont
  {Heinz}(2015)}]{Jeon:2015dfa}%
  \BibitemOpen
  \bibfield  {author} {\bibinfo {author} {\bibfnamefont {S.}~\bibnamefont
  {Jeon}}\ and\ \bibinfo {author} {\bibfnamefont {U.}~\bibnamefont {Heinz}},\
  }\bibfield  {title} {\bibinfo {title} {{Introduction to Hydrodynamics}},\
  }\href {https://doi.org/10.1142/S0218301315300106} {\bibfield  {journal}
  {\bibinfo  {journal} {Int. J. Mod. Phys. E}\ }\textbf {\bibinfo {volume}
  {24}},\ \bibinfo {pages} {1530010} (\bibinfo {year} {2015})},\ \Eprint
  {https://arxiv.org/abs/1503.03931} {arXiv:1503.03931 [hep-ph]} \BibitemShut
  {NoStop}%
\bibitem [{\citenamefont {Romatschke}\ and\ \citenamefont
  {Romatschke}(2019)}]{Romatschke:2017ejr}%
  \BibitemOpen
  \bibfield  {author} {\bibinfo {author} {\bibfnamefont {P.}~\bibnamefont
  {Romatschke}}\ and\ \bibinfo {author} {\bibfnamefont {U.}~\bibnamefont
  {Romatschke}},\ }\href@noop {} {\emph {\bibinfo {title} {Relativistic Fluid
  Dynamics In and Out of Equilibrium}}}\ (\bibinfo  {publisher} {Cambridge
  University Press},\ \bibinfo {year} {2019})\BibitemShut {NoStop}%
\bibitem [{\citenamefont {Ciambelli}\ \emph {et~al.}(2018)\citenamefont
  {Ciambelli}, \citenamefont {Marteau}, \citenamefont {Petkou}, \citenamefont
  {Petropoulos},\ and\ \citenamefont {Siampos}}]{Ciambelli:2018xat}%
  \BibitemOpen
  \bibfield  {author} {\bibinfo {author} {\bibfnamefont {L.}~\bibnamefont
  {Ciambelli}}, \bibinfo {author} {\bibfnamefont {C.}~\bibnamefont {Marteau}},
  \bibinfo {author} {\bibfnamefont {A.~C.}\ \bibnamefont {Petkou}}, \bibinfo
  {author} {\bibfnamefont {P.~M.}\ \bibnamefont {Petropoulos}},\ and\ \bibinfo
  {author} {\bibfnamefont {K.}~\bibnamefont {Siampos}},\ }\bibfield  {title}
  {\bibinfo {title} {{Covariant Galilean versus Carrollian hydrodynamics from
  relativistic fluids}},\ }\href {https://doi.org/10.1088/1361-6382/aacf1a}
  {\bibfield  {journal} {\bibinfo  {journal} {Class. Quant. Grav.}\ }\textbf
  {\bibinfo {volume} {35}},\ \bibinfo {pages} {165001} (\bibinfo {year}
  {2018})},\ \Eprint {https://arxiv.org/abs/1802.05286} {arXiv:1802.05286
  [hep-th]} \BibitemShut {NoStop}%
\bibitem [{\citenamefont {Petkou}\ \emph {et~al.}(2022)\citenamefont {Petkou},
  \citenamefont {Petropoulos}, \citenamefont {Betancour},\ and\ \citenamefont
  {Siampos}}]{Petkou:2022bmz}%
  \BibitemOpen
  \bibfield  {author} {\bibinfo {author} {\bibfnamefont {A.~C.}\ \bibnamefont
  {Petkou}}, \bibinfo {author} {\bibfnamefont {P.~M.}\ \bibnamefont
  {Petropoulos}}, \bibinfo {author} {\bibfnamefont {D.~R.}\ \bibnamefont
  {Betancour}},\ and\ \bibinfo {author} {\bibfnamefont {K.}~\bibnamefont
  {Siampos}},\ }\bibfield  {title} {\bibinfo {title} {{Relativistic fluids,
  hydrodynamic frames and their Galilean versus Carrollian avatars}},\ }\href
  {https://doi.org/10.1007/JHEP09(2022)162} {\bibfield  {journal} {\bibinfo
  {journal} {JHEP}\ }\textbf {\bibinfo {volume} {09}},\ \bibinfo {pages}
  {162}},\ \Eprint {https://arxiv.org/abs/2205.09142} {arXiv:2205.09142
  [hep-th]} \BibitemShut {NoStop}%
\bibitem [{\citenamefont {Lévy-Leblond}(1965)}]{LevyLeblond}%
  \BibitemOpen
  \bibfield  {author} {\bibinfo {author} {\bibfnamefont {J.-M.}\ \bibnamefont
  {Lévy-Leblond}},\ }\bibfield  {title} {\bibinfo {title} {Une nouvelle limite
  non-relativiste du groupe de poincaré},\ }\href {http://eudml.org/doc/75509}
  {\bibfield  {journal} {\bibinfo  {journal} {Annales de l'I.H.P. Physique
  Théorique}\ }\textbf {\bibinfo {volume} {3}},\ \bibinfo {pages} {1}
  (\bibinfo {year} {1965})}\BibitemShut {NoStop}%
\bibitem [{\citenamefont {Sen~Gupta}(1966)}]{NDS}%
  \BibitemOpen
  \bibfield  {author} {\bibinfo {author} {\bibfnamefont {N.~D.}\ \bibnamefont
  {Sen~Gupta}},\ }\bibfield  {title} {\bibinfo {title} {{On an Analogue of the
  Galileo Group}},\ }\href {https://doi.org/10.1007/BF02740871} {\bibfield
  {journal} {\bibinfo  {journal} {Il Nuovo Cimento A}\ }\textbf {\bibinfo
  {volume} {44}},\ \bibinfo {pages} {512–517} (\bibinfo {year}
  {1966})}\BibitemShut {NoStop}%
\bibitem [{\citenamefont {Majumdar}(2024)}]{Majumdar:2024rxg}%
  \BibitemOpen
  \bibfield  {author} {\bibinfo {author} {\bibfnamefont {S.}~\bibnamefont
  {Majumdar}},\ }\bibfield  {title} {\bibinfo {title} {{On the Carrollian
  Nature of the Light Front}},\ }\Eprint {https://arxiv.org/abs/2406.10353}
  {arXiv:2406.10353 [hep-th]}  (\bibinfo {year} {2024})\BibitemShut {NoStop}%
\bibitem [{\citenamefont {Bagchi}\ \emph
  {et~al.}(2024{\natexlab{a}})\citenamefont {Bagchi}, \citenamefont
  {Nachiketh},\ and\ \citenamefont {Soni}}]{Bagchi:2024epw}%
  \BibitemOpen
  \bibfield  {author} {\bibinfo {author} {\bibfnamefont {A.}~\bibnamefont
  {Bagchi}}, \bibinfo {author} {\bibfnamefont {M.}~\bibnamefont {Nachiketh}},\
  and\ \bibinfo {author} {\bibfnamefont {P.}~\bibnamefont {Soni}},\ }\bibfield
  {title} {\bibinfo {title} {{Anatomy of null contractions}},\ }\href
  {https://doi.org/10.1007/JHEP09(2024)141} {\bibfield  {journal} {\bibinfo
  {journal} {JHEP}\ }\textbf {\bibinfo {volume} {09}},\ \bibinfo {pages}
  {141}},\ \Eprint {https://arxiv.org/abs/2406.15061} {arXiv:2406.15061
  [hep-th]} \BibitemShut {NoStop}%
\bibitem [{\citenamefont {Armas}\ and\ \citenamefont
  {Have}(2024)}]{Armas:2023dcz}%
  \BibitemOpen
  \bibfield  {author} {\bibinfo {author} {\bibfnamefont {J.}~\bibnamefont
  {Armas}}\ and\ \bibinfo {author} {\bibfnamefont {E.}~\bibnamefont {Have}},\
  }\bibfield  {title} {\bibinfo {title} {{Carrollian Fluids and Spontaneous
  Breaking of Boost Symmetry}},\ }\href
  {https://doi.org/10.1103/PhysRevLett.132.161606} {\bibfield  {journal}
  {\bibinfo  {journal} {Phys. Rev. Lett.}\ }\textbf {\bibinfo {volume} {132}},\
  \bibinfo {pages} {161606} (\bibinfo {year} {2024})},\ \Eprint
  {https://arxiv.org/abs/2308.10594} {arXiv:2308.10594 [hep-th]} \BibitemShut
  {NoStop}%
\bibitem [{\citenamefont {Bjorken}(1983)}]{Bjorken:1982qr}%
  \BibitemOpen
  \bibfield  {author} {\bibinfo {author} {\bibfnamefont {J.~D.}\ \bibnamefont
  {Bjorken}},\ }\bibfield  {title} {\bibinfo {title} {{Highly Relativistic
  Nucleus-Nucleus Collisions: The Central Rapidity Region}},\ }\href
  {https://doi.org/10.1103/PhysRevD.27.140} {\bibfield  {journal} {\bibinfo
  {journal} {Phys. Rev. D}\ }\textbf {\bibinfo {volume} {27}},\ \bibinfo
  {pages} {140} (\bibinfo {year} {1983})}\BibitemShut {NoStop}%
\bibitem [{\citenamefont {Gubser}(2010)}]{Gubser:2010ze}%
  \BibitemOpen
  \bibfield  {author} {\bibinfo {author} {\bibfnamefont {S.~S.}\ \bibnamefont
  {Gubser}},\ }\bibfield  {title} {\bibinfo {title} {{Symmetry constraints on
  generalizations of Bjorken flow}},\ }\href
  {https://doi.org/10.1103/PhysRevD.82.085027} {\bibfield  {journal} {\bibinfo
  {journal} {Phys. Rev. D}\ }\textbf {\bibinfo {volume} {82}},\ \bibinfo
  {pages} {085027} (\bibinfo {year} {2010})},\ \Eprint
  {https://arxiv.org/abs/1006.0006} {arXiv:1006.0006 [hep-th]} \BibitemShut
  {NoStop}%
\bibitem [{\citenamefont {Gubser}\ and\ \citenamefont
  {Yarom}(2011)}]{Gubser:2010ui}%
  \BibitemOpen
  \bibfield  {author} {\bibinfo {author} {\bibfnamefont {S.~S.}\ \bibnamefont
  {Gubser}}\ and\ \bibinfo {author} {\bibfnamefont {A.}~\bibnamefont {Yarom}},\
  }\bibfield  {title} {\bibinfo {title} {{Conformal hydrodynamics in Minkowski
  and de Sitter spacetimes}},\ }\href
  {https://doi.org/10.1016/j.nuclphysb.2011.01.012} {\bibfield  {journal}
  {\bibinfo  {journal} {Nucl. Phys. B}\ }\textbf {\bibinfo {volume} {846}},\
  \bibinfo {pages} {469} (\bibinfo {year} {2011})},\ \Eprint
  {https://arxiv.org/abs/1012.1314} {arXiv:1012.1314 [hep-th]} \BibitemShut
  {NoStop}%
\bibitem [{\citenamefont {Bagchi}\ \emph
  {et~al.}(2023{\natexlab{a}})\citenamefont {Bagchi}, \citenamefont {Kolekar},\
  and\ \citenamefont {Shukla}}]{Bagchi:2023ysc}%
  \BibitemOpen
  \bibfield  {author} {\bibinfo {author} {\bibfnamefont {A.}~\bibnamefont
  {Bagchi}}, \bibinfo {author} {\bibfnamefont {K.~S.}\ \bibnamefont
  {Kolekar}},\ and\ \bibinfo {author} {\bibfnamefont {A.}~\bibnamefont
  {Shukla}},\ }\bibfield  {title} {\bibinfo {title} {{Carrollian Origins of
  Bjorken Flow}},\ }\href {https://doi.org/10.1103/PhysRevLett.130.241601}
  {\bibfield  {journal} {\bibinfo  {journal} {Phys. Rev. Lett.}\ }\textbf
  {\bibinfo {volume} {130}},\ \bibinfo {pages} {241601} (\bibinfo {year}
  {2023}{\natexlab{a}})},\ \Eprint {https://arxiv.org/abs/2302.03053}
  {arXiv:2302.03053 [hep-th]} \BibitemShut {NoStop}%
\bibitem [{\citenamefont {Bagchi}\ \emph
  {et~al.}(2024{\natexlab{b}})\citenamefont {Bagchi}, \citenamefont {Kolekar},
  \citenamefont {Mandal},\ and\ \citenamefont {Shukla}}]{Bagchi:2023rwd}%
  \BibitemOpen
  \bibfield  {author} {\bibinfo {author} {\bibfnamefont {A.}~\bibnamefont
  {Bagchi}}, \bibinfo {author} {\bibfnamefont {K.~S.}\ \bibnamefont {Kolekar}},
  \bibinfo {author} {\bibfnamefont {T.}~\bibnamefont {Mandal}},\ and\ \bibinfo
  {author} {\bibfnamefont {A.}~\bibnamefont {Shukla}},\ }\bibfield  {title}
  {\bibinfo {title} {{Heavy-ion collisions, Gubser flow, and Carroll
  hydrodynamics}},\ }\href {https://doi.org/10.1103/PhysRevD.109.056004}
  {\bibfield  {journal} {\bibinfo  {journal} {Phys. Rev. D}\ }\textbf {\bibinfo
  {volume} {109}},\ \bibinfo {pages} {056004} (\bibinfo {year}
  {2024}{\natexlab{b}})},\ \Eprint {https://arxiv.org/abs/2310.03167}
  {arXiv:2310.03167 [hep-th]} \BibitemShut {NoStop}%
\bibitem [{\citenamefont {Bondi}\ \emph {et~al.}(1962)\citenamefont {Bondi},
  \citenamefont {van~der Burg},\ and\ \citenamefont {Metzner}}]{Bondi:1962px}%
  \BibitemOpen
  \bibfield  {author} {\bibinfo {author} {\bibfnamefont {H.}~\bibnamefont
  {Bondi}}, \bibinfo {author} {\bibfnamefont {M.~G.~J.}\ \bibnamefont {van~der
  Burg}},\ and\ \bibinfo {author} {\bibfnamefont {A.~W.~K.}\ \bibnamefont
  {Metzner}},\ }\bibfield  {title} {\bibinfo {title} {{Gravitational waves in
  general relativity. 7. Waves from axisymmetric isolated systems}},\ }\href
  {https://doi.org/10.1098/rspa.1962.0161} {\bibfield  {journal} {\bibinfo
  {journal} {Proc. Roy. Soc. Lond. A}\ }\textbf {\bibinfo {volume} {269}},\
  \bibinfo {pages} {21} (\bibinfo {year} {1962})}\BibitemShut {NoStop}%
\bibitem [{\citenamefont {Sachs}(1962)}]{Sachs:1962zza}%
  \BibitemOpen
  \bibfield  {author} {\bibinfo {author} {\bibfnamefont {R.}~\bibnamefont
  {Sachs}},\ }\bibfield  {title} {\bibinfo {title} {{Asymptotic symmetries in
  gravitational theory}},\ }\href {https://doi.org/10.1103/PhysRev.128.2851}
  {\bibfield  {journal} {\bibinfo  {journal} {Phys. Rev.}\ }\textbf {\bibinfo
  {volume} {128}},\ \bibinfo {pages} {2851} (\bibinfo {year}
  {1962})}\BibitemShut {NoStop}%
\bibitem [{\citenamefont {Bagchi}(2010)}]{Bagchi:2010zz}%
  \BibitemOpen
  \bibfield  {author} {\bibinfo {author} {\bibfnamefont {A.}~\bibnamefont
  {Bagchi}},\ }\bibfield  {title} {\bibinfo {title} {{Correspondence between
  Asymptotically Flat Spacetimes and Nonrelativistic Conformal Field
  Theories}},\ }\href {https://doi.org/10.1103/PhysRevLett.105.171601}
  {\bibfield  {journal} {\bibinfo  {journal} {Phys. Rev. Lett.}\ }\textbf
  {\bibinfo {volume} {105}},\ \bibinfo {pages} {171601} (\bibinfo {year}
  {2010})},\ \Eprint {https://arxiv.org/abs/1006.3354} {arXiv:1006.3354
  [hep-th]} \BibitemShut {NoStop}%
\bibitem [{\citenamefont {Duval}\ \emph {et~al.}(2014)\citenamefont {Duval},
  \citenamefont {Gibbons},\ and\ \citenamefont {Horvathy}}]{Duval:2014uva}%
  \BibitemOpen
  \bibfield  {author} {\bibinfo {author} {\bibfnamefont {C.}~\bibnamefont
  {Duval}}, \bibinfo {author} {\bibfnamefont {G.~W.}\ \bibnamefont {Gibbons}},\
  and\ \bibinfo {author} {\bibfnamefont {P.~A.}\ \bibnamefont {Horvathy}},\
  }\bibfield  {title} {\bibinfo {title} {{Conformal Carroll groups and BMS
  symmetry}},\ }\href {https://doi.org/10.1088/0264-9381/31/9/092001}
  {\bibfield  {journal} {\bibinfo  {journal} {Class. Quant. Grav.}\ }\textbf
  {\bibinfo {volume} {31}},\ \bibinfo {pages} {092001} (\bibinfo {year}
  {2014})},\ \Eprint {https://arxiv.org/abs/1402.5894} {arXiv:1402.5894
  [gr-qc]} \BibitemShut {NoStop}%
\bibitem [{\citenamefont {Maldacena}(1998)}]{Maldacena:1997re}%
  \BibitemOpen
  \bibfield  {author} {\bibinfo {author} {\bibfnamefont {J.~M.}\ \bibnamefont
  {Maldacena}},\ }\bibfield  {title} {\bibinfo {title} {{The Large N limit of
  superconformal field theories and supergravity}},\ }\href
  {https://doi.org/10.4310/ATMP.1998.v2.n2.a1} {\bibfield  {journal} {\bibinfo
  {journal} {Adv. Theor. Math. Phys.}\ }\textbf {\bibinfo {volume} {2}},\
  \bibinfo {pages} {231} (\bibinfo {year} {1998})},\ \Eprint
  {https://arxiv.org/abs/hep-th/9711200} {arXiv:hep-th/9711200} \BibitemShut
  {NoStop}%
\bibitem [{\citenamefont {Bagchi}\ and\ \citenamefont
  {Fareghbal}(2012)}]{Bagchi:2012cy}%
  \BibitemOpen
  \bibfield  {author} {\bibinfo {author} {\bibfnamefont {A.}~\bibnamefont
  {Bagchi}}\ and\ \bibinfo {author} {\bibfnamefont {R.}~\bibnamefont
  {Fareghbal}},\ }\bibfield  {title} {\bibinfo {title} {{BMS/GCA Redux: Towards
  Flatspace Holography from Non-Relativistic Symmetries}},\ }\href
  {https://doi.org/10.1007/JHEP10(2012)092} {\bibfield  {journal} {\bibinfo
  {journal} {JHEP}\ }\textbf {\bibinfo {volume} {10}},\ \bibinfo {pages}
  {092}},\ \Eprint {https://arxiv.org/abs/1203.5795} {arXiv:1203.5795 [hep-th]}
  \BibitemShut {NoStop}%
\bibitem [{\citenamefont {Barnich}\ \emph {et~al.}(2012)\citenamefont
  {Barnich}, \citenamefont {Gomberoff},\ and\ \citenamefont
  {Gonzalez}}]{Barnich:2012aw}%
  \BibitemOpen
  \bibfield  {author} {\bibinfo {author} {\bibfnamefont {G.}~\bibnamefont
  {Barnich}}, \bibinfo {author} {\bibfnamefont {A.}~\bibnamefont {Gomberoff}},\
  and\ \bibinfo {author} {\bibfnamefont {H.~A.}\ \bibnamefont {Gonzalez}},\
  }\bibfield  {title} {\bibinfo {title} {{The Flat limit of three dimensional
  asymptotically anti-de Sitter spacetimes}},\ }\href
  {https://doi.org/10.1103/PhysRevD.86.024020} {\bibfield  {journal} {\bibinfo
  {journal} {Phys. Rev. D}\ }\textbf {\bibinfo {volume} {86}},\ \bibinfo
  {pages} {024020} (\bibinfo {year} {2012})},\ \Eprint
  {https://arxiv.org/abs/1204.3288} {arXiv:1204.3288 [gr-qc]} \BibitemShut
  {NoStop}%
\bibitem [{\citenamefont {Barnich}(2012)}]{Barnich:2012xq}%
  \BibitemOpen
  \bibfield  {author} {\bibinfo {author} {\bibfnamefont {G.}~\bibnamefont
  {Barnich}},\ }\bibfield  {title} {\bibinfo {title} {{Entropy of
  three-dimensional asymptotically flat cosmological solutions}},\ }\href
  {https://doi.org/10.1007/JHEP10(2012)095} {\bibfield  {journal} {\bibinfo
  {journal} {JHEP}\ }\textbf {\bibinfo {volume} {10}},\ \bibinfo {pages}
  {095}},\ \Eprint {https://arxiv.org/abs/1208.4371} {arXiv:1208.4371 [hep-th]}
  \BibitemShut {NoStop}%
\bibitem [{\citenamefont {Bagchi}\ \emph {et~al.}(2013)\citenamefont {Bagchi},
  \citenamefont {Detournay}, \citenamefont {Fareghbal},\ and\ \citenamefont
  {Sim\'on}}]{Bagchi:2012xr}%
  \BibitemOpen
  \bibfield  {author} {\bibinfo {author} {\bibfnamefont {A.}~\bibnamefont
  {Bagchi}}, \bibinfo {author} {\bibfnamefont {S.}~\bibnamefont {Detournay}},
  \bibinfo {author} {\bibfnamefont {R.}~\bibnamefont {Fareghbal}},\ and\
  \bibinfo {author} {\bibfnamefont {J.}~\bibnamefont {Sim\'on}},\ }\bibfield
  {title} {\bibinfo {title} {{Holography of 3D Flat Cosmological Horizons}},\
  }\href {https://doi.org/10.1103/PhysRevLett.110.141302} {\bibfield  {journal}
  {\bibinfo  {journal} {Phys. Rev. Lett.}\ }\textbf {\bibinfo {volume} {110}},\
  \bibinfo {pages} {141302} (\bibinfo {year} {2013})},\ \Eprint
  {https://arxiv.org/abs/1208.4372} {arXiv:1208.4372 [hep-th]} \BibitemShut
  {NoStop}%
\bibitem [{\citenamefont {Bagchi}\ \emph
  {et~al.}(2016{\natexlab{a}})\citenamefont {Bagchi}, \citenamefont {Basu},
  \citenamefont {Kakkar},\ and\ \citenamefont {Mehra}}]{Bagchi:2016bcd}%
  \BibitemOpen
  \bibfield  {author} {\bibinfo {author} {\bibfnamefont {A.}~\bibnamefont
  {Bagchi}}, \bibinfo {author} {\bibfnamefont {R.}~\bibnamefont {Basu}},
  \bibinfo {author} {\bibfnamefont {A.}~\bibnamefont {Kakkar}},\ and\ \bibinfo
  {author} {\bibfnamefont {A.}~\bibnamefont {Mehra}},\ }\bibfield  {title}
  {\bibinfo {title} {{Flat Holography: Aspects of the dual field theory}},\
  }\href {https://doi.org/10.1007/JHEP12(2016)147} {\bibfield  {journal}
  {\bibinfo  {journal} {JHEP}\ }\textbf {\bibinfo {volume} {12}},\ \bibinfo
  {pages} {147}},\ \Eprint {https://arxiv.org/abs/1609.06203} {arXiv:1609.06203
  [hep-th]} \BibitemShut {NoStop}%
\bibitem [{\citenamefont {Donnay}\ \emph {et~al.}(2022)\citenamefont {Donnay},
  \citenamefont {Fiorucci}, \citenamefont {Herfray},\ and\ \citenamefont
  {Ruzziconi}}]{Donnay:2022aba}%
  \BibitemOpen
  \bibfield  {author} {\bibinfo {author} {\bibfnamefont {L.}~\bibnamefont
  {Donnay}}, \bibinfo {author} {\bibfnamefont {A.}~\bibnamefont {Fiorucci}},
  \bibinfo {author} {\bibfnamefont {Y.}~\bibnamefont {Herfray}},\ and\ \bibinfo
  {author} {\bibfnamefont {R.}~\bibnamefont {Ruzziconi}},\ }\bibfield  {title}
  {\bibinfo {title} {{Carrollian Perspective on Celestial Holography}},\ }\href
  {https://doi.org/10.1103/PhysRevLett.129.071602} {\bibfield  {journal}
  {\bibinfo  {journal} {Phys. Rev. Lett.}\ }\textbf {\bibinfo {volume} {129}},\
  \bibinfo {pages} {071602} (\bibinfo {year} {2022})},\ \Eprint
  {https://arxiv.org/abs/2202.04702} {arXiv:2202.04702 [hep-th]} \BibitemShut
  {NoStop}%
\bibitem [{\citenamefont {Bagchi}\ \emph {et~al.}(2022)\citenamefont {Bagchi},
  \citenamefont {Banerjee}, \citenamefont {Basu},\ and\ \citenamefont
  {Dutta}}]{Bagchi:2022emh}%
  \BibitemOpen
  \bibfield  {author} {\bibinfo {author} {\bibfnamefont {A.}~\bibnamefont
  {Bagchi}}, \bibinfo {author} {\bibfnamefont {S.}~\bibnamefont {Banerjee}},
  \bibinfo {author} {\bibfnamefont {R.}~\bibnamefont {Basu}},\ and\ \bibinfo
  {author} {\bibfnamefont {S.}~\bibnamefont {Dutta}},\ }\bibfield  {title}
  {\bibinfo {title} {{Scattering Amplitudes: Celestial and Carrollian}},\
  }\href {https://doi.org/10.1103/PhysRevLett.128.241601} {\bibfield  {journal}
  {\bibinfo  {journal} {Phys. Rev. Lett.}\ }\textbf {\bibinfo {volume} {128}},\
  \bibinfo {pages} {241601} (\bibinfo {year} {2022})},\ \Eprint
  {https://arxiv.org/abs/2202.08438} {arXiv:2202.08438 [hep-th]} \BibitemShut
  {NoStop}%
\bibitem [{\citenamefont {Donnay}\ \emph {et~al.}(2023)\citenamefont {Donnay},
  \citenamefont {Fiorucci}, \citenamefont {Herfray},\ and\ \citenamefont
  {Ruzziconi}}]{Donnay:2022wvx}%
  \BibitemOpen
  \bibfield  {author} {\bibinfo {author} {\bibfnamefont {L.}~\bibnamefont
  {Donnay}}, \bibinfo {author} {\bibfnamefont {A.}~\bibnamefont {Fiorucci}},
  \bibinfo {author} {\bibfnamefont {Y.}~\bibnamefont {Herfray}},\ and\ \bibinfo
  {author} {\bibfnamefont {R.}~\bibnamefont {Ruzziconi}},\ }\bibfield  {title}
  {\bibinfo {title} {{Bridging Carrollian and celestial holography}},\ }\href
  {https://doi.org/10.1103/PhysRevD.107.126027} {\bibfield  {journal} {\bibinfo
   {journal} {Phys. Rev. D}\ }\textbf {\bibinfo {volume} {107}},\ \bibinfo
  {pages} {126027} (\bibinfo {year} {2023})},\ \Eprint
  {https://arxiv.org/abs/2212.12553} {arXiv:2212.12553 [hep-th]} \BibitemShut
  {NoStop}%
\bibitem [{\citenamefont {Bagchi}\ \emph
  {et~al.}(2023{\natexlab{b}})\citenamefont {Bagchi}, \citenamefont
  {Dhivakar},\ and\ \citenamefont {Dutta}}]{Bagchi:2023fbj}%
  \BibitemOpen
  \bibfield  {author} {\bibinfo {author} {\bibfnamefont {A.}~\bibnamefont
  {Bagchi}}, \bibinfo {author} {\bibfnamefont {P.}~\bibnamefont {Dhivakar}},\
  and\ \bibinfo {author} {\bibfnamefont {S.}~\bibnamefont {Dutta}},\ }\bibfield
   {title} {\bibinfo {title} {{AdS Witten diagrams to Carrollian
  correlators}},\ }\href {https://doi.org/10.1007/JHEP04(2023)135} {\bibfield
  {journal} {\bibinfo  {journal} {JHEP}\ }\textbf {\bibinfo {volume} {04}},\
  \bibinfo {pages} {135}},\ \Eprint {https://arxiv.org/abs/2303.07388}
  {arXiv:2303.07388 [hep-th]} \BibitemShut {NoStop}%
\bibitem [{\citenamefont {Saha}(2023)}]{Saha:2023hsl}%
  \BibitemOpen
  \bibfield  {author} {\bibinfo {author} {\bibfnamefont {A.}~\bibnamefont
  {Saha}},\ }\bibfield  {title} {\bibinfo {title} {{Carrollian approach to 1 +
  3D flat holography}},\ }\href {https://doi.org/10.1007/JHEP06(2023)051}
  {\bibfield  {journal} {\bibinfo  {journal} {JHEP}\ }\textbf {\bibinfo
  {volume} {06}},\ \bibinfo {pages} {051}},\ \Eprint
  {https://arxiv.org/abs/2304.02696} {arXiv:2304.02696 [hep-th]} \BibitemShut
  {NoStop}%
\bibitem [{\citenamefont {Adami}\ \emph {et~al.}(2023)\citenamefont {Adami},
  \citenamefont {Parvizi}, \citenamefont {Sheikh-Jabbari}, \citenamefont
  {Taghiloo},\ and\ \citenamefont {Yavartanoo}}]{Adami:2023fbm}%
  \BibitemOpen
  \bibfield  {author} {\bibinfo {author} {\bibfnamefont {H.}~\bibnamefont
  {Adami}}, \bibinfo {author} {\bibfnamefont {A.}~\bibnamefont {Parvizi}},
  \bibinfo {author} {\bibfnamefont {M.~M.}\ \bibnamefont {Sheikh-Jabbari}},
  \bibinfo {author} {\bibfnamefont {V.}~\bibnamefont {Taghiloo}},\ and\
  \bibinfo {author} {\bibfnamefont {H.}~\bibnamefont {Yavartanoo}},\ }\bibfield
   {title} {\bibinfo {title} {{Hydro \& thermo dynamics at causal boundaries,
  examples in 3d gravity}},\ }\href {https://doi.org/10.1007/JHEP07(2023)038}
  {\bibfield  {journal} {\bibinfo  {journal} {JHEP}\ }\textbf {\bibinfo
  {volume} {07}},\ \bibinfo {pages} {038}},\ \Eprint
  {https://arxiv.org/abs/2305.01009} {arXiv:2305.01009 [hep-th]} \BibitemShut
  {NoStop}%
\bibitem [{\citenamefont {Nguyen}\ and\ \citenamefont
  {West}(2023)}]{Nguyen:2023vfz}%
  \BibitemOpen
  \bibfield  {author} {\bibinfo {author} {\bibfnamefont {K.}~\bibnamefont
  {Nguyen}}\ and\ \bibinfo {author} {\bibfnamefont {P.}~\bibnamefont {West}},\
  }\bibfield  {title} {\bibinfo {title} {{Carrollian Conformal Fields and Flat
  Holography}},\ }\href {https://doi.org/10.3390/universe9090385} {\bibfield
  {journal} {\bibinfo  {journal} {Universe}\ }\textbf {\bibinfo {volume} {9}},\
  \bibinfo {pages} {385} (\bibinfo {year} {2023})},\ \Eprint
  {https://arxiv.org/abs/2305.02884} {arXiv:2305.02884 [hep-th]} \BibitemShut
  {NoStop}%
\bibitem [{\citenamefont {Bagchi}\ \emph
  {et~al.}(2024{\natexlab{c}})\citenamefont {Bagchi}, \citenamefont
  {Dhivakar},\ and\ \citenamefont {Dutta}}]{Bagchi:2023cen}%
  \BibitemOpen
  \bibfield  {author} {\bibinfo {author} {\bibfnamefont {A.}~\bibnamefont
  {Bagchi}}, \bibinfo {author} {\bibfnamefont {P.}~\bibnamefont {Dhivakar}},\
  and\ \bibinfo {author} {\bibfnamefont {S.}~\bibnamefont {Dutta}},\ }\bibfield
   {title} {\bibinfo {title} {{Holography in flat spacetimes: the case for
  Carroll}},\ }\href {https://doi.org/10.1007/JHEP08(2024)144} {\bibfield
  {journal} {\bibinfo  {journal} {JHEP}\ }\textbf {\bibinfo {volume} {08}},\
  \bibinfo {pages} {144}},\ \Eprint {https://arxiv.org/abs/2311.11246}
  {arXiv:2311.11246 [hep-th]} \BibitemShut {NoStop}%
\bibitem [{\citenamefont {de~Boer}\ \emph {et~al.}(2022)\citenamefont
  {de~Boer}, \citenamefont {Hartong}, \citenamefont {Obers}, \citenamefont
  {Sybesma},\ and\ \citenamefont {Vandoren}}]{deBoer:2021jej}%
  \BibitemOpen
  \bibfield  {author} {\bibinfo {author} {\bibfnamefont {J.}~\bibnamefont
  {de~Boer}}, \bibinfo {author} {\bibfnamefont {J.}~\bibnamefont {Hartong}},
  \bibinfo {author} {\bibfnamefont {N.~A.}\ \bibnamefont {Obers}}, \bibinfo
  {author} {\bibfnamefont {W.}~\bibnamefont {Sybesma}},\ and\ \bibinfo {author}
  {\bibfnamefont {S.}~\bibnamefont {Vandoren}},\ }\bibfield  {title} {\bibinfo
  {title} {{Carroll Symmetry, Dark Energy and Inflation}},\ }\href
  {https://doi.org/10.3389/fphy.2022.810405} {\bibfield  {journal} {\bibinfo
  {journal} {Front. in Phys.}\ }\textbf {\bibinfo {volume} {10}},\ \bibinfo
  {pages} {810405} (\bibinfo {year} {2022})},\ \Eprint
  {https://arxiv.org/abs/2110.02319} {arXiv:2110.02319 [hep-th]} \BibitemShut
  {NoStop}%
\bibitem [{\citenamefont {Bidussi}\ \emph {et~al.}(2022)\citenamefont
  {Bidussi}, \citenamefont {Hartong}, \citenamefont {Have}, \citenamefont
  {Musaeus},\ and\ \citenamefont {Prohazka}}]{Bidussi:2021nmp}%
  \BibitemOpen
  \bibfield  {author} {\bibinfo {author} {\bibfnamefont {L.}~\bibnamefont
  {Bidussi}}, \bibinfo {author} {\bibfnamefont {J.}~\bibnamefont {Hartong}},
  \bibinfo {author} {\bibfnamefont {E.}~\bibnamefont {Have}}, \bibinfo {author}
  {\bibfnamefont {J.}~\bibnamefont {Musaeus}},\ and\ \bibinfo {author}
  {\bibfnamefont {S.}~\bibnamefont {Prohazka}},\ }\bibfield  {title} {\bibinfo
  {title} {{Fractons, dipole symmetries and curved spacetime}},\ }\href
  {https://doi.org/10.21468/SciPostPhys.12.6.205} {\bibfield  {journal}
  {\bibinfo  {journal} {SciPost Phys.}\ }\textbf {\bibinfo {volume} {12}},\
  \bibinfo {pages} {205} (\bibinfo {year} {2022})},\ \Eprint
  {https://arxiv.org/abs/2111.03668} {arXiv:2111.03668 [hep-th]} \BibitemShut
  {NoStop}%
\bibitem [{\citenamefont {Bagchi}\ \emph
  {et~al.}(2023{\natexlab{c}})\citenamefont {Bagchi}, \citenamefont {Banerjee},
  \citenamefont {Basu}, \citenamefont {Islam},\ and\ \citenamefont
  {Mondal}}]{Bagchi:2022eui}%
  \BibitemOpen
  \bibfield  {author} {\bibinfo {author} {\bibfnamefont {A.}~\bibnamefont
  {Bagchi}}, \bibinfo {author} {\bibfnamefont {A.}~\bibnamefont {Banerjee}},
  \bibinfo {author} {\bibfnamefont {R.}~\bibnamefont {Basu}}, \bibinfo {author}
  {\bibfnamefont {M.}~\bibnamefont {Islam}},\ and\ \bibinfo {author}
  {\bibfnamefont {S.}~\bibnamefont {Mondal}},\ }\bibfield  {title} {\bibinfo
  {title} {{Magic fermions: Carroll and flat bands}},\ }\href
  {https://doi.org/10.1007/JHEP03(2023)227} {\bibfield  {journal} {\bibinfo
  {journal} {JHEP}\ }\textbf {\bibinfo {volume} {03}},\ \bibinfo {pages}
  {227}},\ \Eprint {https://arxiv.org/abs/2211.11640} {arXiv:2211.11640
  [hep-th]} \BibitemShut {NoStop}%
\bibitem [{\citenamefont {Kasikci}\ \emph {et~al.}(2024)\citenamefont
  {Kasikci}, \citenamefont {Ozkan}, \citenamefont {Pang},\ and\ \citenamefont
  {Zorba}}]{Kasikci:2023zdn}%
  \BibitemOpen
  \bibfield  {author} {\bibinfo {author} {\bibfnamefont {O.}~\bibnamefont
  {Kasikci}}, \bibinfo {author} {\bibfnamefont {M.}~\bibnamefont {Ozkan}},
  \bibinfo {author} {\bibfnamefont {Y.}~\bibnamefont {Pang}},\ and\ \bibinfo
  {author} {\bibfnamefont {U.}~\bibnamefont {Zorba}},\ }\bibfield  {title}
  {\bibinfo {title} {{Carrollian supersymmetry and SYK-like models}},\ }\href
  {https://doi.org/10.1103/PhysRevD.110.L021702} {\bibfield  {journal}
  {\bibinfo  {journal} {Phys. Rev. D}\ }\textbf {\bibinfo {volume} {110}},\
  \bibinfo {pages} {L021702} (\bibinfo {year} {2024})},\ \Eprint
  {https://arxiv.org/abs/2311.00039} {arXiv:2311.00039 [hep-th]} \BibitemShut
  {NoStop}%
\bibitem [{\citenamefont {Bagchi}(2013)}]{Bagchi:2013bga}%
  \BibitemOpen
  \bibfield  {author} {\bibinfo {author} {\bibfnamefont {A.}~\bibnamefont
  {Bagchi}},\ }\bibfield  {title} {\bibinfo {title} {{Tensionless Strings and
  Galilean Conformal Algebra}},\ }\href
  {https://doi.org/10.1007/JHEP05(2013)141} {\bibfield  {journal} {\bibinfo
  {journal} {JHEP}\ }\textbf {\bibinfo {volume} {05}},\ \bibinfo {pages}
  {141}},\ \Eprint {https://arxiv.org/abs/1303.0291} {arXiv:1303.0291 [hep-th]}
  \BibitemShut {NoStop}%
\bibitem [{\citenamefont {Bagchi}\ \emph
  {et~al.}(2016{\natexlab{b}})\citenamefont {Bagchi}, \citenamefont
  {Chakrabortty},\ and\ \citenamefont {Parekh}}]{Bagchi:2015nca}%
  \BibitemOpen
  \bibfield  {author} {\bibinfo {author} {\bibfnamefont {A.}~\bibnamefont
  {Bagchi}}, \bibinfo {author} {\bibfnamefont {S.}~\bibnamefont
  {Chakrabortty}},\ and\ \bibinfo {author} {\bibfnamefont {P.}~\bibnamefont
  {Parekh}},\ }\bibfield  {title} {\bibinfo {title} {{Tensionless Strings from
  Worldsheet Symmetries}},\ }\href {https://doi.org/10.1007/JHEP01(2016)158}
  {\bibfield  {journal} {\bibinfo  {journal} {JHEP}\ }\textbf {\bibinfo
  {volume} {01}},\ \bibinfo {pages} {158}},\ \Eprint
  {https://arxiv.org/abs/1507.04361} {arXiv:1507.04361 [hep-th]} \BibitemShut
  {NoStop}%
\bibitem [{\citenamefont {Bagchi}\ \emph {et~al.}(2020)\citenamefont {Bagchi},
  \citenamefont {Banerjee}, \citenamefont {Chakrabortty}, \citenamefont
  {Dutta},\ and\ \citenamefont {Parekh}}]{Bagchi:2020fpr}%
  \BibitemOpen
  \bibfield  {author} {\bibinfo {author} {\bibfnamefont {A.}~\bibnamefont
  {Bagchi}}, \bibinfo {author} {\bibfnamefont {A.}~\bibnamefont {Banerjee}},
  \bibinfo {author} {\bibfnamefont {S.}~\bibnamefont {Chakrabortty}}, \bibinfo
  {author} {\bibfnamefont {S.}~\bibnamefont {Dutta}},\ and\ \bibinfo {author}
  {\bibfnamefont {P.}~\bibnamefont {Parekh}},\ }\bibfield  {title} {\bibinfo
  {title} {{A tale of three \textemdash{} tensionless strings and vacuum
  structure}},\ }\href {https://doi.org/10.1007/JHEP04(2020)061} {\bibfield
  {journal} {\bibinfo  {journal} {JHEP}\ }\textbf {\bibinfo {volume} {04}},\
  \bibinfo {pages} {061}},\ \Eprint {https://arxiv.org/abs/2001.00354}
  {arXiv:2001.00354 [hep-th]} \BibitemShut {NoStop}%
\bibitem [{\citenamefont {Bagchi}\ \emph
  {et~al.}(2024{\natexlab{d}})\citenamefont {Bagchi}, \citenamefont
  {Chakraborty}, \citenamefont {Chakrabortty}, \citenamefont {Fredenhagen},
  \citenamefont {Grumiller},\ and\ \citenamefont {Pandit}}]{Bagchi:2024qsb}%
  \BibitemOpen
  \bibfield  {author} {\bibinfo {author} {\bibfnamefont {A.}~\bibnamefont
  {Bagchi}}, \bibinfo {author} {\bibfnamefont {P.}~\bibnamefont {Chakraborty}},
  \bibinfo {author} {\bibfnamefont {S.}~\bibnamefont {Chakrabortty}}, \bibinfo
  {author} {\bibfnamefont {S.}~\bibnamefont {Fredenhagen}}, \bibinfo {author}
  {\bibfnamefont {D.}~\bibnamefont {Grumiller}},\ and\ \bibinfo {author}
  {\bibfnamefont {P.}~\bibnamefont {Pandit}},\ }\bibfield  {title} {\bibinfo
  {title} {{Boundary Carrollian CFTs and Open Null Strings}},\ }\Eprint
  {https://arxiv.org/abs/2409.01094} {arXiv:2409.01094 [hep-th]}  (\bibinfo
  {year} {2024}{\natexlab{d}})\BibitemShut {NoStop}%
\bibitem [{\citenamefont {Penna}(2018)}]{Penna:2018gfx}%
  \BibitemOpen
  \bibfield  {author} {\bibinfo {author} {\bibfnamefont {R.~F.}\ \bibnamefont
  {Penna}},\ }\bibfield  {title} {\bibinfo {title} {{Near-horizon Carroll
  symmetry and black hole Love numbers}},\ }\Eprint
  {https://arxiv.org/abs/1812.05643} {arXiv:1812.05643 [hep-th]}  (\bibinfo
  {year} {2018})\BibitemShut {NoStop}%
\bibitem [{\citenamefont {Donnay}\ and\ \citenamefont
  {Marteau}(2019)}]{Donnay:2019jiz}%
  \BibitemOpen
  \bibfield  {author} {\bibinfo {author} {\bibfnamefont {L.}~\bibnamefont
  {Donnay}}\ and\ \bibinfo {author} {\bibfnamefont {C.}~\bibnamefont
  {Marteau}},\ }\bibfield  {title} {\bibinfo {title} {{Carrollian Physics at
  the Black Hole Horizon}},\ }\href {https://doi.org/10.1088/1361-6382/ab2fd5}
  {\bibfield  {journal} {\bibinfo  {journal} {Class. Quant. Grav.}\ }\textbf
  {\bibinfo {volume} {36}},\ \bibinfo {pages} {165002} (\bibinfo {year}
  {2019})},\ \Eprint {https://arxiv.org/abs/1903.09654} {arXiv:1903.09654
  [hep-th]} \BibitemShut {NoStop}%
\bibitem [{\citenamefont {Freidel}\ and\ \citenamefont
  {Jai-akson}(2024)}]{Freidel:2022vjq}%
  \BibitemOpen
  \bibfield  {author} {\bibinfo {author} {\bibfnamefont {L.}~\bibnamefont
  {Freidel}}\ and\ \bibinfo {author} {\bibfnamefont {P.}~\bibnamefont
  {Jai-akson}},\ }\bibfield  {title} {\bibinfo {title} {{Carrollian
  hydrodynamics and symplectic structure on stretched horizons}},\ }\href
  {https://doi.org/10.1007/JHEP05(2024)135} {\bibfield  {journal} {\bibinfo
  {journal} {JHEP}\ }\textbf {\bibinfo {volume} {05}},\ \bibinfo {pages}
  {135}},\ \Eprint {https://arxiv.org/abs/2211.06415} {arXiv:2211.06415
  [gr-qc]} \BibitemShut {NoStop}%
\bibitem [{\citenamefont {Redondo-Yuste}\ and\ \citenamefont
  {Lehner}(2023)}]{Redondo-Yuste:2022czg}%
  \BibitemOpen
  \bibfield  {author} {\bibinfo {author} {\bibfnamefont {J.}~\bibnamefont
  {Redondo-Yuste}}\ and\ \bibinfo {author} {\bibfnamefont {L.}~\bibnamefont
  {Lehner}},\ }\bibfield  {title} {\bibinfo {title} {{Non-linear black hole
  dynamics and Carrollian fluids}},\ }\href
  {https://doi.org/10.1007/JHEP02(2023)240} {\bibfield  {journal} {\bibinfo
  {journal} {JHEP}\ }\textbf {\bibinfo {volume} {02}},\ \bibinfo {pages}
  {240}},\ \Eprint {https://arxiv.org/abs/2212.06175} {arXiv:2212.06175
  [gr-qc]} \BibitemShut {NoStop}%
\bibitem [{\citenamefont {Bagchi}\ \emph
  {et~al.}(2023{\natexlab{d}})\citenamefont {Bagchi}, \citenamefont
  {Grumiller}, \citenamefont {Sheikh-Jabbari},\ and\ \citenamefont
  {Sheikh-Jabbari}}]{Bagchi:2022iqb}%
  \BibitemOpen
  \bibfield  {author} {\bibinfo {author} {\bibfnamefont {A.}~\bibnamefont
  {Bagchi}}, \bibinfo {author} {\bibfnamefont {D.}~\bibnamefont {Grumiller}},
  \bibinfo {author} {\bibfnamefont {S.}~\bibnamefont {Sheikh-Jabbari}},\ and\
  \bibinfo {author} {\bibfnamefont {M.~M.}\ \bibnamefont {Sheikh-Jabbari}},\
  }\bibfield  {title} {\bibinfo {title} {{Horizon strings as 3D black hole
  microstates}},\ }\href {https://doi.org/10.21468/SciPostPhys.15.5.210}
  {\bibfield  {journal} {\bibinfo  {journal} {SciPost Phys.}\ }\textbf
  {\bibinfo {volume} {15}},\ \bibinfo {pages} {210} (\bibinfo {year}
  {2023}{\natexlab{d}})},\ \Eprint {https://arxiv.org/abs/2210.10794}
  {arXiv:2210.10794 [hep-th]} \BibitemShut {NoStop}%
\bibitem [{\citenamefont {Ecker}\ \emph {et~al.}(2023)\citenamefont {Ecker},
  \citenamefont {Grumiller}, \citenamefont {Hartong}, \citenamefont {P\'erez},
  \citenamefont {Prohazka},\ and\ \citenamefont {Troncoso}}]{Ecker:2023uwm}%
  \BibitemOpen
  \bibfield  {author} {\bibinfo {author} {\bibfnamefont {F.}~\bibnamefont
  {Ecker}}, \bibinfo {author} {\bibfnamefont {D.}~\bibnamefont {Grumiller}},
  \bibinfo {author} {\bibfnamefont {J.}~\bibnamefont {Hartong}}, \bibinfo
  {author} {\bibfnamefont {A.}~\bibnamefont {P\'erez}}, \bibinfo {author}
  {\bibfnamefont {S.}~\bibnamefont {Prohazka}},\ and\ \bibinfo {author}
  {\bibfnamefont {R.}~\bibnamefont {Troncoso}},\ }\bibfield  {title} {\bibinfo
  {title} {{Carroll black holes}},\ }\href
  {https://doi.org/10.21468/SciPostPhys.15.6.245} {\bibfield  {journal}
  {\bibinfo  {journal} {SciPost Phys.}\ }\textbf {\bibinfo {volume} {15}},\
  \bibinfo {pages} {245} (\bibinfo {year} {2023})},\ \Eprint
  {https://arxiv.org/abs/2308.10947} {arXiv:2308.10947 [hep-th]} \BibitemShut
  {NoStop}%
\bibitem [{\citenamefont {Bagchi}\ \emph
  {et~al.}(2024{\natexlab{e}})\citenamefont {Bagchi}, \citenamefont {Banerjee},
  \citenamefont {Hartong}, \citenamefont {Have}, \citenamefont {Kolekar},\ and\
  \citenamefont {Mandlik}}]{Bagchi:2023cfp}%
  \BibitemOpen
  \bibfield  {author} {\bibinfo {author} {\bibfnamefont {A.}~\bibnamefont
  {Bagchi}}, \bibinfo {author} {\bibfnamefont {A.}~\bibnamefont {Banerjee}},
  \bibinfo {author} {\bibfnamefont {J.}~\bibnamefont {Hartong}}, \bibinfo
  {author} {\bibfnamefont {E.}~\bibnamefont {Have}}, \bibinfo {author}
  {\bibfnamefont {K.~S.}\ \bibnamefont {Kolekar}},\ and\ \bibinfo {author}
  {\bibfnamefont {M.}~\bibnamefont {Mandlik}},\ }\bibfield  {title} {\bibinfo
  {title} {{Strings near black holes are Carrollian}},\ }\href
  {https://doi.org/10.1103/PhysRevD.110.086009} {\bibfield  {journal} {\bibinfo
   {journal} {Phys. Rev. D}\ }\textbf {\bibinfo {volume} {110}},\ \bibinfo
  {pages} {086009} (\bibinfo {year} {2024}{\natexlab{e}})},\ \Eprint
  {https://arxiv.org/abs/2312.14240} {arXiv:2312.14240 [hep-th]} \BibitemShut
  {NoStop}%
\bibitem [{\citenamefont {Aggarwal}\ \emph {et~al.}(2024)\citenamefont
  {Aggarwal}, \citenamefont {Ecker}, \citenamefont {Grumiller},\ and\
  \citenamefont {Vassilevich}}]{Aggarwal:2024yxy}%
  \BibitemOpen
  \bibfield  {author} {\bibinfo {author} {\bibfnamefont {A.}~\bibnamefont
  {Aggarwal}}, \bibinfo {author} {\bibfnamefont {F.}~\bibnamefont {Ecker}},
  \bibinfo {author} {\bibfnamefont {D.}~\bibnamefont {Grumiller}},\ and\
  \bibinfo {author} {\bibfnamefont {D.}~\bibnamefont {Vassilevich}},\
  }\bibfield  {title} {\bibinfo {title} {{Carroll-Hawking effect}},\ }\href
  {https://doi.org/10.1103/PhysRevD.110.L041506} {\bibfield  {journal}
  {\bibinfo  {journal} {Phys. Rev. D}\ }\textbf {\bibinfo {volume} {110}},\
  \bibinfo {pages} {L041506} (\bibinfo {year} {2024})},\ \Eprint
  {https://arxiv.org/abs/2403.00073} {arXiv:2403.00073 [hep-th]} \BibitemShut
  {NoStop}%
\bibitem [{\citenamefont {Bagchi}\ \emph
  {et~al.}(2024{\natexlab{f}})\citenamefont {Bagchi}, \citenamefont {Banerjee},
  \citenamefont {Hartong}, \citenamefont {Have},\ and\ \citenamefont
  {Kolekar}}]{Bagchi:2024rje}%
  \BibitemOpen
  \bibfield  {author} {\bibinfo {author} {\bibfnamefont {A.}~\bibnamefont
  {Bagchi}}, \bibinfo {author} {\bibfnamefont {A.}~\bibnamefont {Banerjee}},
  \bibinfo {author} {\bibfnamefont {J.}~\bibnamefont {Hartong}}, \bibinfo
  {author} {\bibfnamefont {E.}~\bibnamefont {Have}},\ and\ \bibinfo {author}
  {\bibfnamefont {K.~S.}\ \bibnamefont {Kolekar}},\ }\bibfield  {title}
  {\bibinfo {title} {{Strings near black holes are Carrollian. Part II}},\
  }\href {https://doi.org/10.1007/JHEP11(2024)024} {\bibfield  {journal}
  {\bibinfo  {journal} {JHEP}\ }\textbf {\bibinfo {volume} {11}},\ \bibinfo
  {pages} {024}},\ \Eprint {https://arxiv.org/abs/2407.12911} {arXiv:2407.12911
  [hep-th]} \BibitemShut {NoStop}%
\bibitem [{\citenamefont {Tadros}\ and\ \citenamefont
  {Kolàr}(2024)}]{Tadros:2024bev}%
  \BibitemOpen
  \bibfield  {author} {\bibinfo {author} {\bibfnamefont {P.}~\bibnamefont
  {Tadros}}\ and\ \bibinfo {author} {\bibfnamefont {I.}~\bibnamefont
  {Kolàr}},\ }\bibfield  {title} {\bibinfo {title} {{Carroll black holes in
  (A)dS and their higher-derivative modifications}},\ }\Eprint
  {https://arxiv.org/abs/2408.01836} {arXiv:2408.01836 [gr-qc]}  (\bibinfo
  {year} {2024})\BibitemShut {NoStop}%
\bibitem [{\citenamefont {Bagchi}\ \emph {et~al.}(2025)\citenamefont {Bagchi},
  \citenamefont {Banerjee}, \citenamefont {Dhivakar}, \citenamefont {Mondal},\
  and\ \citenamefont {Shukla}}]{Bagchi:2025vri}%
  \BibitemOpen
  \bibfield  {author} {\bibinfo {author} {\bibfnamefont {A.}~\bibnamefont
  {Bagchi}}, \bibinfo {author} {\bibfnamefont {A.}~\bibnamefont {Banerjee}},
  \bibinfo {author} {\bibfnamefont {P.}~\bibnamefont {Dhivakar}}, \bibinfo
  {author} {\bibfnamefont {S.}~\bibnamefont {Mondal}},\ and\ \bibinfo {author}
  {\bibfnamefont {A.}~\bibnamefont {Shukla}},\ }\bibfield  {title} {\bibinfo
  {title} {{The Carrollian Kaleidoscope}},\ }\href@noop {} {\  (\bibinfo {year}
  {2025})},\ \Eprint {https://arxiv.org/abs/2506.16164} {arXiv:2506.16164
  [hep-th]} \BibitemShut {NoStop}%
\bibitem [{\citenamefont {de~Boer}\ \emph {et~al.}(2023)\citenamefont
  {de~Boer}, \citenamefont {Hartong}, \citenamefont {Obers}, \citenamefont
  {Sybesma},\ and\ \citenamefont {Vandoren}}]{deBoer:2023fnj}%
  \BibitemOpen
  \bibfield  {author} {\bibinfo {author} {\bibfnamefont {J.}~\bibnamefont
  {de~Boer}}, \bibinfo {author} {\bibfnamefont {J.}~\bibnamefont {Hartong}},
  \bibinfo {author} {\bibfnamefont {N.~A.}\ \bibnamefont {Obers}}, \bibinfo
  {author} {\bibfnamefont {W.}~\bibnamefont {Sybesma}},\ and\ \bibinfo {author}
  {\bibfnamefont {S.}~\bibnamefont {Vandoren}},\ }\bibfield  {title} {\bibinfo
  {title} {{Carroll stories}},\ }\href
  {https://doi.org/10.1007/JHEP09(2023)148} {\bibfield  {journal} {\bibinfo
  {journal} {JHEP}\ }\textbf {\bibinfo {volume} {09}},\ \bibinfo {pages}
  {148}},\ \Eprint {https://arxiv.org/abs/2307.06827} {arXiv:2307.06827
  [hep-th]} \BibitemShut {NoStop}%
\bibitem [{\citenamefont {Parkkila}\ \emph {et~al.}(2021)\citenamefont
  {Parkkila}, \citenamefont {Onnerstad},\ and\ \citenamefont
  {Kim}}]{Parkkila:2021tqq}%
  \BibitemOpen
  \bibfield  {author} {\bibinfo {author} {\bibfnamefont {J.~E.}\ \bibnamefont
  {Parkkila}}, \bibinfo {author} {\bibfnamefont {A.}~\bibnamefont
  {Onnerstad}},\ and\ \bibinfo {author} {\bibfnamefont {D.~J.}\ \bibnamefont
  {Kim}},\ }\bibfield  {title} {\bibinfo {title} {{Bayesian estimation of the
  specific shear and bulk viscosity of the quark-gluon plasma with additional
  flow harmonic observables}},\ }\href
  {https://doi.org/10.1103/PhysRevC.104.054904} {\bibfield  {journal} {\bibinfo
   {journal} {Phys. Rev. C}\ }\textbf {\bibinfo {volume} {104}},\ \bibinfo
  {pages} {054904} (\bibinfo {year} {2021})},\ \Eprint
  {https://arxiv.org/abs/2106.05019} {arXiv:2106.05019 [hep-ph]} \BibitemShut
  {NoStop}%
\bibitem [{\citenamefont {Bhattacharyya}(2012)}]{Bhattacharyya:2012nq}%
  \BibitemOpen
  \bibfield  {author} {\bibinfo {author} {\bibfnamefont {S.}~\bibnamefont
  {Bhattacharyya}},\ }\bibfield  {title} {\bibinfo {title} {{Constraints on the
  second order transport coefficients of an uncharged fluid}},\ }\href
  {https://doi.org/10.1007/JHEP07(2012)104} {\bibfield  {journal} {\bibinfo
  {journal} {JHEP}\ }\textbf {\bibinfo {volume} {07}},\ \bibinfo {pages}
  {104}},\ \Eprint {https://arxiv.org/abs/1201.4654} {arXiv:1201.4654 [hep-th]}
  \BibitemShut {NoStop}%
\bibitem [{\citenamefont {Baier}\ \emph {et~al.}(2008)\citenamefont {Baier},
  \citenamefont {Romatschke}, \citenamefont {Son}, \citenamefont {Starinets},\
  and\ \citenamefont {Stephanov}}]{Baier:2007ix}%
  \BibitemOpen
  \bibfield  {author} {\bibinfo {author} {\bibfnamefont {R.}~\bibnamefont
  {Baier}}, \bibinfo {author} {\bibfnamefont {P.}~\bibnamefont {Romatschke}},
  \bibinfo {author} {\bibfnamefont {D.~T.}\ \bibnamefont {Son}}, \bibinfo
  {author} {\bibfnamefont {A.~O.}\ \bibnamefont {Starinets}},\ and\ \bibinfo
  {author} {\bibfnamefont {M.~A.}\ \bibnamefont {Stephanov}},\ }\bibfield
  {title} {\bibinfo {title} {{Relativistic viscous hydrodynamics, conformal
  invariance, and holography}},\ }\href
  {https://doi.org/10.1088/1126-6708/2008/04/100} {\bibfield  {journal}
  {\bibinfo  {journal} {JHEP}\ }\textbf {\bibinfo {volume} {04}},\ \bibinfo
  {pages} {100}},\ \Eprint {https://arxiv.org/abs/0712.2451} {arXiv:0712.2451
  [hep-th]} \BibitemShut {NoStop}%
\bibitem [{\citenamefont {Busza}\ \emph {et~al.}(2018)\citenamefont {Busza},
  \citenamefont {Rajagopal},\ and\ \citenamefont {van~der
  Schee}}]{Busza:2018rrf}%
  \BibitemOpen
  \bibfield  {author} {\bibinfo {author} {\bibfnamefont {W.}~\bibnamefont
  {Busza}}, \bibinfo {author} {\bibfnamefont {K.}~\bibnamefont {Rajagopal}},\
  and\ \bibinfo {author} {\bibfnamefont {W.}~\bibnamefont {van~der Schee}},\
  }\bibfield  {title} {\bibinfo {title} {{Heavy Ion Collisions: The Big
  Picture, and the Big Questions}},\ }\href
  {https://doi.org/10.1146/annurev-nucl-101917-020852} {\bibfield  {journal}
  {\bibinfo  {journal} {Ann. Rev. Nucl. Part. Sci.}\ }\textbf {\bibinfo
  {volume} {68}},\ \bibinfo {pages} {339} (\bibinfo {year} {2018})},\ \Eprint
  {https://arxiv.org/abs/1802.04801} {arXiv:1802.04801 [hep-ph]} \BibitemShut
  {NoStop}%
\bibitem [{\citenamefont {Dusling}\ and\ \citenamefont
  {Teaney}(2008)}]{Dusling:2007gi}%
  \BibitemOpen
  \bibfield  {author} {\bibinfo {author} {\bibfnamefont {K.}~\bibnamefont
  {Dusling}}\ and\ \bibinfo {author} {\bibfnamefont {D.}~\bibnamefont
  {Teaney}},\ }\bibfield  {title} {\bibinfo {title} {{Simulating elliptic flow
  with viscous hydrodynamics}},\ }\href
  {https://doi.org/10.1103/PhysRevC.77.034905} {\bibfield  {journal} {\bibinfo
  {journal} {Phys. Rev. C}\ }\textbf {\bibinfo {volume} {77}},\ \bibinfo
  {pages} {034905} (\bibinfo {year} {2008})},\ \Eprint
  {https://arxiv.org/abs/0710.5932} {arXiv:0710.5932 [nucl-th]} \BibitemShut
  {NoStop}%
\bibitem [{\citenamefont {Kurkela}\ \emph {et~al.}(2020)\citenamefont
  {Kurkela}, \citenamefont {van~der Schee}, \citenamefont {Wiedemann},\ and\
  \citenamefont {Wu}}]{Kurkela:2019set}%
  \BibitemOpen
  \bibfield  {author} {\bibinfo {author} {\bibfnamefont {A.}~\bibnamefont
  {Kurkela}}, \bibinfo {author} {\bibfnamefont {W.}~\bibnamefont {van~der
  Schee}}, \bibinfo {author} {\bibfnamefont {U.~A.}\ \bibnamefont
  {Wiedemann}},\ and\ \bibinfo {author} {\bibfnamefont {B.}~\bibnamefont
  {Wu}},\ }\bibfield  {title} {\bibinfo {title} {{Early- and Late-Time Behavior
  of Attractors in Heavy-Ion Collisions}},\ }\href
  {https://doi.org/10.1103/PhysRevLett.124.102301} {\bibfield  {journal}
  {\bibinfo  {journal} {Phys. Rev. Lett.}\ }\textbf {\bibinfo {volume} {124}},\
  \bibinfo {pages} {102301} (\bibinfo {year} {2020})},\ \Eprint
  {https://arxiv.org/abs/1907.08101} {arXiv:1907.08101 [hep-ph]} \BibitemShut
  {NoStop}%
\bibitem [{\citenamefont {Nijs}\ \emph {et~al.}(2021)\citenamefont {Nijs},
  \citenamefont {van~der Schee}, \citenamefont {G\"ursoy},\ and\ \citenamefont
  {Snellings}}]{Nijs:2020roc}%
  \BibitemOpen
  \bibfield  {author} {\bibinfo {author} {\bibfnamefont {G.}~\bibnamefont
  {Nijs}}, \bibinfo {author} {\bibfnamefont {W.}~\bibnamefont {van~der Schee}},
  \bibinfo {author} {\bibfnamefont {U.}~\bibnamefont {G\"ursoy}},\ and\
  \bibinfo {author} {\bibfnamefont {R.}~\bibnamefont {Snellings}},\ }\bibfield
  {title} {\bibinfo {title} {{Bayesian analysis of heavy ion collisions with
  the heavy ion computational framework Trajectum}},\ }\href
  {https://doi.org/10.1103/PhysRevC.103.054909} {\bibfield  {journal} {\bibinfo
   {journal} {Phys. Rev. C}\ }\textbf {\bibinfo {volume} {103}},\ \bibinfo
  {pages} {054909} (\bibinfo {year} {2021})},\ \Eprint
  {https://arxiv.org/abs/2010.15134} {arXiv:2010.15134 [nucl-th]} \BibitemShut
  {NoStop}%
\bibitem [{\citenamefont {van~der Schee}\ and\ \citenamefont
  {Schenke}(2015)}]{vanderSchee:2015rta}%
  \BibitemOpen
  \bibfield  {author} {\bibinfo {author} {\bibfnamefont {W.}~\bibnamefont
  {van~der Schee}}\ and\ \bibinfo {author} {\bibfnamefont {B.}~\bibnamefont
  {Schenke}},\ }\bibfield  {title} {\bibinfo {title} {{Rapidity dependence in
  holographic heavy ion collisions}},\ }\href
  {https://doi.org/10.1103/PhysRevC.92.064907} {\bibfield  {journal} {\bibinfo
  {journal} {Phys. Rev. C}\ }\textbf {\bibinfo {volume} {92}},\ \bibinfo
  {pages} {064907} (\bibinfo {year} {2015})},\ \Eprint
  {https://arxiv.org/abs/1507.08195} {arXiv:1507.08195 [nucl-th]} \BibitemShut
  {NoStop}%
\bibitem [{\citenamefont {Hansen}\ \emph {et~al.}(2022)\citenamefont {Hansen},
  \citenamefont {Obers}, \citenamefont {Oling},\ and\ \citenamefont
  {Sogaard}}]{Hansen:2021fxi}%
  \BibitemOpen
  \bibfield  {author} {\bibinfo {author} {\bibfnamefont {D.}~\bibnamefont
  {Hansen}}, \bibinfo {author} {\bibfnamefont {N.~A.}\ \bibnamefont {Obers}},
  \bibinfo {author} {\bibfnamefont {G.}~\bibnamefont {Oling}},\ and\ \bibinfo
  {author} {\bibfnamefont {B.~T.}\ \bibnamefont {Sogaard}},\ }\bibfield
  {title} {\bibinfo {title} {{Carroll Expansion of General Relativity}},\
  }\href {https://doi.org/10.21468/SciPostPhys.13.3.055} {\bibfield  {journal}
  {\bibinfo  {journal} {SciPost Phys.}\ }\textbf {\bibinfo {volume} {13}},\
  \bibinfo {pages} {055} (\bibinfo {year} {2022})},\ \Eprint
  {https://arxiv.org/abs/2112.12684} {arXiv:2112.12684 [hep-th]} \BibitemShut
  {NoStop}%
\bibitem [{\citenamefont {Loganayagam}(2008)}]{Loganayagam:2008is}%
  \BibitemOpen
  \bibfield  {author} {\bibinfo {author} {\bibfnamefont {R.}~\bibnamefont
  {Loganayagam}},\ }\bibfield  {title} {\bibinfo {title} {{Entropy Current in
  Conformal Hydrodynamics}},\ }\href
  {https://doi.org/10.1088/1126-6708/2008/05/087} {\bibfield  {journal}
  {\bibinfo  {journal} {JHEP}\ }\textbf {\bibinfo {volume} {05}},\ \bibinfo
  {pages} {087}},\ \Eprint {https://arxiv.org/abs/0801.3701} {arXiv:0801.3701
  [hep-th]} \BibitemShut {NoStop}%
\bibitem [{\citenamefont {Romatschke}(2010)}]{Romatschke:2009kr}%
  \BibitemOpen
  \bibfield  {author} {\bibinfo {author} {\bibfnamefont {P.}~\bibnamefont
  {Romatschke}},\ }\bibfield  {title} {\bibinfo {title} {{Relativistic Viscous
  Fluid Dynamics and Non-Equilibrium Entropy}},\ }\href
  {https://doi.org/10.1088/0264-9381/27/2/025006} {\bibfield  {journal}
  {\bibinfo  {journal} {Class. Quant. Grav.}\ }\textbf {\bibinfo {volume}
  {27}},\ \bibinfo {pages} {025006} (\bibinfo {year} {2010})},\ \Eprint
  {https://arxiv.org/abs/0906.4787} {arXiv:0906.4787 [hep-th]} \BibitemShut
  {NoStop}%
\bibitem [{\citenamefont
  {Bhattacharyya}(2014{\natexlab{a}})}]{Bhattacharyya:2013lha}%
  \BibitemOpen
  \bibfield  {author} {\bibinfo {author} {\bibfnamefont {S.}~\bibnamefont
  {Bhattacharyya}},\ }\bibfield  {title} {\bibinfo {title} {{Entropy current
  and equilibrium partition function in fluid dynamics}},\ }\href
  {https://doi.org/10.1007/JHEP08(2014)165} {\bibfield  {journal} {\bibinfo
  {journal} {JHEP}\ }\textbf {\bibinfo {volume} {08}},\ \bibinfo {pages}
  {165}},\ \Eprint {https://arxiv.org/abs/1312.0220} {arXiv:1312.0220 [hep-th]}
  \BibitemShut {NoStop}%
\bibitem [{\citenamefont
  {Bhattacharyya}(2014{\natexlab{b}})}]{Bhattacharyya:2014bha}%
  \BibitemOpen
  \bibfield  {author} {\bibinfo {author} {\bibfnamefont {S.}~\bibnamefont
  {Bhattacharyya}},\ }\bibfield  {title} {\bibinfo {title} {{Entropy Current
  from Partition Function: One Example}},\ }\href
  {https://doi.org/10.1007/JHEP07(2014)139} {\bibfield  {journal} {\bibinfo
  {journal} {JHEP}\ }\textbf {\bibinfo {volume} {07}},\ \bibinfo {pages}
  {139}},\ \Eprint {https://arxiv.org/abs/1403.7639} {arXiv:1403.7639 [hep-th]}
  \BibitemShut {NoStop}%
\bibitem [{\citenamefont {Bhattacharyya}\ \emph {et~al.}(2008)\citenamefont
  {Bhattacharyya}, \citenamefont {Hubeny}, \citenamefont {Loganayagam},
  \citenamefont {Mandal}, \citenamefont {Minwalla}, \citenamefont {Morita},
  \citenamefont {Rangamani},\ and\ \citenamefont
  {Reall}}]{Bhattacharyya:2008xc}%
  \BibitemOpen
  \bibfield  {author} {\bibinfo {author} {\bibfnamefont {S.}~\bibnamefont
  {Bhattacharyya}}, \bibinfo {author} {\bibfnamefont {V.~E.}\ \bibnamefont
  {Hubeny}}, \bibinfo {author} {\bibfnamefont {R.}~\bibnamefont {Loganayagam}},
  \bibinfo {author} {\bibfnamefont {G.}~\bibnamefont {Mandal}}, \bibinfo
  {author} {\bibfnamefont {S.}~\bibnamefont {Minwalla}}, \bibinfo {author}
  {\bibfnamefont {T.}~\bibnamefont {Morita}}, \bibinfo {author} {\bibfnamefont
  {M.}~\bibnamefont {Rangamani}},\ and\ \bibinfo {author} {\bibfnamefont
  {H.~S.}\ \bibnamefont {Reall}},\ }\bibfield  {title} {\bibinfo {title}
  {{Local Fluid Dynamical Entropy from Gravity}},\ }\href
  {https://doi.org/10.1088/1126-6708/2008/06/055} {\bibfield  {journal}
  {\bibinfo  {journal} {JHEP}\ }\textbf {\bibinfo {volume} {06}},\ \bibinfo
  {pages} {055}},\ \Eprint {https://arxiv.org/abs/0803.2526} {arXiv:0803.2526
  [hep-th]} \BibitemShut {NoStop}%
\bibitem [{\citenamefont {Hansen}\ \emph {et~al.}(2019)\citenamefont {Hansen},
  \citenamefont {Hartong},\ and\ \citenamefont {Obers}}]{Hansen:2019vqf}%
  \BibitemOpen
  \bibfield  {author} {\bibinfo {author} {\bibfnamefont {D.}~\bibnamefont
  {Hansen}}, \bibinfo {author} {\bibfnamefont {J.}~\bibnamefont {Hartong}},\
  and\ \bibinfo {author} {\bibfnamefont {N.~A.}\ \bibnamefont {Obers}},\
  }\bibfield  {title} {\bibinfo {title} {{Gravity between Newton and
  Einstein}},\ }\href {https://doi.org/10.1142/S0218271819440103} {\bibfield
  {journal} {\bibinfo  {journal} {Int. J. Mod. Phys. D}\ }\textbf {\bibinfo
  {volume} {28}},\ \bibinfo {pages} {1944010} (\bibinfo {year} {2019})},\
  \Eprint {https://arxiv.org/abs/1904.05706} {arXiv:1904.05706 [gr-qc]}
  \BibitemShut {NoStop}%
\bibitem [{\citenamefont {Bergshoeff}\ \emph {et~al.}(2019)\citenamefont
  {Bergshoeff}, \citenamefont {Izquierdo}, \citenamefont {Ort\'\i{}n},\ and\
  \citenamefont {Romano}}]{Bergshoeff:2019ctr}%
  \BibitemOpen
  \bibfield  {author} {\bibinfo {author} {\bibfnamefont {E.}~\bibnamefont
  {Bergshoeff}}, \bibinfo {author} {\bibfnamefont {J.~M.}\ \bibnamefont
  {Izquierdo}}, \bibinfo {author} {\bibfnamefont {T.}~\bibnamefont
  {Ort\'\i{}n}},\ and\ \bibinfo {author} {\bibfnamefont {L.}~\bibnamefont
  {Romano}},\ }\bibfield  {title} {\bibinfo {title} {{Lie Algebra Expansions
  and Actions for Non-Relativistic Gravity}},\ }\href
  {https://doi.org/10.1007/JHEP08(2019)048} {\bibfield  {journal} {\bibinfo
  {journal} {JHEP}\ }\textbf {\bibinfo {volume} {08}},\ \bibinfo {pages}
  {048}},\ \Eprint {https://arxiv.org/abs/1904.08304} {arXiv:1904.08304
  [hep-th]} \BibitemShut {NoStop}%
\bibitem [{\citenamefont {Gomis}\ \emph {et~al.}(2019)\citenamefont {Gomis},
  \citenamefont {Kleinschmidt},\ and\ \citenamefont
  {Palmkvist}}]{Gomis:2019fdh}%
  \BibitemOpen
  \bibfield  {author} {\bibinfo {author} {\bibfnamefont {J.}~\bibnamefont
  {Gomis}}, \bibinfo {author} {\bibfnamefont {A.}~\bibnamefont
  {Kleinschmidt}},\ and\ \bibinfo {author} {\bibfnamefont {J.}~\bibnamefont
  {Palmkvist}},\ }\bibfield  {title} {\bibinfo {title} {{Galilean free Lie
  algebras}},\ }\href {https://doi.org/10.1007/JHEP09(2019)109} {\bibfield
  {journal} {\bibinfo  {journal} {JHEP}\ }\textbf {\bibinfo {volume} {09}},\
  \bibinfo {pages} {109}},\ \Eprint {https://arxiv.org/abs/1907.00410}
  {arXiv:1907.00410 [hep-th]} \BibitemShut {NoStop}%
\bibitem [{\citenamefont {Gomis}\ \emph {et~al.}(2020)\citenamefont {Gomis},
  \citenamefont {Kleinschmidt}, \citenamefont {Palmkvist},\ and\ \citenamefont
  {Salgado-Rebolledo}}]{Gomis:2019sqv}%
  \BibitemOpen
  \bibfield  {author} {\bibinfo {author} {\bibfnamefont {J.}~\bibnamefont
  {Gomis}}, \bibinfo {author} {\bibfnamefont {A.}~\bibnamefont {Kleinschmidt}},
  \bibinfo {author} {\bibfnamefont {J.}~\bibnamefont {Palmkvist}},\ and\
  \bibinfo {author} {\bibfnamefont {P.}~\bibnamefont {Salgado-Rebolledo}},\
  }\bibfield  {title} {\bibinfo {title} {{Symmetries of post-Galilean
  expansions}},\ }\href {https://doi.org/10.1103/PhysRevLett.124.081602}
  {\bibfield  {journal} {\bibinfo  {journal} {Phys. Rev. Lett.}\ }\textbf
  {\bibinfo {volume} {124}},\ \bibinfo {pages} {081602} (\bibinfo {year}
  {2020})},\ \Eprint {https://arxiv.org/abs/1910.13560} {arXiv:1910.13560
  [hep-th]} \BibitemShut {NoStop}%
\bibitem [{\citenamefont {Hansen}\ \emph {et~al.}(2020)\citenamefont {Hansen},
  \citenamefont {Hartong},\ and\ \citenamefont {Obers}}]{Hansen:2020pqs}%
  \BibitemOpen
  \bibfield  {author} {\bibinfo {author} {\bibfnamefont {D.}~\bibnamefont
  {Hansen}}, \bibinfo {author} {\bibfnamefont {J.}~\bibnamefont {Hartong}},\
  and\ \bibinfo {author} {\bibfnamefont {N.~A.}\ \bibnamefont {Obers}},\
  }\bibfield  {title} {\bibinfo {title} {{Non-Relativistic Gravity and its
  Coupling to Matter}},\ }\href {https://doi.org/10.1007/JHEP06(2020)145}
  {\bibfield  {journal} {\bibinfo  {journal} {JHEP}\ }\textbf {\bibinfo
  {volume} {06}},\ \bibinfo {pages} {145}},\ \Eprint
  {https://arxiv.org/abs/2001.10277} {arXiv:2001.10277 [gr-qc]} \BibitemShut
  {NoStop}%
\bibitem [{\citenamefont {Hartong}(2015)}]{Hartong:2015xda}%
  \BibitemOpen
  \bibfield  {author} {\bibinfo {author} {\bibfnamefont {J.}~\bibnamefont
  {Hartong}},\ }\bibfield  {title} {\bibinfo {title} {{Gauging the Carroll
  Algebra and Ultra-Relativistic Gravity}},\ }\href
  {https://doi.org/10.1007/JHEP08(2015)069} {\bibfield  {journal} {\bibinfo
  {journal} {JHEP}\ }\textbf {\bibinfo {volume} {08}},\ \bibinfo {pages}
  {069}},\ \Eprint {https://arxiv.org/abs/1505.05011} {arXiv:1505.05011
  [hep-th]} \BibitemShut {NoStop}%
\bibitem [{\citenamefont {Kovtun}\ and\ \citenamefont
  {Shukla}(2018)}]{Kovtun:2018dvd}%
  \BibitemOpen
  \bibfield  {author} {\bibinfo {author} {\bibfnamefont {P.}~\bibnamefont
  {Kovtun}}\ and\ \bibinfo {author} {\bibfnamefont {A.}~\bibnamefont
  {Shukla}},\ }\bibfield  {title} {\bibinfo {title} {{Kubo formulas for
  thermodynamic transport coefficients}},\ }\href
  {https://doi.org/10.1007/JHEP10(2018)007} {\bibfield  {journal} {\bibinfo
  {journal} {JHEP}\ }\textbf {\bibinfo {volume} {10}},\ \bibinfo {pages}
  {007}},\ \Eprint {https://arxiv.org/abs/1806.05774} {arXiv:1806.05774
  [hep-th]} \BibitemShut {NoStop}%
\bibitem [{\citenamefont {Kovtun}\ and\ \citenamefont
  {Shukla}(2020)}]{Kovtun:2019wjz}%
  \BibitemOpen
  \bibfield  {author} {\bibinfo {author} {\bibfnamefont {P.}~\bibnamefont
  {Kovtun}}\ and\ \bibinfo {author} {\bibfnamefont {A.}~\bibnamefont
  {Shukla}},\ }\bibfield  {title} {\bibinfo {title}
  {{Einstein\textquoteright{}s equations in matter}},\ }\href
  {https://doi.org/10.1103/PhysRevD.101.104051} {\bibfield  {journal} {\bibinfo
   {journal} {Phys. Rev. D}\ }\textbf {\bibinfo {volume} {101}},\ \bibinfo
  {pages} {104051} (\bibinfo {year} {2020})},\ \Eprint
  {https://arxiv.org/abs/1907.04976} {arXiv:1907.04976 [gr-qc]} \BibitemShut
  {NoStop}%
\bibitem [{\citenamefont {Shukla}(2019)}]{Shukla:2019shf}%
  \BibitemOpen
  \bibfield  {author} {\bibinfo {author} {\bibfnamefont {A.}~\bibnamefont
  {Shukla}},\ }\bibfield  {title} {\bibinfo {title} {{Equilibrium thermodynamic
  susceptibilities for a dense degenerate Dirac field}},\ }\href
  {https://doi.org/10.1103/PhysRevD.100.096010} {\bibfield  {journal} {\bibinfo
   {journal} {Phys. Rev. D}\ }\textbf {\bibinfo {volume} {100}},\ \bibinfo
  {pages} {096010} (\bibinfo {year} {2019})},\ \Eprint
  {https://arxiv.org/abs/1906.02334} {arXiv:1906.02334 [hep-th]} \BibitemShut
  {NoStop}%
\bibitem [{\citenamefont {Grieninger}\ and\ \citenamefont
  {Shukla}(2021)}]{Grieninger:2021rxd}%
  \BibitemOpen
  \bibfield  {author} {\bibinfo {author} {\bibfnamefont {S.}~\bibnamefont
  {Grieninger}}\ and\ \bibinfo {author} {\bibfnamefont {A.}~\bibnamefont
  {Shukla}},\ }\bibfield  {title} {\bibinfo {title} {{Second order equilibrium
  transport in strongly coupled $ \mathcal{N} $ = 4 supersymmetric SU(N$_{c}$)
  Yang-Mills plasma via holography}},\ }\href
  {https://doi.org/10.1007/JHEP08(2021)108} {\bibfield  {journal} {\bibinfo
  {journal} {JHEP}\ }\textbf {\bibinfo {volume} {08}},\ \bibinfo {pages}
  {108}},\ \Eprint {https://arxiv.org/abs/2105.08673} {arXiv:2105.08673
  [hep-th]} \BibitemShut {NoStop}%
\end{thebibliography}%
%%%%%%%%%%%%%%%%%%%%%%%%%%%%%%%%%%%

\end{document}